\documentclass[a4paper,fleqn,usenatbib]{mnras}

\usepackage[T1]{fontenc}
\usepackage{ae,aecompl}
\usepackage[normalem]{ulem}
\usepackage{widetext}
\usepackage{flushend}
\usepackage{cuted}
\usepackage{multirow}
\usepackage{lipsum}

\usepackage{graphicx}	
\usepackage{amssymb}	

\usepackage{epsfig}
\usepackage{amsmath,bm}
\usepackage{color}

\title[Bispectrum from the DR12 BOSS galaxies]{The clustering of galaxies in the SDSS-III Baryon Oscillation Spectroscopic Survey: RSD measurement from the power spectrum and bispectrum of the DR12 BOSS galaxies}

\author[H. Gil-Mar\'in et al.]{H\'ector Gil-Mar\'in$^{1,2,3}$\thanks{hector.gilmarin@lpnhe.in2p3.fr},  Will J. Percival$^{3}$, Licia Verde$^{4,5,6}$, Joel R. Brownstein$^7$, \and Chia-Hsun Chuang$^{8,9}$, Francisco-Shu Kitaura$^{9,10}$, Sergio A. Rodr\'iguez-Torres$^{8,11}$, \and Matthew D. Olmstead$^{12}$\\
$^1$ Sorbonne Universit\'es, Institut Lagrange de Paris (ILP), 98 bis Boulevard Arago, 75014 Paris, France \\
$^2$ Laboratoire de Physique Nucléaire et de Hautes Energies, Universit\'e Pierre et Marie Curie, 4 Place Jussieu, 75005 Paris, France \\
 $^{3}$ Institute of Cosmology \& Gravitation, University of Portsmouth, Dennis Sciama Building, Portsmouth PO1 3FX, UK\\ 
 $^{4}$ ICREA \& Institut de Ci\`encies del Cosmos,  Universitat  de  Barcelona  (IEEC-UB),  Mart\'i  i  Franqu\`es,  1,  Barcelona  08028,
Spain \\
 $^{5}$ Institute of Theoretical Astrophysics, University of Oslo, Oslo 0315, Norway\\ 
 $^{6}$ Radcliffe Institute for Advanced Study, Harvard University, MA 02138, USA  \\
$^{7}$Department of Physics and Astronomy, University of Utah, 115 S 1400 E, Salt Lake City, UT 84112, USA \\
$^{8}$ Instituto de F\'{\i}sica Te\'orica, (UAM/CSIC), Universidad Aut\'onoma de Madrid, Cantoblanco, E-28049 Madrid, Spain \\
$^{9}$ Leibniz-Institut f\"ur Astrophysik Potsdam (AIP), An der Sternwarte 16, 14482 Potsdam, Germany\\
$^{10}$ Lawrence Berkeley National Lab \& Department of Physics and Astronomy, UC Berkeley, Berkeley CA 94720, USA\\
  $^{11}$ Campus of International Excellence UAM+CSIC \& Departamento de F\'isica Te\'orica, UAM, Cantoblanco, E-28049 Madrid, Spain\\
 $^{12}$  Department of Chemistry and Physics, King's College, 133 North River St, Wilkes Barre, PA 18711, USA }
\date{Accepted XXX. Received YYY; in original form \today}

\pubyear{2016}

\begin{document}
\label{firstpage}
\pagerange{\pageref{firstpage}--\pageref{lastpage}}
\maketitle
\begin{abstract}
We measure and analyse the bispectrum of the final, Data Release 12, galaxy sample provided by the Baryon Oscillation Spectroscopic Survey, splitting by selection algorithm into LOWZ and CMASS galaxies. The LOWZ sample contains 361\,762 galaxies with an effective redshift of $z_{\rm LOWZ}=0.32$, and the CMASS sample 777\,202 galaxies with an effective redshift of $z_{\rm CMASS}=0.57$.  Combining the power spectrum, measured relative to the line-of-sight, with the spherically averaged bispectrum, we are able to constrain the product of the  growth of structure parameter, $f$, and the amplitude of dark matter density fluctuations, $\sigma_8$, along with the  geometric Alcock-Paczynski parameters, the product of the Hubble constant and the comoving sound horizon at the baryon drag epoch, $H(z)r_s(z_d)$, and the angular distance parameter divided by the sound horizon, $D_A(z)/r_s(z_d)$. After combining pre-reconstruction RSD analyses of the power spectrum monopole, quadrupole and bispectrum monopole; with post-reconstruction analysis of the BAO power spectrum monopole and quadrupole, we find $f(z_{\rm LOWZ})\sigma_8(z_{\rm LOWZ})=0.427\pm 0.056$, $D_A(z_{\rm LOWZ})/r_s(z_d)=6.60 \pm 0.13$, $H(z_{\rm LOWZ})r_s(z_d)=(11.55\pm 0.38)10^3\,{\rm kms}^{-1}$ for the LOWZ sample, and $f(z_{\rm CMASS})\sigma_8(z_{\rm CMASS})=0.426\pm 0.029$, $D_A(z_{\rm CMASS})/r_s(z_d)=9.39 \pm 0.10$, $H(z_{\rm CMASS})r_s(z_d)=(14.02\pm 0.22)10^3\,{\rm kms}^{-1}$ for the CMASS sample. We find general agreement with previous BOSS DR11 and DR12 measurements. Combining our dataset with {\it Planck15} we perform a null test of General Relativity (GR) through the $\gamma$-parametrisation finding $\gamma=0.733^{+0.068}_{-0.069}$, which is $\sim2.7\sigma$ away from the GR predictions. 
\end{abstract}

\quad

\section{Introduction}

The projected distribution of galaxies on large scales is a key observable for understanding the matter and energy content of the Universe, as well as for explaining the laws of gravity at scales $\geq100\,{\rm Mpc}$. Cosmic Microwave Background (CMB) observations suggest that the primordial density fluctuations follow a Gaussian random field. Consequently, the 2-point correlation function and its Fourier counterpart, the power spectrum, are used to characterise the spacial distribution of galaxies that grow from these fluctuations. From a galaxy survey, correlation function measurements are usually based on the pair-counting methodology of \citet{Landay:Szalay}, while power spectrum measurements use the methodology introduced by \citet{FKP}, and recently extended to measure moments around the line-of-sight, making the pairwise plane-parallel assumption \citep{Bianchietal:2015,scoccimarro15}. The detection of the baryon signature in the clustering from the 2dFGRS \citep{Percivaletal:2001,Coleetal:2005} and SDSS \citep{Eis05} surveys provided the impetus to consider using the projected position of the Baryon Acoustic Oscillations (BAO) from these 2-point measurements to extract geometrical information. The large-scale and sharpness of the BAO features means that measurements are robust to galaxy bias, the relationship between the galaxies and the matter field \citep{Growth_Peebles:1977,PH85,BBKS86,Fry86,Fry_Gazta:1993,Fry94}. Information about the growth of structure comes from the Redshift Space Distortions (RSD) \citep{Kaiser:1987,Hamilton:1992}, which alter the clustering signal along the line of sight.

The evolution of gravitational instabilities amplifies the initial perturbations and generates a non-Gaussian signal in the distribution of galaxies. Therefore, in order to fully extract the information contained in the galaxy field, we need to additionally measure statistics of higher order than the power spectrum. The 3-point correlation function, and its Fourier counter part, the bispectrum, add significant additional information to state-of-art analyses based on the 2-point correlation function and power spectrum. 

Historically, the first measurements of the 3-point correlation function (3PCF) and bispectrum of an observed galaxy sample were performed by \cite{Peebles_Groth:1975,Growth_Peebles:1977} and \cite{Fry_Seldner:1982}; followed by its cosmological interpretation  performed by \cite{Fry:1984} who, using cosmological perturbation theory showed how the bispectrum shape and amplitude were affected by galaxy bias. At that time the effects that galaxy bias and RSD have on the galaxy bispectrum signal were still not well understood. It was not until the late 1990s when perturbative models were developed \citep{MVH97,Scoccimarroetal:1998,HMV98,VHMM98, Scoccetal99,Scocc00} that enabled the galaxy bispectrum to be used for accurate cosmological predictions. At the beginning of 2000s, many 3PCF and bispectrum analyses benefited from these works and the creation of large galaxy redshift surveys (2df Galaxy Redshift Survey: \citealt{Verdeetal:2002}, IRAS redshift catalogue: \citealt{Scoccimarroetal:2001}, PSCz galaxy redshift survey: \citealt{Feldmanetal:2001}). The progress in the creation of galaxy surveys, and the development of theoretical models and analyses methods has continued, with recent measurements and interpretations including: 3PCF of WiggleZ \citep{Marinetal:2013}; bispectrum and 3PCF of DR11 CMASS BOSS  \citep{hector_bispectrum1,hector_bispectrum2,Guoetal:2015} and 3PCF of DR12 CMASS BOSS \citep{Slepianetal:2015,Slepianetal:2016}.

In addition to analyses designed to measure the standard RSD and bias signal, the bispectrum and the 3-point correlation function have been proposed as key statistics to measure potential deviations from GR \citep{Borisovetal:2009,Bernardeau_Brax:2011,HGMfR,Bartoloetal:2013,Bellinietal:2015,Sabiuetal:2016}  and to put constrains on the strength of the primordial non-Gaussian signal, for example through $f_{\rm  NL}$ \citep{Ganguietal:1994,Fry_Scherrer:1994,Verdeetal:2000,Scoccimarroetal:2004,Sefusatti_Komatsu:2007,Sefusatti:2009,Sefusattietal:2010,Sefusattietal:2012,Scoccimarroetal:2012,matteo,Tellarinietal:2016,Wellingetal:2016}.  Although such phenomena also have an impact on the other statistics, the bispectrum is especially sensitive to modifications of GR and primordial non-Gaussian features, as it is essentially a non-linear quantity, even at large scales. 

In previous papers \citep{hector_bispectrum1,hector_bispectrum2}, we computed the angle-averaged bispectrum of the DR11 CMASS BOSS galaxies and analysed it, in combination with the power spectrum monopole, to measure the galaxy bias and constrain the growth of structure. In this paper, we apply a new measurement pipeline to the data from the final Data Release 12, improving the analysis and reducing the systematics and statistical errors of the measured cosmological parameters. The differences between the DR11 and DR12 samples are small, and the main difference between the previous and the present results are due to the new analysis procedure. Here, we list the main points we have improved:
\begin{itemize}
\item We include bispectrum measurements based on more triangular shapes. In the DR11 analysis we only included triangles with $k_1/k_2=1,\,2$, whereas now all the triangular shapes are included. By using all the shapes we gain statistical signal.  
\item  We perform the analysis using a full-covariance of the power spectrum and bispectrum. In the DR11 analysis only the diagonal terms of the covariance were used and the error-bars of the parameters were inferred from the dispersion of similar measurements recovered from mock catalogues. In this paper we use 2048 galaxy mocks to estimate the full covariance of the bispectrum and power spectrum measurements, and hence the error-bars of all measured parameters. 
\item We combine the bispectrum with the power spectrum monopole and quadrupole. This allows us to constrain $f\sigma_8$, and also to break the degeneracy between $f$ and $\sigma_8$, estimating these two parameters separately.  
\item We allow for Alcock-Paczynski geometrical distortions in the analysis. This allows us to constrain not only $f$, and $\sigma_8$, but also $D_A$ and $H(z)$. 
\item We measure the signal in both LOWZ and CMASS samples. Therefore we provide measurements of the clustering in two redshift bins. 
\end{itemize}
Among these changes the more relevant ones are the inclusion of all triangular shapes above a certain scale and the full fit to the monopole and quadrupole power spectrum along with the bispectrum monopole.   

The paper is organised as follow. In \S\ref{sec:data} we briefly present the data and mocks used for the analysis. In \S\ref{sec:algorithm} we describe the algorithm used for measuring the bispectrum in both data and mocks. In \S\ref{sec:measurement} we present the measurements of the bispectrum.  In \S\ref{sec:model} we describe the modelling used for extracting cosmological information from the measured quantities. In \S\ref{sec:covariances} we display the covariance matrices used for the analysis. In \S\ref{sec:tests} we present the systematic tests performed using the galaxy mocks and $N$-body simulations. In \S\ref{sec:results} we present the final results and compare them with results from other works. Additionally, we also show the outcome of combining the results from this paper with complementary analyses based on the same dataset. Finally in \S\ref{sec:conclusions} we present the conclusion of this work. 

\section{Data and mocks}\label{sec:data}

\subsection{The SDSS III BOSS data}

As part of the Sloan Digital Sky Survey III \citep{EISetal:2011} the Baryon Oscillation Spectroscopic Survey (BOSS) \citep{Dawsonetal:2012}  measured spectroscopic redshifts \citep{Bolton12,Smee13} for more than 1 million galaxies and over 200\,000 quasars. The galaxies were selected from multi-colour SDSS imaging \citep{Fukugitaetal:1996,Gunnetal:1998,Smithetal:2002,Gunnetal:2006,Doietal:2010} focussing on the redshift range of $0.15\leq z \leq0.70$. The galaxy survey used two primary target algorithms, selecting samples called LOWZ, with 361\,762 galaxies in the final data release DR12 \citep{dr12} between $0.15\leq z \leq0.43$ and CMASS, with 777\,202 galaxies in DR12 between $0.43\leq z \leq 0.70$. The full targeting algorithms used and the galaxy and random catalogues are presented in \citet{Reidetal:2015}. The samples jointly cover a cosmic volume corresponding to $V_{\rm eff}=7.4\,{\rm Gpc}^3$ (for a fiducial value of $h=0.677$ assumed), with a number density of galaxies that ensures that the shot noise does not dominate at BAO scales at all redshifts.

In order to correct for observational artefacts in the catalogues, the CMASS and LOWZ galaxy samples include a set of weights designed to counteract the effects of redshift failure $w_{\rm rf}$, fibre collision $w_{\rm fc}$, and target density variations $w_{\rm sys}$ (CMASS only) combing variations in seeing and stellar density \citep{Rossetal:2012,Andersonetal:2014,Reidetal:2015}. Hence, each observed galaxy contributes to our estimate of the underlying true galaxy density field by
\begin{equation}
\label{eq:wc}w_c\equiv w_{\rm sys}(w_{\rm rf}+w_{\rm fc}-1).
\end{equation}
The redshift failure weights account for galaxies that have been observed, but whose redshifts have not been measured: nearby galaxies, which are approximated as being ``equivalent'' in terms of properties, are up-weighted to remove any bias in the resulting field. The fibre collision weight similarly corrects for galaxies that could not be observed as there was another target within $62''$, a physical limitation of the spectrograph (see \cite{Rossetal:2012} for details). The systematic weight accounts for fluctuations in the target density caused by changes in the imaging observational efficiency. This effect is only corrected in the CMASS sample, which relies on deeper imaging data;  such a weight is not required for the brighter LOWZ sample \citep{Tojeiro14}.

Additionally, we adopt the standard weight to the over-density (i.e. to both galaxies and randoms) to optimise measurements across regions of high and low density, $w_{\rm FKP}({\bf r})=1/[1+n({\bf r})P_0]$, where $n$ is the mean number density of galaxies and $P_0$ is the amplitude of the galaxy power spectrum at the scale where the error is minimised. We assume $P_0=10^4\,h^{-3}\,{\rm Mpc}^3$, which corresponds to the amplitude of the power spectrum at scales $k\sim0.10\,h\,{\rm Mpc}^{-1}$ \citep{Reidetal:2015}.

\subsection{The mock survey catalogues}

Mock  catalogs are a fundamental component of a modern cosmological analysis of galaxy survey data. They are used to test the modelling of the large-scale structure and they help to uncover and quantify potential systematic errors induced by potentially incomplete or imperfect modelling. Most of the relevant large-scale physics is captured by approximate methods such as second-order Lagrangian perturbation theory (2LPT) \citep{Scoccimarro_Sheth:2002,Maneraetal:2013}, or augmented Lagrangian perturbation theory (ALPT) \citep{Kitaura_Hess:2013}, so we do not necessarily need to base mock catalogues on full $N$-body cosmological simulations; small numbers of $N$-body simulations can instead be used to calibrate a more efficient scheme. 

In this paper we use 2048 realisations of the  MultiDark-Patchy BOSS DR12 mocks\footnote{http://data.sdss3.org/datamodel/index-files.html} (hereafter \textsc{MD-Patchy} mocks) \citep{Kitauraetal:2015} for computing the covariance matrices and testing the potential systematics of the modelling used. These mocks incorporate observational effects including the survey selection window, veto mask and fibre collisions. Since the covariance matrix is estimated from a set of mocks, its inverse is biased due to the limited number of realisations. We account this effect by applying the correction proposed by \cite{Hartlap07}. Details on how the \textsc{MD-Patchy} mocks are produced can be found in \cite{Kitauraetal:2015}. The underling cosmology for these mocks has been chosen to be ${\boldsymbol \Omega}^{\rm MD-\textsc{Patchy}}=(\Omega_\Lambda,\, \Omega_m,\,\Omega_b,\,\sigma_8,\,h,\,n_s)=(0.692885,\,0.307115,\,0.048,\,0.8288,\,0.6777,\,0.96)$, being very close to the best-fitting values of the last release of {\it Planck15} \citep{Planck_cosmology15}. For this cosmological model, the sound horizon at drag redshift is $r_s(z_d) = 147.66\,{\rm Mpc}$.

Additionally, we also use the set of 1000 realisations of Quick-Particle-Mesh mocks (hereafter \textsc{qpm}) \citep{QPMmocks} to test the potential systematics introduced by the collided galaxies in the bispectrum signal (Appendix~\ref{appendixb}). As for  \textsc{MD-Patchy} mocks, \textsc{qpm} mocks  incorporate observational effects including the survey selection window, veto mask and fibre collisions. The \textsc{qpm} mocks are based on low-resolution particle mesh simulations that accurately reproduce the large-scale dark matter density field, in combination with the halo occupation distribution technique (HOD) to populate the resolved haloes with galaxies. For the \textsc{qpm} mocks, the  snapshots are at the effective redshift of, $z_{\rm eff}=0.55$ for CMASS and $z_{\rm eff}=0.40$ for LOWZ.  The underling cosmology for these mocks has been chosen to be ${\boldsymbol \Omega}^{\rm \textsc{qpm}}=(\Omega_\Lambda,\, \Omega_m,\,\Omega_b,\,\sigma_8,\,h,\,n_s)=(0.71,\,0.29,\,0.0458,\,0.80,\,0.7,\,0.97)$. For this cosmological model the sound horizon at drag redshift is $r_s(z_d) = 147.13\,{\rm Mpc}$.
As we reported in \cite{gil-marin15_rsd}, \textsc{qpm} mocks do not represent sufficiently accurately the effect of redshift space distortions on the power spectrum multipoles, and we consequently do not use them to test theoretical models. However, the fibre collision effects are more realistic than in the  \textsc{MD-Patchy} mocks, which underestimate their effect in the CMASS sample (see the discussion in appendix B of  \citealt{gil-marin15_rsd}). Therefore, we use \textsc{qpm} mocks instead of \textsc{MD-Patchy} mocks to test the effects of this. 

We also analyse the 20 catalogues of dark matter haloes, each drawn from an $N$-body simulation, that were used in previous works \citep{Whiteetal:2011,Reid_White:2011} to test the validity of the bispectrum theoretical model \citep{hector_bispectrum0,hector_bispectrum1}. These catalogues do not incorporate any survey selection function nor any observational effect. In contrast to the galaxy mocks they are drawn from full $N$-body simulations, and are therefore expected to reproduce more realistically the expected bispectrum signal in real and redshift space for dark matter and for halos. Along with the MD-\textsc{patchy} mocks, we use these simulations to test the validity of the theoretical model in \S\ref{sec:tests}.  Full details on these catalogues can be found in section 3.2 of \cite{hector_bispectrum0}. To summarise, each set of haloes was drawn from a periodic-box simulation with side length $1.5\,{\rm Gpc}h^{-1}$ and particle mass $m_p= 7.6\times10^{10}\,M_\odot h^{-1}$. The halo catalogues are generated by the Friends of Friends algorithm \citep{Davisetal:1985} with a linking length of 0.168
times the mean inter-particle spacing.  The underling cosmology for the $N$-body haloes is ${\boldsymbol \Omega}^{\rm N-body Haloes}=(\Omega_\Lambda,\, \Omega_m,\,\Omega_b,\,\sigma_8,\,h,\,n_s)=(0.726,\,0.274,\,0.0457,\,0.80,\,0.7,\,0.95)$.

\subsection{Fiducial Cosmology}\label{sec:fiducial_cosmology}

In order to compress the information from a galaxy survey into a usable form, we work in comoving space, and transform the observed angular separations and redshifts into comoving coordinates, which is a model dependent transformation. Therefore we need to assume {\it a priori} a cosmological model, which we call the fiducial model. For a flat $\Lambda$CDM cosmology, this corresponds to specifying a value for the matter density of the Universe, $\Omega_m$. The modelling of the power spectrum and bispectrum used in this paper take into account that a fiducial cosmology was used to make this transformation. In particular, if the fiducial model does not match that to be compared with the data, we would expect to see a geometrical stretching of the intrinsic clustering predicted by the model, corresponding to Alcock-Paczynski dilations (see \S5.3 for further explanation). On the other hand, in our analysis the covariance is computed at a fixed cosmology. However, as long as we do not explore models very disfavoured from the data, the choice of cosmology for the covariance is expected to have a negligible impact \citep{WP15}. 

We have opted to analyse both mocks and data using the same cosmological model. The fiducial value assumed for this is $\Omega^{\rm fid}_m=0.31$, which is in agreement with the last {\it Planck15} release \citep{Planck_cosmology15}. As a consequence we will analyse the mocks using a value of $\Omega_m$ that is different than their true values. As discussed above, this will introduce an extra geometrical distortion, which will be accounted for by the Alcock-Paczynski (AP) scaling relations  presented in \S\ref{sec:AP}. The rest of the cosmological parameters in the fiducial cosmology are ${\boldsymbol \Omega}^{\rm fid}\equiv(\Omega_\Lambda^{\rm fid},\, \Omega_m^{\rm fid},\,\Omega_b^{\rm fid},\,\sigma_8^{\rm fid},\,h^{\rm fid},\,n_s^{\rm fid})=(0.69,\,0.31,\,0.049,\,0.8475,\,0.6711,\,0.9624)$.  For this cosmology the sound horizon at drag redshift is $r_s(z_d)=148.11\,{\rm Mpc}$. 

\section{Computing the Bispectrum}\label{sec:algorithm}

We start from the weighted field of density fluctuations, as used for previous measurements of the power spectrum and bispectrum \citep{FKP},
\begin{equation}
F_\lambda({\bf r})=\frac{w_{\rm FKP}({\bf r})}{I_\lambda^{1/2}}\left[ w_c({\bf r}) n({\bf r})-\alpha n_s({\bf r}) \right],
\label{F3equation}
\end{equation}
where $n$ and $n_s$ are, respectively, the observed number density of galaxies and the number density of objects in a synthetic catalogue. In the synthetic catalog, objects represent a Poisson sampled distribution of points with the same survey mask and radial selection function as the galaxies, but with no other cosmological correlation. The functions $w_{\rm FKP}$ and $w_c$ were defined in \S\ref{sec:data}. The factor $\alpha$ is the ratio between weighted number of observed galaxies over the weighted number of objects in the synthetic catalogues. The factor $I_\lambda$  normalises the amplitude of the observed power in accordance with its definition in a galaxy distribution with no survey selection,
\begin{equation}
I_\lambda \equiv \int d^3{\bf r}\, w_{\rm FKP}^\lambda\langle n w_c\rangle^\lambda (\bf r),
\end{equation}
where $\lambda=2$ for the power spectrum and $\lambda=3$ for the bispectrum, which ensures the correct normalisation for each statistic. The power spectrum monopole and quadrupole are estimated as described in \cite{gil-marin15_rsd,gil-marin15_bao,Bianchietal:2015}, which account for the effect of a varying line-of-sight in the quadrupole. 

We have developed a new bispectrum measurement pipeline, which is significantly faster than that used in \cite{hector_bispectrum1}, based on the estimator described in \cite{Sefusatti:2005,Baldaufetal:2015}. Here, we present a brief description for completeness. In this paper we focus on measuring the isotropic component of the bispectrum (i.e. the monopole). An extension of this estimator based the Yamamoto estimator \citep{Yamamotoetal:2006} for the quadrupole of the bispectrum would be possible (see \citealt{scoccimarro15}), although for simplicity (especially on the modelling) we do not consider this quantity here. 

The bispectrum estimator is defined as the angle-average of closed triangles defined by the $\bf k$-modes, ${\bf k}_1,\,{\bf k}_2,\,{\bf k}_3$,
\begin{eqnarray}
\nonumber &&\langle F_3({\bf k}_1)F_3({\bf k}_2) F_3({\bf k}_3)\rangle=\frac{k_f^3}{V_{123}}\int_{\mathcal{S}_1} d{\bf q}_1\, F_3({\bf q}_1)\\
\nonumber&\times&\int_{\mathcal{S}_2} d{\bf q}_2\, F_3({\bf q}_2)\int_{\mathcal{S}_3} d{\bf q}_3\, F_3({\bf q}_3) \delta^D({\bf q}_1+{\bf q}_2+{\bf q}_3),\\
\label{eq:bis1}
\end{eqnarray}
where, $F_3({\bf q})$ is the Fourier transform of Eq. \ref{F3equation} with $\lambda=3$, $k_f$ is the fundamental frequency, $k_f=2\pi/L_{\rm box}$, $L_{\rm box}$ the size of the box in which the galaxies are embedded and   $V_{123}$ is the number of fundamental triangles,
\begin{equation}
 V_{123}\equiv\int_{\mathcal{S}_1} d{\bf q}_1\, \int_{\mathcal{S}_2} d{\bf q}_2\, \int_{\mathcal{S}_3} d{\bf q}_3\, \delta^D({\bf q}_1+{\bf q}_2+{\bf q}_3),
\end{equation}
inside the shell defined by $\mathcal{S}_1$, $\mathcal{S}_2$ and $\mathcal{S}_3$, where $\mathcal{S}_i\equiv \mathcal{S}(k_i|\Delta k)$ is the $k$-region contained by $k_i-\Delta k/2 \leq k \leq k_i+ \Delta k/2$, given a $k$-bin, $\Delta k$. $\delta^D$ is the Dirac delta distribution that ensures the condition of only including closed triangles. 

Writing the Dirac delta as the exponential expression $\delta^D({\bf x})\equiv\int d{\bf y}\, e^{i{\bf x}\cdot{\bf y}}$, we can re-write Eq.~\ref{eq:bis1} as a separate product of Fourier Transforms,
\begin{equation}
 \langle F_3({\bf k}_1)F_3({\bf k}_2) F_3({\bf k}_3)\rangle=\frac{k_f^3}{V_{123}}\int d^3{\bf r}\, \mathcal{D}_{\mathcal{S}_1}({\bf r}) \mathcal{D}_{\mathcal{S}_2}({\bf r}) \mathcal{D}_{\mathcal{S}_3}({\bf r})
 \label{eq:bis2}
\end{equation}
where,
\begin{equation}
 \mathcal{D}_{\mathcal{S}_j}({\bf r})\equiv \int_{\mathcal{S}_j} d{\bf q}_j\, F_3({\bf q}_j)e^{i{\bf q}_j\cdot{\bf r}}.
\end{equation}
Without any lost of generality we propose $k_1\leq k_2 \leq k_3$.
The total number of non-equivalent triangular configurations is $\sim N_b^3/2$, where $N_b$ is the number of bins considered between the fundamental frequency, $k_f$, and the Nyquist frequency, $k_{\rm Ny}\equiv N_{\rm grid}^{1/3}/2k_f$. Here, $N_{\rm grid}$ is the total number of cartesian grid-cells chosen for performing the FT (Fourier Transform). In the computation of Eq. \ref{eq:bis2}, the bispectrum of a triplet $\{k_1,k_2,k_3\}$, requires only 3 FT and most triplets share one or two $k_i$-vectors, which means that they share as well the output of the FT of the shared vectors. Consequently, the estimator of Eq. \ref{eq:bis2}  is much faster than the naive implementation of Eq. \ref{eq:bis1}, used in \cite{hector_bispectrum1}. 

As the implementation of the power spectrum and bispectrum estimators is based on FTs, we are required to determine $F_\lambda({\bf r})$ (Eq.~\ref{F3equation}) on a cartesian grid, that serves as an input for the FTs. We use a random catalogue of number density of $\bar{n}_s({\bf r})=\alpha^{-1} \bar{n}({\bf r})$ with $\alpha^{-1}\simeq50$. We place the LOWZ and CMASS galaxy and random samples on $N_{\rm grid}=512^3$ grids, of box side $L_b=2300\,h^{-1}{\rm Mpc}$ for the LOWZ sample,  yielding to a $k_{\rm Ny}=0.70\,h{\rm Mpc}^{-1}$, and $L_b=3500\,h^{-1}{\rm Mpc}$  to fit the CMASS sample, yielding to $k_{\rm Ny}=0.46\,h{\rm Mpc}^{-1}$. This corresponds to a grid-cell resolution of $3.42\,h^{-1}{\rm Mpc}$ for the CMASS sample and $2.25\,h^{-1}{\rm Mpc}$ for the LOWZ sample. The fundamental frequencies are therefore $k_f=1.795\cdot10^{-3}\,h{\rm Mpc}^{-1}$ and $k_f=2.732\cdot10^{-3}\,h{\rm Mpc}^{-1}$ for the CMASS and LOWZ samples, respectively.  We have checked that for $k\leq 1/2 k_{\rm Ny}$, doubling the number of grid-cells per side, from $512$ to $1024$, produces a negligible change in the bispectrum, $\ll1\%$. This result indicates that using $512^3$ grid-cells provides sufficient resolution at the scales of interest. We apply the Cloud-in-Cells scheme (CiC) to associate galaxies to grid-cells and we apply the corresponding grid-deconvolution correction once the FT is performed \citep{Jing:2005}. We measure the power spectrum monopole and quadrupole modes using a logarithmic bin, as described in \cite{gil-marin15_rsd}. For the bispectrum we opt to bin in $k$-shells of 6$k_f$. By doing this we limit the total number of fundamental triangles to be $825$, up to $k_{\rm Ny}/2$. By reducing the size of the $k$-shells we would obtain more triangle shapes and potentially winning more signal (for instance, choosing a $k$-shells of 3$k_f$, we would obtain 6391 fundamental triangles). However, we want to keep the total number of bins significantly smaller than the total number of mocks (2048 realisations), so the covariance matrices estimated from the mocks are reliable. 

We limit the range of large scales as follows: for the power spectrum monopole and quadrupole we discard modes with $k<0.02\,h{\rm Mpc}^{-1}$ and $k<0.04\,h{\rm Mpc}^{-1}$, respectively. This reduces the impact of the large scale systematic weights on the analysis as described in appendix~B of \cite{gil-marin15_rsd}. In short, by discarding the large scale modes, we reduce the potential inaccuracies of the large scale systematic corrections to be less than $5\%$ of the signal. 
For the isotropic bispectrum signal we discard those triangles with at least one $k$-vector $k\leq0.03\,h{\rm Mpc}^{-1}$. As described in \cite{hector_bispectrum1}, the survey window geometry used in the analysis is not able to account for those shapes, in the sense that the actual effect of the window function selection is not well described by this approach. Although it would be possible to generalise the window survey treatment to incorporate those large scales triangles, we would also have to account for other artefacts, such as the impact of the large scale systematic weights on those triangles, expected to be larger than for the power spectrum, which is a difficult task that goes beyond the scope of this paper (see Appendix~\ref{appendixa} for further details).  

 The larger the maximum $k_i$   included in the cosmological analysis, the more triangles are used and therefore the smaller the statistical errors of the estimated parameters. However, small scales are poorly modelled in comparison to large scales because of dark matter non-linearities; and the effect of shot noise starts to be dominant. Because of this, we expect the systematic errors on the modelling to grow as the minimum scale decreases. Therefore, we empirically find a compromise between these two effects such that the systematic offset induced by the poorly modelled non-linear behaviour is smaller than the statistical error. To do so, we analyse the galaxy mocks and full $N$-body halo catalogues at different minimum scales and check that the best-fitting parameters of interest does not change significantly (compared to the statistical errors) as a function of this minimum scale. This is described in detail in \S\ref{sec:tests}.

\section{The Power Spectrum and Bispectrum measurements}\label{sec:measurement}

\begin{figure*}
\includegraphics[scale=0.3]{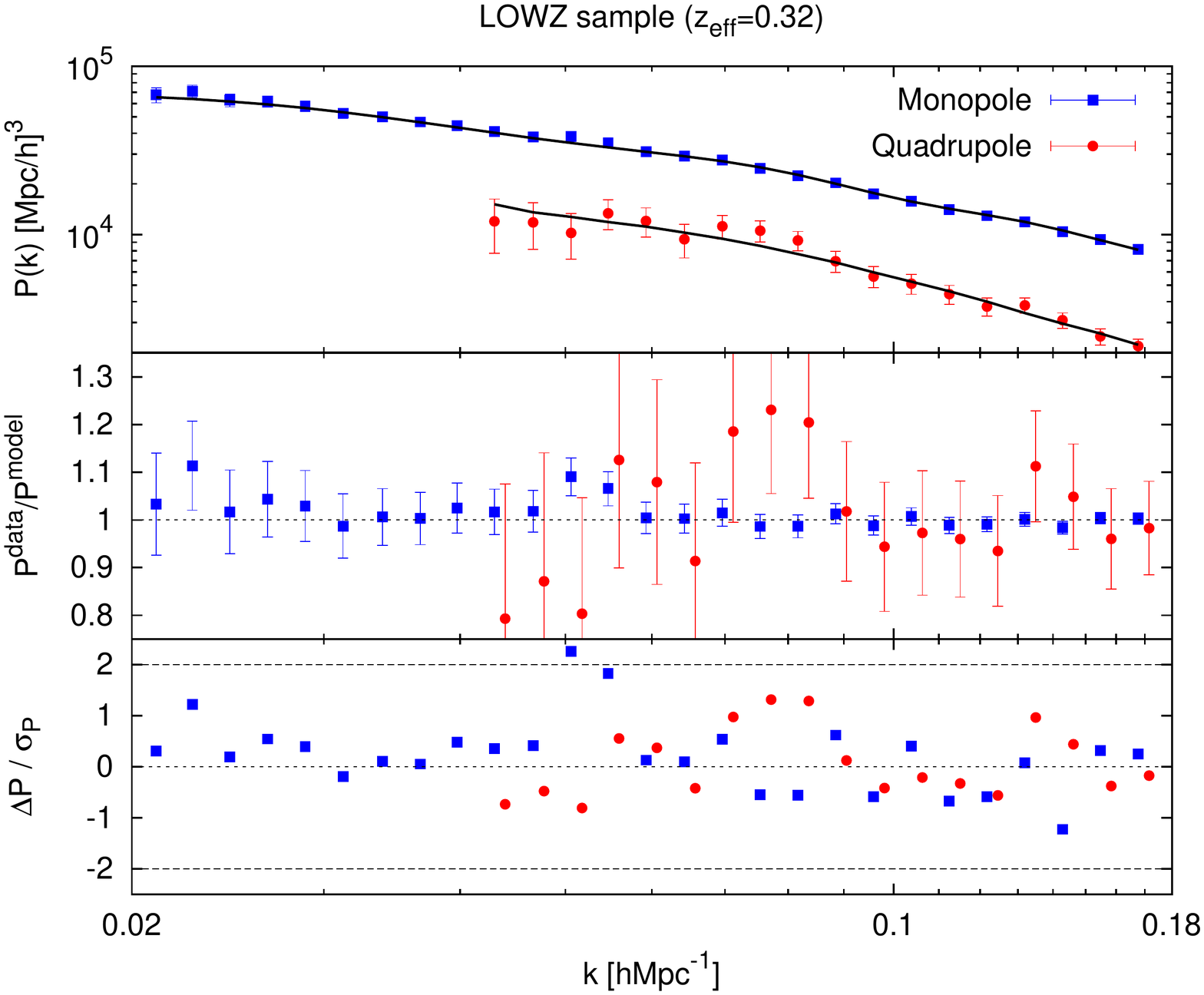}
\includegraphics[scale=0.3]{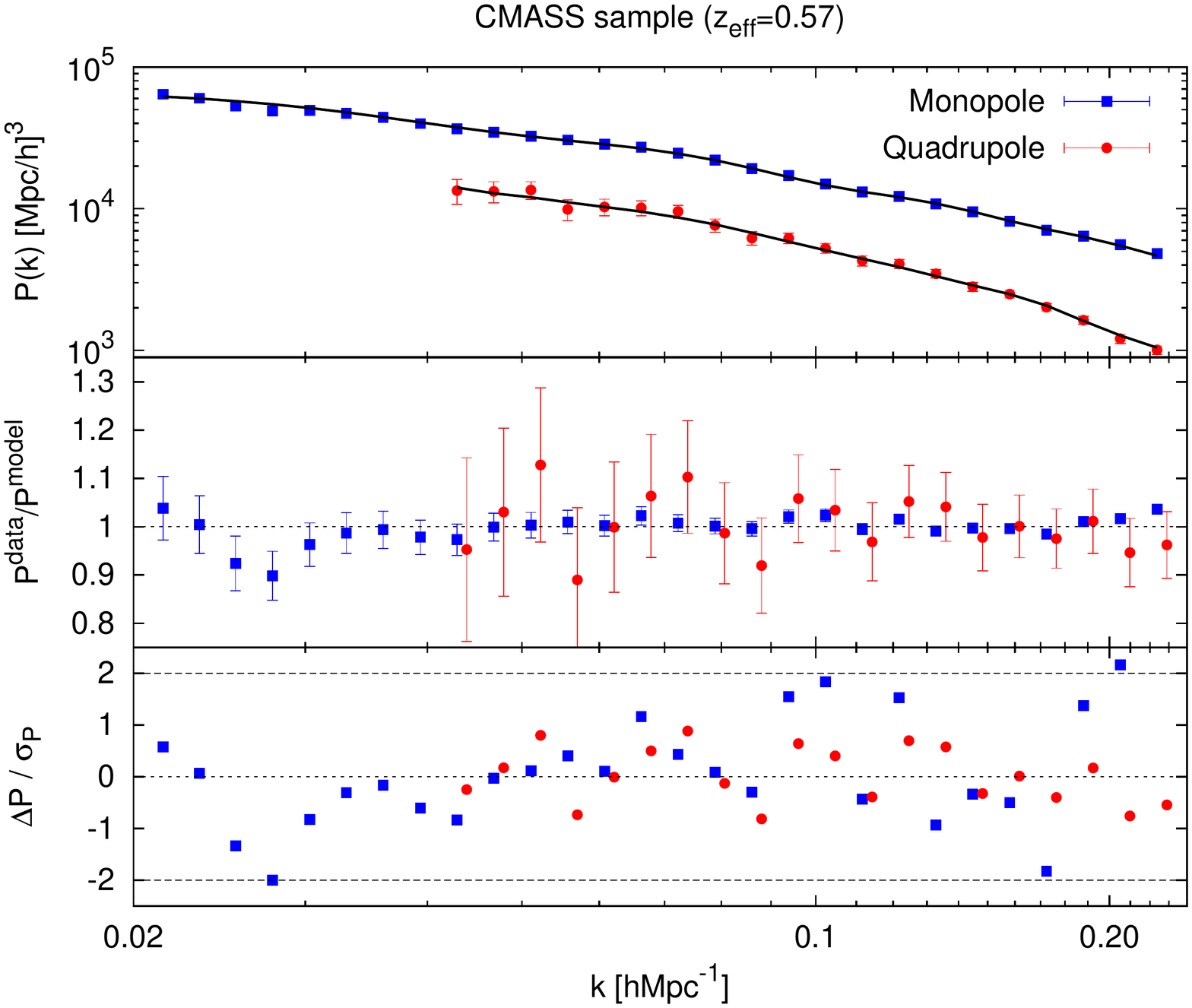}
\caption{Power spectrum data: the top sub-panels display the measured LOWZ- (left panel) and CMASS-DR12 (right panel), monopole (blue symbols) and quadrupole (red circles) power spectra. For both cases, the measurements correspond to a combination of the northern and southern  galaxy caps according to their effective areas as presented in Eq. \ref{eq:NGCSGC}. The error-bars correspond to the dispersion among 2048 realisations of the MD-\textsc{Patchy} mocks. The black solid lines correspond to the best-fitting model, calculated from a full fit to the power spectrum and bispectrum moments, with parameters as listed in Table~\ref{table:results}. The middle sub-panel shows the ratio between the power spectrum multipole measurements and the best-fitting models. In the bottom sub-panel the difference between the data and the model, $\Delta P\equiv P^{\rm data}-P^{\rm model}$, relative to the statistical error of the data, $\sigma_P$, is presented. The black dashed lines represent a $2\sigma$ deviation (95.4\%) confidence level. In the middle and bottom sub-panel the quadrupole measurements have been horizontally displaced for clarity.  }
\label{fig:powerspectrum_data}
\end{figure*}

\begin{figure*}
\includegraphics[scale=0.5]{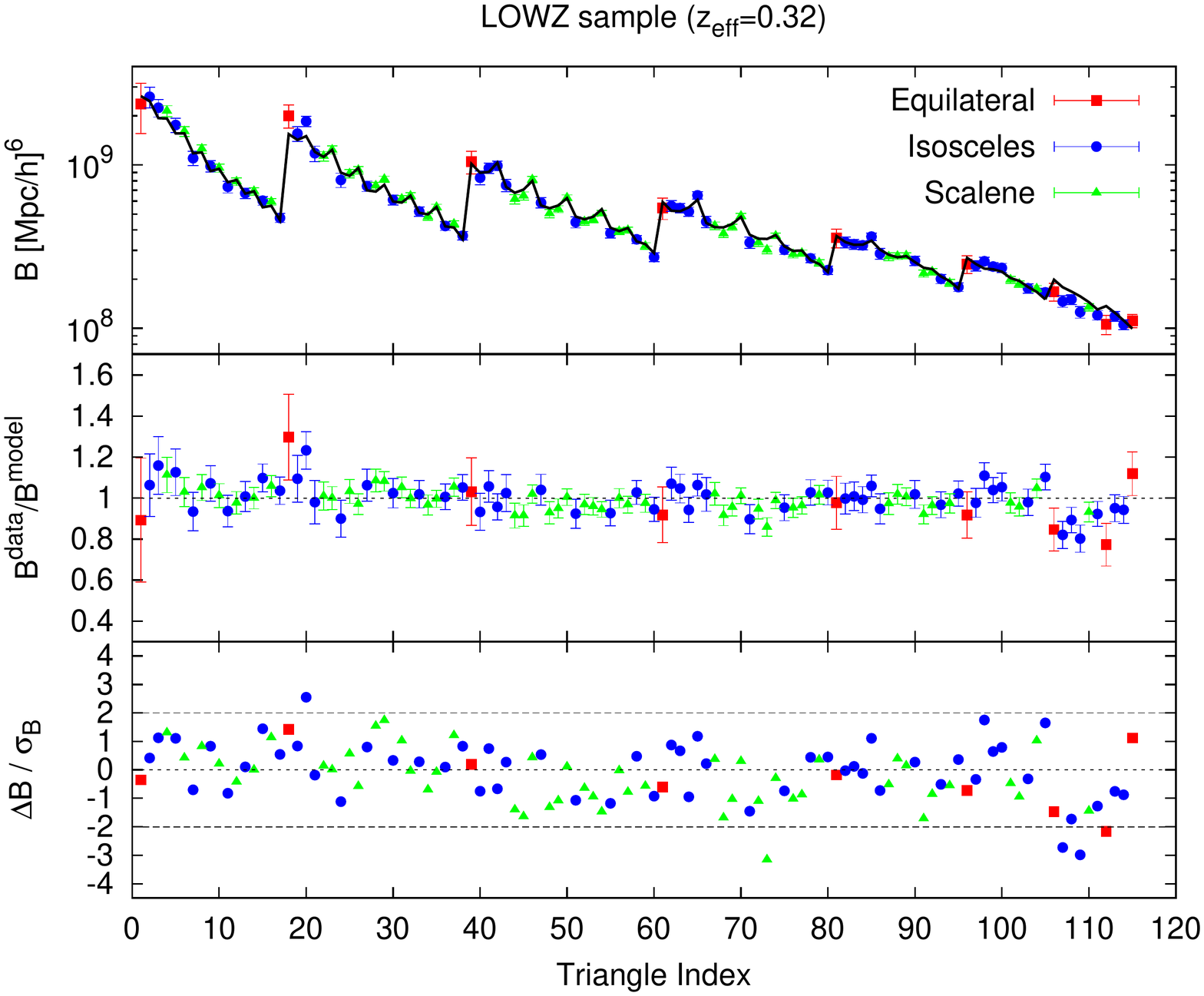}
\includegraphics[scale=0.5]{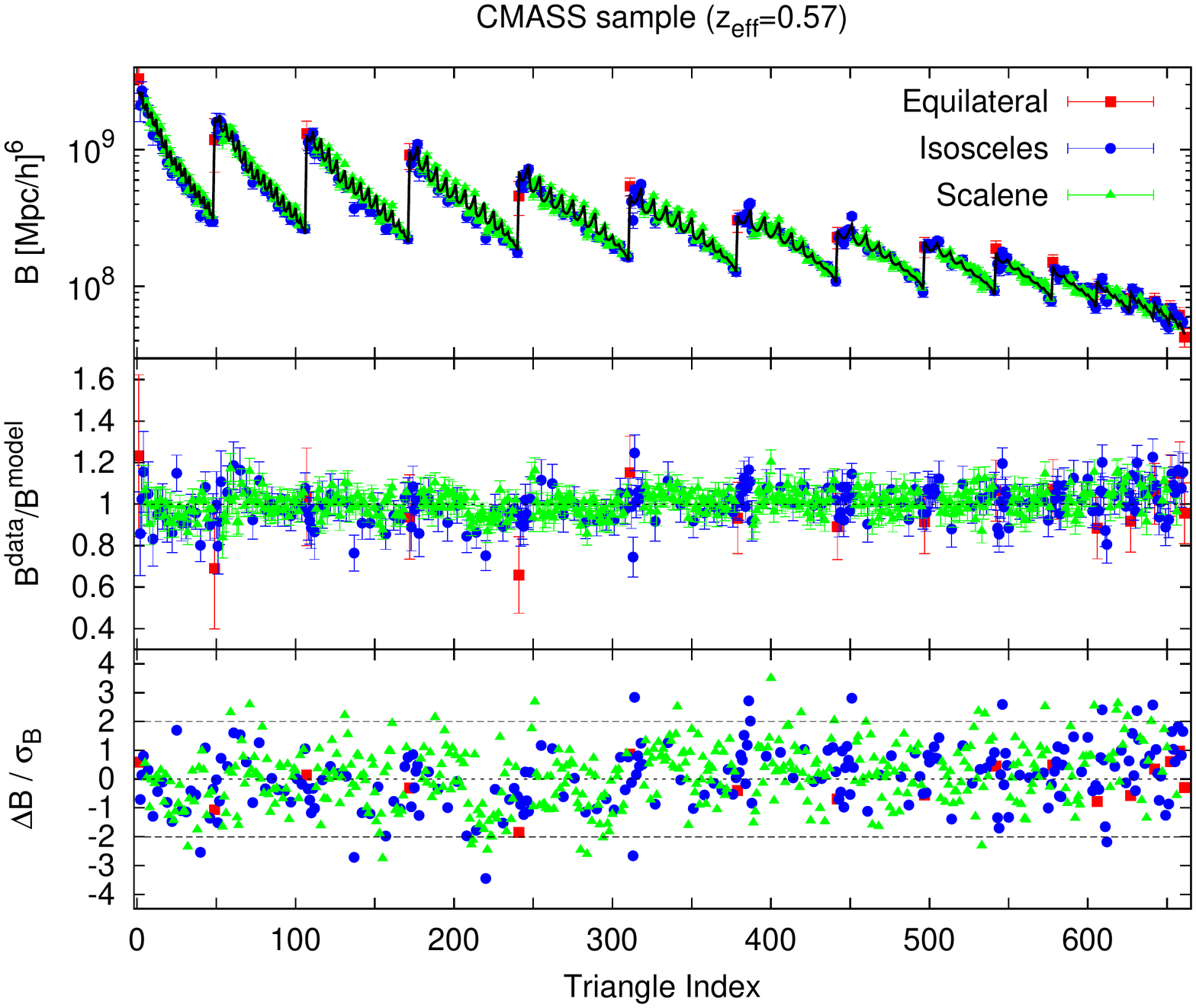}
\caption{Bispectrum data: the top sub-panels display the measured LOWZ- (top panel) and CMASS-DR12 (bottom panel) bispectrum monopole for different triangular shapes: equilateral triangles (red squares), isosceles triangles (blue circles) and scalene triangles (green triangles), ordered sequentially in $k_1$, $k_2$ and $k_3$ (see text for details of the ordering), and covering $0.03\leq k_i\,[h{\rm Mpc}^{-1}]\leq 0.18$ for the LOWZ sample and $0.03\leq k_i\,[h{\rm Mpc}^{-1}]\leq 0.22$ for the CMASS sample. As for the power spectrum, the measurements correspond to a combination of the northern and southern  galactic caps, described by Eq. \ref{eq:NGCSGC}. The displayed error-bars correspond to the dispersion among 2048 realisations of the MD-\textsc{Patchy} mocks. The black solid line represent the best-fitting model using the parameters of Table~\ref{table:results}. The middle and the bottom sub-panel show the deviation of the model respect to the data,  as it is shown in Fig.~\ref{fig:powerspectrum_data} for the power spectrum.}
\label{fig:bispectrum_data}
\end{figure*}

The top panel of Fig.~\ref{fig:powerspectrum_data} shows the measured power spectrum monopole (blue squares), and quadrupole (red circles), while the top-panel of Fig.~\ref{fig:bispectrum_data} shows the bispectrum monopole (all symbols). For both the power spectrum and bispectrum we have combined the NGC (Northern Galactic Cap) and SGC (Southern Galactic Cap) by weighting each statistic by the fraction of their effective area,
\begin{equation}
\mathcal{P}=(\mathcal{P}_{\rm NGC} A_{\rm NGC} + \mathcal{P}_{\rm SGC} A_{\rm SGC})/(A_{\rm NGC} +A_{\rm SGC} )
\label{eq:NGCSGC}
\end{equation}
where $\mathcal{P}$ represent the statistic (power spectrum monopole, quadrupole or bispectrum) and $A_{\rm NGC}$ and $A_{\rm SGC}$ the effective areas of NGC and SGC, respectively. The values for these areas are $A_{\rm NGC}^{\rm LOWZ}=5836\,{\rm deg}^2$, $A_{\rm SGC}^{\rm LOWZ}=2501\,{\rm deg}^2$ for the LOWZ sample and $A_{\rm NGC}^{\rm CMASS}=6851\,{\rm deg}^2$ and $A_{\rm SGC}^{\rm CMASS}=2525\,{\rm deg}^2$ for the CMASS sample. 

Although 3D plots of the bispectrum are possible, we have adopted a simple approach where we order the binned bispectrum in $k_1$, $k_2$ and $k_3$. The first value plotted corresponds to an equilateral triangle of side $N_0\Delta k$, where $\Delta k$ is the bin-size used (in this case $\Delta k=6k_f$) and $N_0$ is an integer corresponding to the first bin-size considered (in this case $N_0=2$ for LOWZ and  $N_0=3$ for CMASS because the triangles with $k_i\leq0.03\,h{\rm Mpc}^{-1}$ are not considered).  We then plot the bispectrum for triangular bins where we sequentially loop through all possible sets of values of $k_1$, $k_2$ and $k_3$, with $k_3$ in the inner most loop, and $k_1$ in the outer most increasing loop, where the loops go from $N_0\Delta k$ to the maximum value considered, either $k_{\rm Ny}/2$, a truncation scale set by our constraints $k_1\leq k_2 \leq k_3$ and $k_i< k_1 + k_2$, or the maximum $k$-value considered. For the bispectrum displayed in Fig.~\ref{fig:bispectrum_data}, the data points have been coloured according to the type of triangular shape they represent. Equilateral triangles are displayed by red squares, isosceles by blue circles and scalene by green triangles. 

The power spectrum data cover $0.02\,h{\rm Mpc}^{-1} \leq k\leq0.18\,h{\rm Mpc}^{-1}$ for the LOWZ monopole and $0.04\,h{\rm Mpc}^{-1} \leq k\leq0.18\,h{\rm Mpc}^{-1}$  for the LOWZ quadrupole; and $0.02\,h{\rm Mpc}^{-1} \leq k\leq0.22\,h{\rm Mpc}^{-1}$ for the CMASS monopole and $0.04\,h{\rm Mpc}^{-1} \leq k\leq0.22\,h{\rm Mpc}^{-1}$  for the CMASS quadrupole. In \S\ref{sec:tests} we will discuss how we select the maximum $k$-value used. In a similar way, the bispectrum of Fig.~\ref{fig:bispectrum_data} represent triangles whose $k$-vectors are contained by $0.03\,h{\rm Mpc}^{-1} \leq k\leq0.18\,h{\rm Mpc}^{-1}$ for LOWZ sample and $0.03\,h{\rm Mpc}^{-1} \leq k\leq0.22\,h{\rm Mpc}^{-1}$ for the CMASS sample.  In total, for the LOWZ sample we have 160 bins, whereas for the CMASS sample 707.

The black solid lines in the upper panels of Fig.~\ref{fig:powerspectrum_data} and~\ref{fig:bispectrum_data} show the best-fitting model for the appropriate power spectrum and bispectrum moments. The details about the models are presented in \S\ref{sec:model}, and the way the fit has been performed is described in \S\ref{sec:covariances}, and the best-fitting parameters are reported in Table~\ref{table:results}. The middle and bottom panels demonstrate how well the best-fitting theoretical model describes the data. In the middle panel the ratio between the data points and the model is presented, whereas in the bottom panel the difference between the data and the model divided by the diagonal component of the covariance matrix is displayed. In the bottom panel the $2\sigma$ deviation (95.4\% confidence level) is shown in black dashed lines. 

\section{Modelling the bispectrum}\label{sec:model}

In this section we introduce the model used to describe the power spectrum and bispectrum measurements presented in \S\ref{sec:measurement}. The same model has been used in previous works \citep{hector_bispectrum1,gil-marin15_rsd}, so here we only present a brief description. The model is based in the following 4 steps: \S\ref{sec:bias}  the galaxy bias model, \S\ref{sec:RSD} the dark matter clustering and the redshift space distortions model, \S\ref{sec:AP} the Alcock-Paczynski effect, \S\ref{sec:mask} the modelling of the survey window mask. 

\subsection{The Bias Model}\label{sec:bias}
We assume an Eulerian non-linear and non-local bias model, as proposed by \cite{McDonaldRoy:2009} and used in previous analysis of the power spectrum and bispectrum of BOSS data \citep{Beutleretal:2013,hector_bispectrum1, gil-marin15_rsd}. The non-local bias model proposed in  \cite{McDonaldRoy:2009}  depends on 4 bias parameters: the linear bias parameter $b_1$, the non-linear bias parameter $b_2$ and the non-local bias parameters $b_{s^2}$ and $b_{3{\rm nl}}$. Here we constrain the values of the non-local bias parameters by assuming that the bias model is local in Lagrangian space. This allows us to set the values of $b_{s^2}$ and $b_{3{\rm nl}}$ as functions of $b_1$ \citep{Baldaufetal:2012,Chan12,Saitoetal:2014}, $b_{s^2}=-4/7(b_1-1)$ and $b_{3{\rm nl}}=32/315(b_1-1)$, meaning that only two bias parameters need to be fitted. This approach has been validated by $N$-body simulations and has been shown to provide consistent results between the bias parameter obtained from the power spectrum and bispectrum (see section 5.5.1 and fig. 15 of  \citealt{hector_bispectrum1} for a further details). 

\subsection{Redshift Space Distortions}\label{sec:RSD}
We describe the effects of redshift space distortions on the power spectrum as in the model presented in \cite{Taruyaetal:2010,Nishimichietal:2011} (hereafter TNS model), which has been used in previous analysis of the power spectrum multipoles of BOSS dataset \citep{Beutleretal:2013,hector_bispectrum1,gil-marin15_rsd}. The TNS model provides a description of redshift space quantities in terms of real space quantities: the density and velocity power spectrum components of the matter power spectrum. For this work we assume that there is no velocity bias on the scales of interest, so that the dark matter and galaxy velocity fields are the same. The non-linear real space components used in this paper are as presented in \cite{gil-marin15_rsd}. In short, the non-linear matter quantities are obtained using resummed perturbation theory at 2-loop level as described in \cite{HGMetal:2012} and the necessary linear power spectrum input is computed using CAMB \citep{camb}, corresponding to the  fiducial cosmological parameters of \S\ref{sec:fiducial_cosmology}. The Fingers-of-God (hereafter FoG) are accounted through a one-free parameter Lorentzian damping function as described in eq. B.19 of \cite{hector_bispectrum1}, where the free parameter is referred as $\sigma_{\rm FoG}^P$. Although this factor aims to parametrise the expected non-linear damping due to the dispersion of satellite galaxies inside the host haloes, in practice, we treat this parameter as an effective free parameter that encodes our poor understanding of the non-linear component of RSD.

Modelling the bispectrum, both in real and redshift space, is a more challenging task than modelling the power spectrum. In this paper we use the same phenomenological model as our previous analysis of the DR11 BOSS CMASS data. This model relies on 18 parameters that are calibrated using dark matter $N$-body simulations which modify the SPT kernels $\mathcal{F}^{\rm SPT}_2$ and $\mathcal{G}^{\rm SPT}_2$ into an effective kernels, $\mathcal{F}^{\rm eff}_2$ \citep{HGMetal:2011} and $\mathcal{G}^{\rm eff}_2$ \citep{hector_bispectrum0}, and extend the linear behaviour of the predictions of SPT up to the weakly non-linear regime (for details concerning the bispectrum model see \S 3.6 and appendix C of \citealt{hector_bispectrum1}). As for the power spectrum, we account for FoG in the bispectrum, through a Lorentzian damping function, as described in eq. C15 of \cite{hector_bispectrum1}, with one free parameter, $\sigma_{\rm FoG}^B$, which accounts for the non-linear dispersion of galaxies inside the host haloes as well as for the poor understanding of the non-linear components of RSD. 

In this paper we describe the amplitude of the shot noise, both in the power spectrum and bispectrum, through a free parameter $A_{\rm noise}$, which  parametrises the shot noise deviation with respect to the Poisson prediction. For the power spectrum this is $P_0^{\rm noise}=(1-A_{\rm noise})P_0^{\rm Poisson}$ and for the bispectrum $B_0^{\rm noise}(k_1,k_2,k_3)=(1-A_{\rm noise})B_0^{\rm Poisson}(k_1,k_2,k_3)$, where $P_0^{\rm Poisson}$ and $B_0^{\rm Poisson}$ are the Poisson prediction for the power spectrum and bispectrum\footnote{For the expression of the Poisson shot noise prediction see eq. A3 and eq. A10 of \cite{hector_bispectrum1} }.  For simplicity we still assume that $A_{\rm noise}$ do not depend on the scale. 

\subsection{The Alcock-Paczynski effect}\label{sec:AP}
The AP effect \citep{AP} leads to observed distortions in the clustering signal about the line-of-sight, resulting from converting redshifts into distances using a different cosmological model than the actual one. By measuring this signal in both isotropic moments and anisotropic moments about the line-of-sight, we can constrain the Hubble parameter $H(z)$, which is inversely proportional to the distortion in the radial direction,  and the angular diameter distance parameter $D_A(z)$, which depends on the angular distortion. Practically, we assume a fiducial model for converting redshift into distances, so the AP signal can be described by dilation scale factors, 
\begin{eqnarray}
\label{alphapara}\alpha_\parallel(z)&\equiv&\frac{H^{\rm fid}(z)r_s^{\rm fid}(z_d)}{H(z)r_s(z_d)},\\
\label{alphaperp}\alpha_\perp(z)&\equiv&\frac{D_A(z)r_s^{\rm fid}(z_d)}{D_A^{\rm fid}(z)r_s(z_d)}.
\end{eqnarray}
Here $\alpha_\parallel$ and $\alpha_\perp$ are the parallel- and perpendicular-to-the-line-of-sight dilation scales, respectively; $r_s^{\rm fid}(z_d)$ is the fiducial sound horizon at the baryon drag redshift and $H^{\rm fid}$ and $D_A^{\rm fid}$ are the fiducial values of the Hubble parameter and the angular diameter distance, respectively. Assuming the fiducial cosmological model presented in \S\ref{sec:fiducial_cosmology}, the values for the Hubble parameter and the angular diameter distances are, $H^{\rm fid}(z_{\rm LOWZ})=79.49\,{\rm km}{s}^{-1}{\rm Mpc}^{-1}$, $D_A^{\rm fid}(z_{\rm LOWZ})=999.23\,{\rm Mpc}$ for the LOWZ sample at $z_{\rm LOWZ}=0.32$, and  $H^{\rm fid}(z_{\rm CMASS})=92.25\,{\rm km}{s}^{-1}{\rm Mpc}^{-1}$, $D_A^{\rm fid}(z_{\rm CMASS})=1398.43\,{\rm Mpc}$ for the CMASS sample at $z_{\rm CMASS}=0.57$. The value for the fiducial sound horizon distance is $r_s^{\rm fid}(z_d)=148.11\, {\rm Mpc}$. Unlike a BAO-only analyses, for a RSD analysis Eqs.~\ref{alphapara}-\ref{alphaperp} do not match the degeneracy between recombination BAO position give by $r_s$, and its projection. However, we opt to keep the $r_s^{\rm fid}(z_d)/r_s(z_d)$ scaling as it will be simpler to combine the results from RSD and BAO analyses in terms of the same variables.  

The AP dilation scales relate the observed modes ${\bf k}$ to those in a cosmological model to be tested ${\bf q}$, which is assumed to match the underlying Universe: $k_\parallel=\alpha_\parallel q_\parallel$, $k_\perp=\alpha_\perp q_\perp$. So, for example, if we project angular modes to be larger than correct distances ($\alpha_\perp<1$), they are observed on larger scales than in the comoving model ($k_\perp<q_\perp$).
When we express these dilations in terms of the modulus of the frequency vector, $k\equiv |\bf k|$ and $q\equiv |\bf q|$,
and the cosine of its  angle with respect to the line-of-sight (LOS), $\mu\equiv \hat{\bf k}\cdot \hat x_{\rm LOS}$ and 
$\nu\equiv \hat{\bf q}\cdot \hat x_{\rm LOS}$ \citep{Beutleretal:2013},
\begin{eqnarray}
\label{AP:k}q&=&\frac{k}{\alpha_\perp}\left[ 1+\mu^2\left( F^{-2}-1 \right)  \right]^{1/2},\\
\label{AP:LOS}\nu&=&\frac{\mu}{F}\left[ 1+\mu^2\left( F^{-2}-1  \right) \right]^{-1/2},
\end{eqnarray}
where $F\equiv\alpha_\parallel/\alpha_\perp$.  
For the power spectrum,  the $\ell$-multipole in terms of the observed frame, $k$, can be related to the line-of-sight dependent power spectrum in the frame of the theoretical model, i.e. as a function of $q^{\rm }$ and $\nu^{\rm }$,
\begin{equation}
P_\ell(k^{\rm })=\frac{2\ell+1}{2\alpha_\perp^2\alpha_\parallel}\int_{-1}^{1} d\mu\, \mathcal{L}_\ell(\mu)P(q^{\rm },\nu^{\rm }).
\end{equation}
This formalism can be easily generalised in terms of the bispectrum \citep{songetal:2015},
\begin{equation}
B_0(k_1,k_2,k_3)=\frac{1}{2\alpha_\parallel^2\alpha_\perp^4}\int_{-1}^{1}d\mu_1\int_0^{2\pi}d\varphi\, B(q_1,q_2,q_3,\nu_1,\nu_2),
\end{equation}
where $\varphi$ has defined to be $\mu_2=\mu_1 \eta_{12}-\sqrt{1-\mu_1^2}\sqrt{1-\eta_{12}^2}\cos\varphi$; with $\eta_{12}\equiv ({\bf k}_1\cdot{\bf k}_2)/(k_1 k_2)$, the cosine of the angle between ${\bf k}_1$ and ${\bf k}_2$. In this case $k_i$ and $q_i$ are the actual and the fiducial vectors of the triangle, respectively; and $\mu_i$ and $\nu_i$ are the actual and fiducial cosines of the angle of the vector with respect to the LOS, for each side of the triangle. Analogously to Eq. \ref{AP:k} and \ref{AP:LOS}, the fiducial and observed components of the triangles are related as, 
\begin{eqnarray}
q_i&=&\frac{k_i}{\alpha_\perp}\left[ 1+\mu_i^2(F^{-2}-1) \right]^{1/2},\\
\nu_i&=&\frac{\mu_i}{F}\left[ 1+\mu_i^2(F^{-2}-1) \right]^{-1/2}.
\end{eqnarray}
Additionally, the cosine of the angle between the vectors $k_1$ and $k_2$ is also distorted (because of the distortion of $k_3$) as,
\begin{eqnarray}
\nonumber\upsilon_{12}&=&\left[ \eta_{12}+\mu_1\mu_2(F^{-2}-1) \right]\left[1+\mu_1^2(F^{-2}-1)  \right]^{-1/2}\\
&\times&\left[1+\mu_2^2(F^{-2}-1)  \right]^{-1/2}.
\end{eqnarray}
Note that by construction $-1\leq \eta_{12} \leq +1$, but this is not necessary true for $\upsilon_{12}$: the AP distortions can break the condition of closed triangle. This happens when $q_3>q_1+q_2$. If this is the case the theoretical bispectrum is predicted to be 0.

\subsection{Survey Geometry}\label{sec:mask}
The estimator of the Fourier space statistics presented in \S\ref{sec:algorithm} provides a measurement that is affected by the survey geometry: at intermediate and large scales, the measured power spectrum and bispectrum are affected by the angular and radial selection function of the survey. It is important to account for these effects in the theoretical models in order to avoid spurious results. For the power spectrum monopole and quadrupole the method for doing this is fully described in section 5.4 of \cite{gil-marin15_rsd}. In this section we describe the effects on the bispectrum monopole. 

The measured  quantity $\langle F_3({\bf k}_1)  F_3({\bf k}_2)  F_3({\bf k}_3) \rangle$ in Eq. \ref{eq:bis2} is related to the underlying galaxy bispectrum, $B_{\rm gal}({\bf k}_1,\,{\bf k}_2)$, through \citep{hector_bispectrum1},
\begin{eqnarray}
\nonumber&& \langle F_3({\bf k}_1) F_3({\bf k}_2) F_3({\bf k}_3)\rangle=\int \frac{d^3{\bf k'}}{(2\pi)^3} \frac{d^3{\bf k''}}{(2\pi)^3}\, B_{\rm gal}({\bf k}',{\bf k}'')\\
\label{Bmask}&\times&W_3({\bf k}_1-{\bf k}',{\bf k}_2-{\bf k}'')+B_{\rm noise}({\bf k}_1,{\bf k}_2),
\end{eqnarray}
where $W_3$ is the window function for the bispectrum defined in eq.  12 of \cite{hector_bispectrum1}, and $B_{\rm noise}$ is the shot noise of the measurement due to the discreteness of objects (see eq. A10 of  \citealt{hector_bispectrum1} for the Poisson prediction). Since performing the double convolution described by the integral of Eq. \ref{Bmask} is computationally expensive, we work in the same approximative regime described in \cite{hector_bispectrum1}. This consists in writing the  theoretical bispectrum as $B_{\rm gal}(k_1,k_2,k_3)\sim P(k_1)P(k_2) {\cal Q}(k_1,k_2,k_3) + {\rm cyc}$, where ${\cal Q}$ can be any function of the 3 $k$-vectors. When we ignore the effects of the window function on ${\cal Q}$, the integral of Eq.~\ref{Bmask} is separable, and can be written as a mask integrals of the power spectrum. As a consequence, the integral of Eq. \ref{Bmask} becomes,
\begin{eqnarray}
\label{bispectrum_approximation}
[P\otimes W_2](k_1)	\times[P\otimes W_2](k_2) \times {\cal Q}(k_1,k_2,k_3),
\end{eqnarray}
where we have defined,
 \begin{eqnarray}
\label{combolucion_window} [P\otimes W_2](k_i)\equiv\int \frac{d^3{\bf k}'}{(2\pi)^3}\, P({\bf k}')|W_2({\bf k}_i-{\bf k}')|^2, 
\end{eqnarray}
with $W_2$ given by eq. 9 of \cite{hector_bispectrum1}. This approximation works reasonably well  for modes that are not too close to the size of the survey, i.e. {\it all} three sides of the $k$-triangle are sufficiently large. The approximation  fails to reproduce accurately the correct bispectrum shape when  (at least) one of the $k_i$ is close to the fundamental frequency. In previous works \citep{hector_bispectrum1} we tested effect of the window selection function in the bispectrum. By using mocks with and without the effect of the survey geometry we determined that at scales of $k_i\sim0.03\,h{\rm Mpc}^{-1}$ the full deviation caused by the survey geometry in the measured bispectrum was within $5\%$ percent level, and the difference between applying the full window and our approximation is less than 1\% in the measured parameters when the scales used were between $k_{\rm max}=0.03\,h{\rm Mpc}^{-1}$ and  $k_{\rm max}=0.15\,h{\rm Mpc}^{-1}$.  In particular, for the survey geometry of CMASS and LOWZ samples, this limitation only applies to  triangle configurations  where the modulus of one $k$-vector is much shorter than the other two ($k_3\ll k1\sim k2$, the so-called {\it squeezed} configuration) and the shortest $k$ is $\lesssim0.03\, h{\rm Mpc}^{-1}$. Therefore, we do not include these triangles in our analysis, as we have mentioned at the end of \S\ref{sec:measurement}. 

\section{Parameter estimation}\label{sec:covariances}

In this section we describe the estimation of the parameters of interests and their errors.
We also present the measurement of the bispectrum on the \textsc {MD-Patchy} and \textsc{qpm} mocks and how it compares to the actual data.  
 
\subsection{Covariance Matrices}

We aim to account for the full covariance matrix between the power spectrum monopole, quadrupole and bispectrum monopole for both LOWZ and CMASS samples. We measure each statistic for the 2048 \textsc{MD-Patchy} mock catalogues, using the same method described for the actual dataset in \S\ref{sec:measurement}. Since the total number of bins (including the power spectrum monopole, quadrupole and bispectrum) is significantly fewer than the total number of realisations, we can obtain a reliable covariance matrix. The full covariance matrix $C$ consists of 6 sub-covariances that describe the auto correlation terms of the power spectrum multipoles and bispectrum monopole, and the corresponding cross terms,
\begin{equation}
C =
 \begin{pmatrix}
  P^{(0)}\times P^{(0)} & P^{(0)}\times P^{(2)} & P^{(0)}\times B^{(0)} \\
  & P^{(2)}\times P^{(2)} & P^{(2)}\times B^{(0)} \\
  &  & B^{(0)}\times B^{(0)} \\
 \end{pmatrix}.
 \end{equation}
 We correct the inverse covariance matrix $C^{-1}$ by the bias produced due to the limited number of realisations \citep{Hartlap07}. 
 
\begin{figure*}
\includegraphics[trim = 60mm 0mm 0mm 0mm, clip=false,scale=0.27]{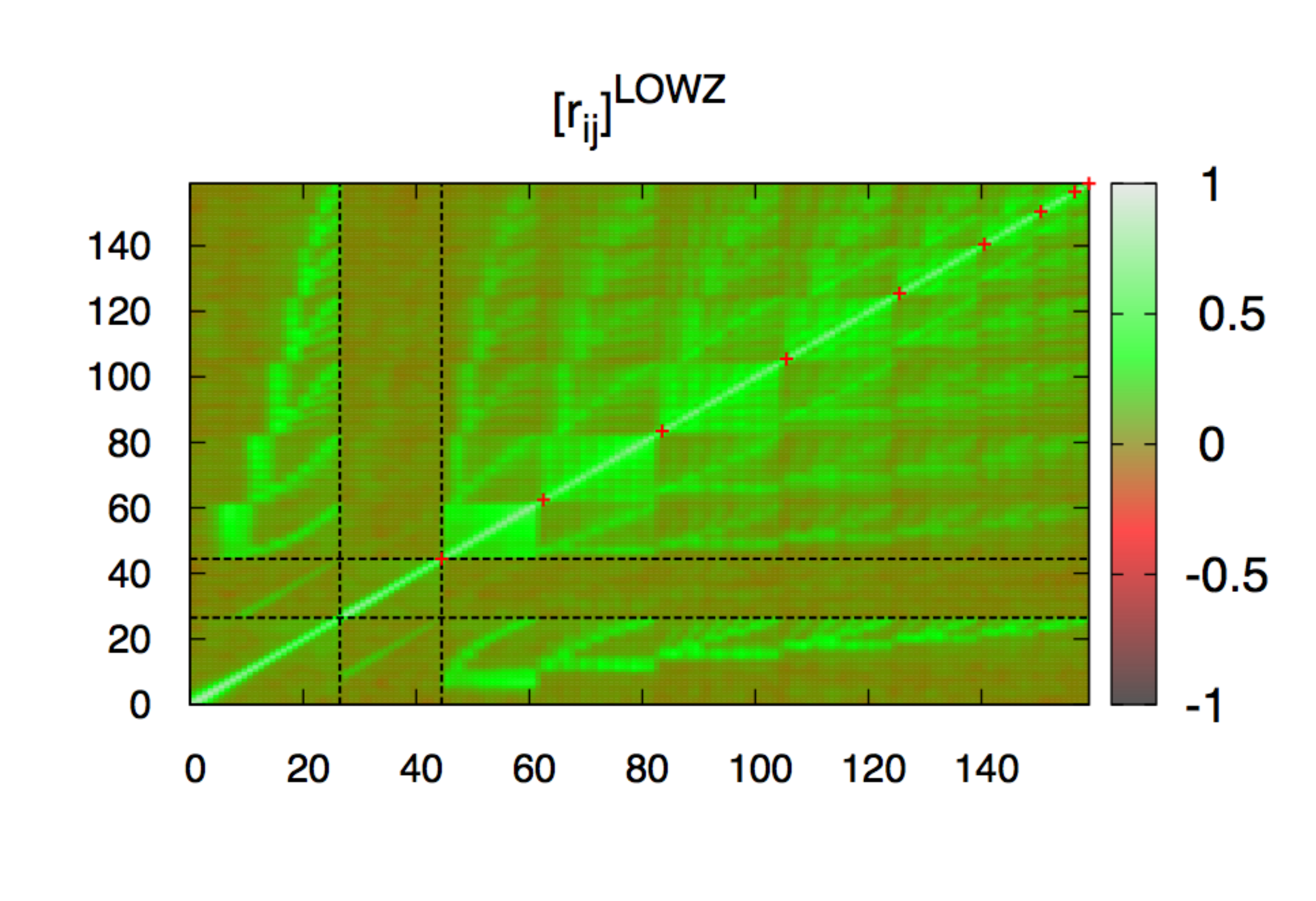}
\includegraphics[trim = 40mm  0mm 50mm 0mm, clip=false,scale=0.27]{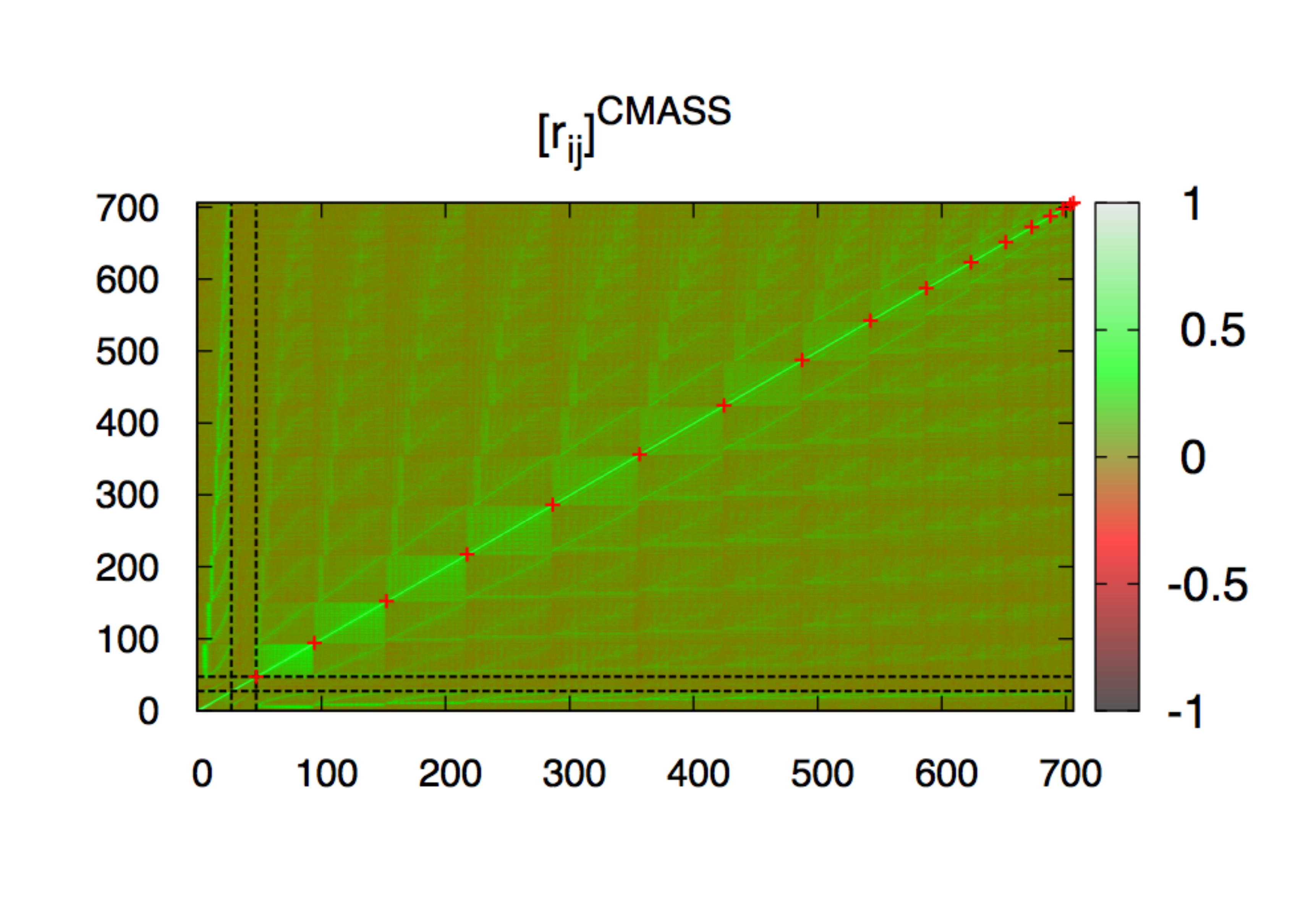}
\caption{Correlation coefficients for the power spectrum monopole - power spectrum quadrupole - bispectrum monopole from the LOWZ-DR12 sample (left panels) and from the CMASS-DR12 sample (right panels), extracted from the 2048 realisations of the \textsc{MD-Patchy} galaxy mocks. The dashed black lines marks the limit between the auto-correlation and the cross-correlation parts of the covariance. For the LOWZ sample the covariance contains the elements with $0.02\leq k\, [h{\rm Mpc}^{-1}]\leq 0.18$ for the power spectrum monopole,  $0.04\leq k\, [h{\rm Mpc}^{-1}] \leq 0.18$ for the power spectrum quadrupole and  $0.03\leq k_i\, [h{\rm Mpc}^{-1}] \leq 0.18$ for the bispectrum monopole. For the CMASS sample the covariance contains the elements with $0.02\leq k\, [h{\rm Mpc}^{-1}]\leq 0.22$ for the power spectrum monopole,  $0.04\leq k\, [h{\rm Mpc}^{-1}] \leq 0.22$ for the power spectrum quadrupole and  $0.03\leq k_i\, [h{\rm Mpc}^{-1}] \leq 0.22$ for the bispectrum monopole. Red crosses mark the position of equilateral bispectra. The ordering of the triangles follows that presented in Fig.~\ref{fig:bispectrum_data}, and therefore, the bispectra corresponding to the indices between crosses share the same value of $k_1$. }
\label{fig:covariances}
\end{figure*}

The panels of Fig.~\ref{fig:covariances} display the correlation coefficients $r_{ij}\equiv C_{ij}/[C_{ii}C_{jj}]^{1/2}$ obtained from the \textsc{MD-Patchy} mocks for the LOWZ and CMASS samples, as labeled. The black dashed lines separate the correlation coefficient elements corresponding to the auto-correlation (sub-panels along the diagonal) and to the cross-correlation (sub-panels outside the diagonal) among power spectrum monopole, quadrupole and bispectrum monopole. Along the diagonal from the bottom left to the top right: the power spectrum monopole, the power spectrum quadrupole and the bispectrum monopole correlation coefficients\footnote{The correlation coefficient elements corresponding to the power spectrum monopole and quadrupole are the same as those of fig. 3 of \cite{gil-marin15_rsd}}. The red crosses mark the equilateral bispectra indices for clarity. Since the ordering of triangles is the same as described in Fig.~\ref{fig:bispectrum_data}, the bispectra corresponding to the indices between the red crosses share the same value for $k_1$. We observe that the correlation between bispectrum elements is high, especially on those triangles with a close triangular index. These are typically triangles which share at least $k_1$, in some cases also $k_2$ and with a close value on $k_3$ (see the index ordering described in \S\ref{sec:measurement}). As we compare triangles with a more different triangle indices, the correlation decreases, although some pattern is observed. This is due to a correlation between triangles with similar shapes but different scales: different  triangular indices can contain triangles with similar shape and relatively close scales as it is the case for the equilaterals triangles. 
There is a significant cross-correlation between the bispectrum monopole and the power spectrum monopole. This correlation is higher when the $k$-vectors of the triangle are close to the $k$-vector of the power spectrum monopole. On the other hand, the cross-correlation between the power spectrum quadrupole and bispectrum monopole is low and very consistent with 0 for all the shapes and scales studied. 

\begin{figure*}
\includegraphics[scale=0.3]{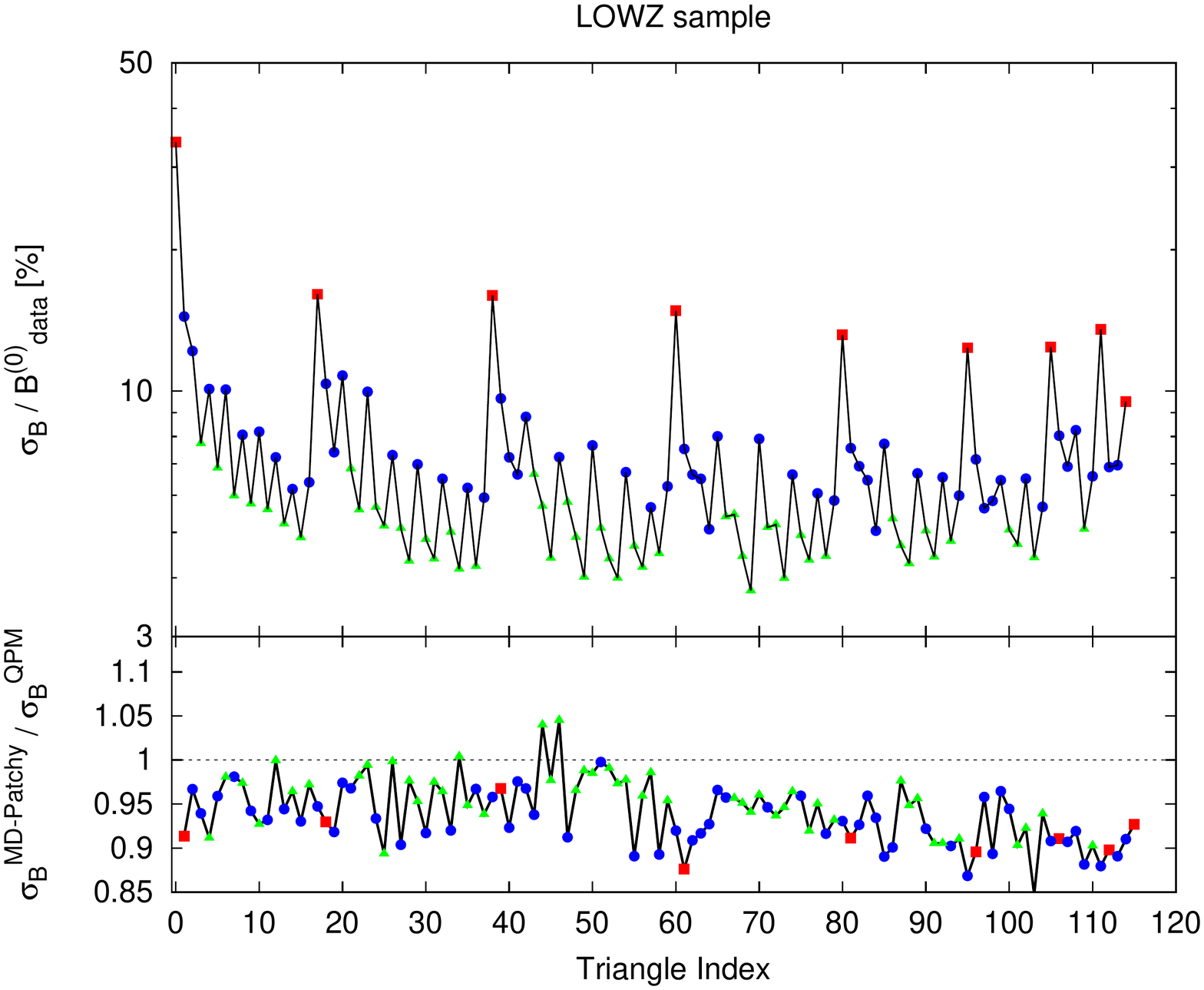}
\includegraphics[scale=0.3]{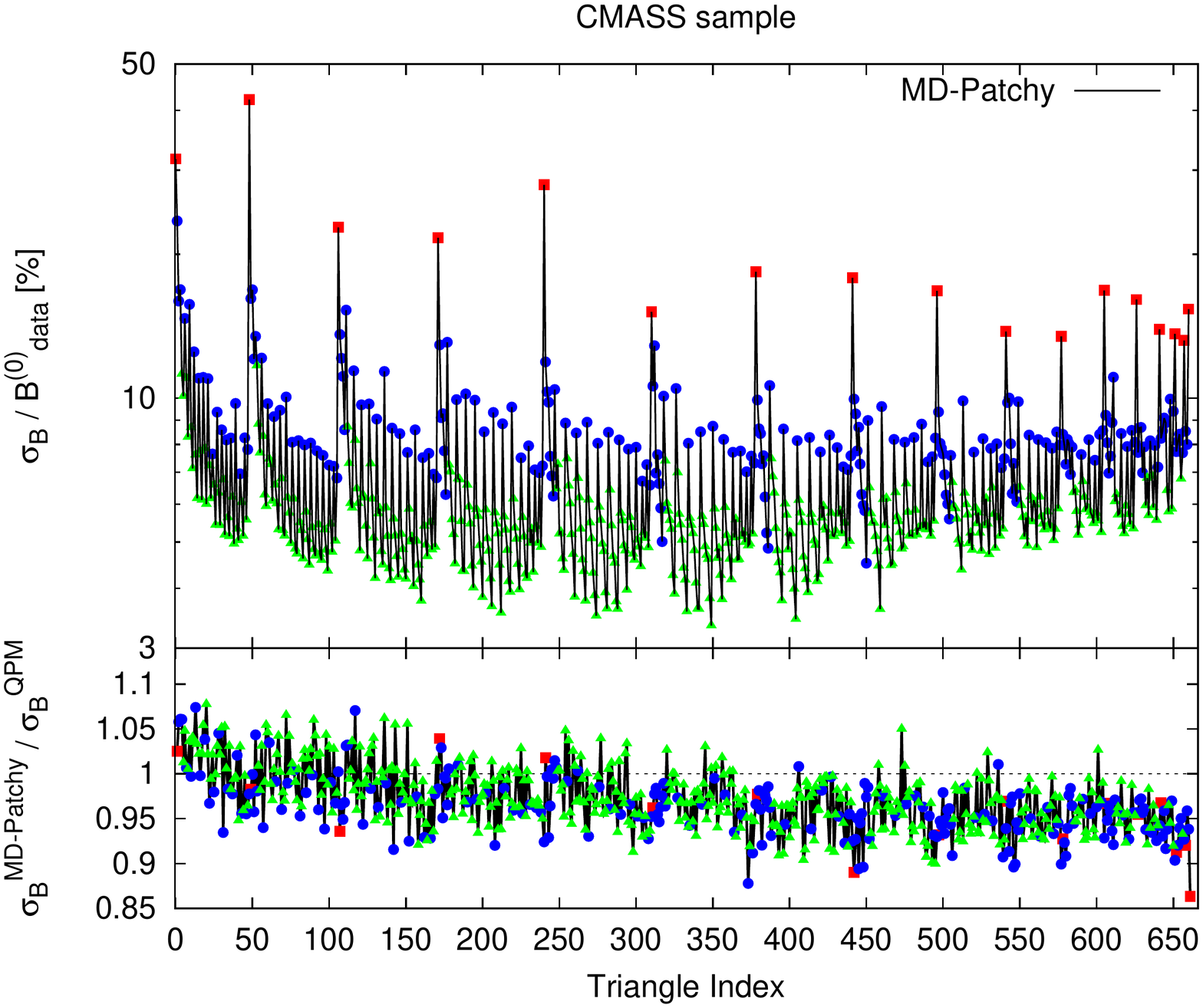}
\caption{Statistical diagonal errors for the measured LOWZ and CMASS bispectrum, as labeled, as a function of the same triangle index presented in Fig.~\ref{fig:bispectrum_data}. The top sub-panels show in solid black line the diagonal statistical errors computed from the {\it rms} of the 2048 realisations of MD-\textsc{Patchy} mocks, divided by the measured bispectrum from the data. The different coloured symbols on the top of the black line represent the different triangle shapes with the same colour and symbol notation than in Fig.~\ref{fig:bispectrum_data}. In the bottom sub-panels the ratio between the diagonal errors computed from the 1000 realisations of the \textsc{qpm} mocks and the 2048 realisations of the MD-\textsc{Patchy} mocks is displayed.  }
\label{fig:error}
\end{figure*}

The top sub-panels of Fig.~\ref{fig:error} display the diagonal  errors ({\it rms} among MD-\textsc{Patchy} mocks realisations) corresponding to the triangle shapes shown in Fig.~\ref{fig:bispectrum_data}, for the LOWZ sample (left panel) and CMASS sample (right panel). The different coloured symbol represent the different triangular shapes with the same notation that in Fig.~\ref{fig:bispectrum_data}. We observe that in both samples for a given scale, the equilateral triangles have a larger statistical error, followed by the isosceles triangles and scalene.  This is certainly the expected behaviour. Assuming that the Fourier components are Gaussian-distributed (which should be true at least at sufficiently large scales) the effective Fourier volume is reduced by a symmetry factor of 6 for equilateral, and 2 for isosceles, respect to the scalene triangles of similar scales \citep{Scoccimarroetal:1998,Sefusattietal:2006}. On average, the statistical error on the measurement represent about $10\%$ of the signal, $\sigma_B/B^{(0)}_{\rm data}\sim0.1$, both for the LOWZ and CMASS sample. It may surprise that the statistical errors are similar for both samples, taking into account that CMASS has about 3 times the volume of LOWZ. However, we have to bear in mind that for both samples we have used a bin corresponding to $6k_f$. Since the CMASS sample is embedded in a box whose side is $\simeq1.5$ times larger than the LOWZ sample box, the CMASS $k_f$ is also 1.5 smaller for CMASS respect to LOWZ, and so the binning applied for the bispectrum, which scales as $\Delta k^3$.

The lower sub-panels of Fig.~\ref{fig:error} show the ratio between the diagonal errors extracted from the MD-\textsc{Patchy} mocks (shown also in the upper sub-panel) and the \textsc{qpm} mocks. For the CMASS sample the difference in these errors is very small at large scales and we observe no trend that depends on triangle shapes. At small scales ($k\gtrsim0.15\,h{\rm Mpc}^{-1}$),  \textsc{qpm} mocks predict larger error-bars, which are about $5-10\%$ larger than those predicted by the MD-\textsc{Patchy} mocks. For the LOWZ sample we observe that \textsc{qpm} mocks predict larger error-bars than  MD-\textsc{Patchy} mocks for all scales and shapes, which vary between $5-10\%$, independent of shape, and slightly increasing with the scale.  The observed difference between the bispectrum error of the \textsc{qpm} and MD-\textsc{Patchy} mocks is probably related to the difference in the clustering signal of the bispectrum. This is displayed later in \S\ref{sec:mocksdata}, in Fig.~\ref{fig:mocks}. The error of the bispectrum is related to the 6-point correlation function\footnote{This is strictly true only when the Fourier components are Gaussian-distributed, which is true at least at large scales.}, therefore, differences in the  6-point correlation clustering signal function between \textsc{qpm} and  MD-\textsc{Patchy} would produce such differences. We have not checked explicitly the  6-point correlation signal in the \textsc{qpm} and  MD-\textsc{Patchy}, as it would be computationally very challenging, but given that the bispectrum of these two is already significantly different, specially for the CMASS sample, it is likely that the 6-point correlation signal also differs.  

\subsection{Best-fitting and error estimation}
We model the shape of the power spectrum multipoles in combination with the bispectrum monopole through 9 free parameters ${\bf \Psi}=\{b_1,b_2,A_{\rm noise}, \sigma_{\rm FoG}^P, \sigma_{\rm FoG}^B, f,\sigma_8, \alpha_\parallel,\alpha_\perp  \}$. Among these parameters, $b_1$, $b_2$, $A_{\rm noise}$ are related to the way the galaxies populate the haloes; $f$, $\sigma_8$, $\alpha_\perp$ and $\alpha_\parallel$ are related to the cosmology; and $\sigma_{\rm FoG}^P$, $\sigma_{\rm FoG}^B$ are nuisance parameters of the model. We briefly describe them below.
\begin{enumerate}

\item The galaxy bias model is parametrised through two free bias parameters, $b_1$ and $b_2$ as described in \S\ref{sec:bias}. The non-local bias parameters $b_{s^2}$ and $b_{3\rm nl}$ are set to the values predicted by the local Lagrangian bias assumption.  The deviation of the amplitude of the shot noise with respect to the Poisson prediction, as described in \S\ref{sec:RSD}

\item The logarithmic growth factor $f(z)\equiv{[d\log D(z)}]/[{d\log a(z)}]$; where $D(z)$ is the linear growth factor and $a$ the scale factor. This parameter can be predicted for a specific cosmological model when a theory of gravity is assumed: $f\simeq\Omega_m^\gamma(z)$ with $\gamma=0.55$ for GR. In this work we consider $f$ as a free parameter in order to account for potential deviations from GR. 

\item The amplitude of primordial dark matter power spectrum fluctuations times the growth factor: $\sigma_8(z)\equiv D(z){\sigma_8}_0$; where ${\sigma_8}_0$ is the value of $\sigma_8$ at some fiducial time.  

\item The AP parameters, $\alpha_\parallel$ and $\alpha_\perp$ described in \S\ref{sec:AP}. 

\item The FoG parameters, $\sigma^P_{\rm FoG}$ and $\sigma^B_{\rm FoG}$ introduced in  \S\ref{sec:RSD}. 
\end{enumerate}
The rest of cosmological parameters, including the Hubble parameter $h$ and the spectral index $n_s$ are kept fixed during the analysis at the fiducial values described in \S\ref{sec:fiducial_cosmology}. Therefore the results of such analysis could {\it a priori} depend on the fiducial value chosen for this parameters. We quantified in previous works \citep{hector_bispectrum1,gil-marin15_rsd} the impact of such assumption, assuming that these parameters can oscillate $\pm1\sigma$ about the {\it Planck15} measurements. We found that the shifts in the cosmological parameters of interest was of order of $\sim1/5$ of the statistical errors (see for example fig. 10 of \citealt{gil-marin15_rsd}). 

We perform the parameter estimation assuming that the power spectrum and bispectrum are drawn from a multivariate Gaussian distribution and use,
\begin{equation}
\chi^2({\bf \Psi})=[\Delta D({\bf \Psi})] [{\widetilde{C}^{-1}}][\Delta D({\bf \Psi})]^t,
\end{equation}
where, $\Delta D({\bf \Psi})$ is the vector whose elements contain the difference between the data and the model for the power spectrum monopole, quadrupole and bispectrum monopole; and $\widetilde{C^{-1}}$ is the estimator of the inverse covariance matrix. 
By minimising the $\chi^2$ function respect to ${\bf \Psi}$ we obtain  the best-fitting set of parameters. The errors associated to each parameter are computed by performing Markov chains (\textsc{mcmc}-chains) using as a starting point the best-fitting solution found by the minimisation procedure.
We use a simple Metropolis-Hasting algorithm and we ensure its convergence performing the Gelman-Rubin convergence test. 

We do not opt for fitting the bispectrum and power spectrum signals separately. Minimising the $\chi^2$ function only based on the bispectrum signal would generate a set of results in terms of cosmological parameters, $f$, $\sigma_8$, $H$ and $D_A$ that would be difficult to combine a posteriori with the results from the power spectrum only fits, even with the appropriate parameter covariance. One of the reasons is that the bispectrum signal is not sensitive to the $f\times\sigma_8$ quantity, but a more complex combination of these two variables, as it can be inferred from the tree-level expansion (see eq. C19-C26 as well as fig. 4 of \citealt{hector_bispectrum1}). Because of this, we think that the best approach to this problem is to build a full $\chi^2$ function, including both power spectrum and bispectrum, whose degeneration is dominated by $f\times\sigma_8$ and the contours for the cosmological parameters of interests are very close to those drawn from a multivariate Gaussian distribution, as we will later show in Fig. \ref{cosmo_covariance2}.

\subsection{Comparing the mocks and Halo $N$-body simulations to the data}\label{sec:mocksdata}

\begin{figure*}
\centering
\includegraphics[scale=0.3]{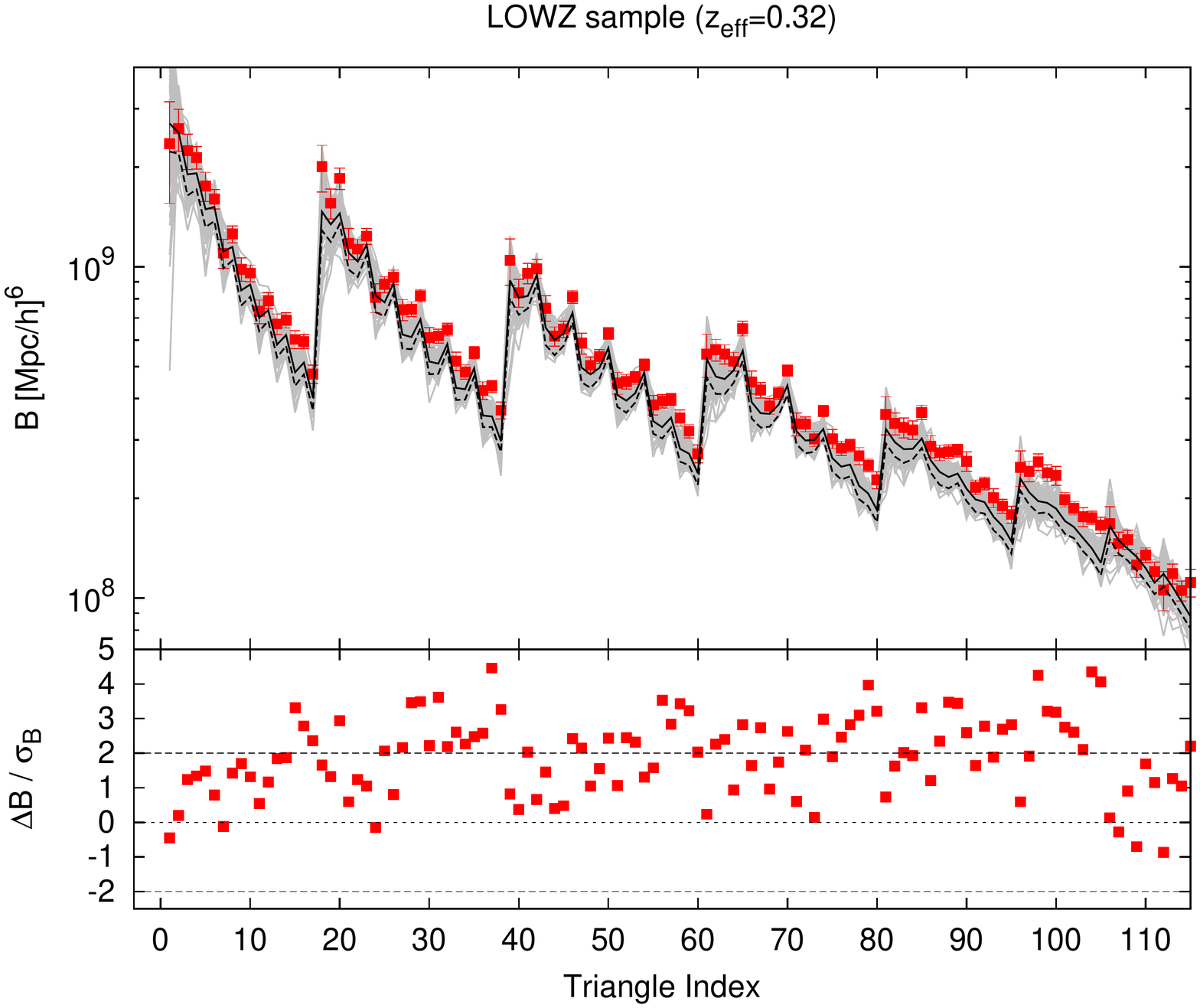}
\includegraphics[scale=0.3]{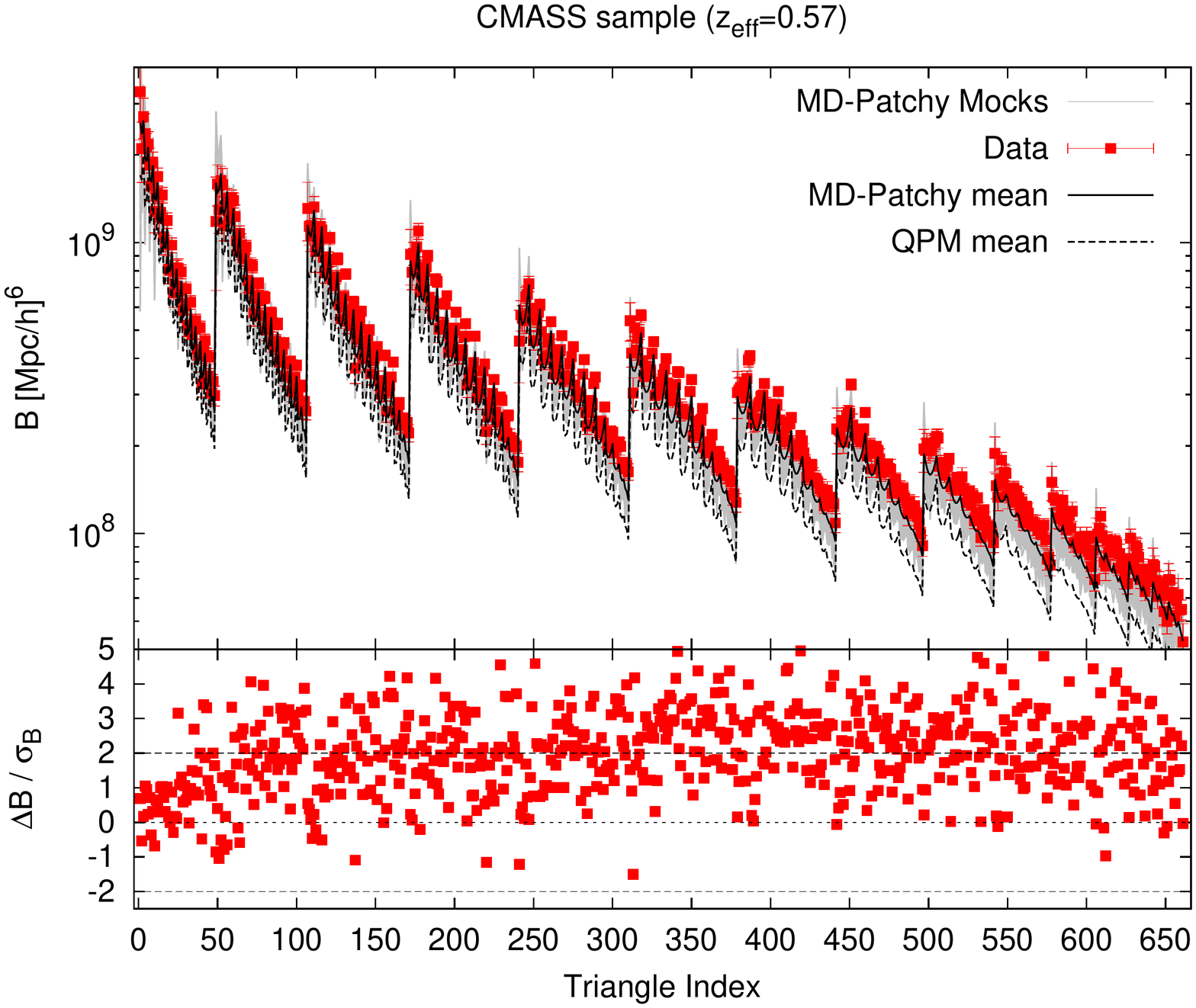}
\caption{The top sub-panels display the performance of the bispectrum of the  \textsc{MD-Patchy} mocks (black solid line for the mean of the mocks and grey lines for each of the first 100 mock realisation), the \textsc{qpm} mocks (black dashed line),  compared to the data with the $1\sigma$ error-bars extracted from the $\it rms$ of the MD-\textsc{patchy} mocks (red symbols) as a function of the triangle index (same definition that in Fig.~\ref{fig:bispectrum_data}). The bottom sub-panels display the difference between the data and the mean of the  \textsc{MD-Patchy} mocks relative to the error of the data: $(B^{\rm data}-B^{\rm mocks})/\sigma_B$ The left panel display the results for LOWZ and the right panel for CMASS, as labeled. From the bottom sub-panels we see that, both for LOWZ and CMASS samples, the mocks reproduce well the data at large scales, but as we explore smaller scales (higher triangle index) there is a systematic offset of order of $\sim2\sigma$ between the data and the mocks, having the data a higher amplitude. }
\label{fig:mocks}
\end{figure*}

In this section we compare the bispectrum signal of the data with that extracted from the \textsc{MD-Patchy} galaxy and $N$-body dark matter halo simulations mocks to check whether they are realistic representation of the actual data. The top sub-panels of Fig.~\ref{fig:mocks}  displays the bispectrum of the data (red symbols),  of the  \textsc{MD-Patchy} mocks (black solid line for the mean and grey lines for each of the first 100 realisations) and of the \textsc{qpm} mocks (black dashed line). The left and right panels correspond to the LOWZ and CMASS sample, respectively. The error-bars of the data, $\sigma_B$, are the {\it rms} from the MD-\textsc{patchy} mocks. 
In the bottom sub-panel we show the difference between the bispectrum of the data and the bispectrum of the mean of the  \textsc{MD-Patchy} mocks relative to $\sigma_B$. At large scales the difference between the data and the  \textsc{MD-Patchy} mocks is well within the $2\sigma$ dispersion. As we move to smaller scales, and consequently to higher values of the triangle index, the data show a constant offset with respect to the mean of the  \textsc{MD-Patchy} mocks of about $\sim2\sigma$. We can quantify this deviation by performing a $\chi^2$ of the mean to each realisation and to the data. We obtain that for LOWZ the combined power spectrum and bispectrum reduced $\chi^2$ up to $k_{\rm max}=0.15\,h{\rm Mpc}^{-1}$ is 1.30 for LOWZ and 1.63 for CMASS. As we move to larger scales the reduced $\chi^2$ values approach 1: for e.g. at $k_{\rm max}=0.12\,h{\rm Mpc}^{-1}$ we obtain 1.12 for LOWZ and 1.15 for CMASS.  These differences can be due to differences in the bias parameters between mocks and data.
We check in the next section whether we can recover the input cosmological parameters, $f$, $\sigma_8$, $\alpha_\parallel$ and $\alpha_\perp$ from these mocks.

The \textsc{qpm} mocks present a significantly different bispectrum than the  \textsc{MD-Patchy} mocks for the CMASS sample, especially at small scales, having a significant lower amplitude. We believe that this strong difference between \textsc{qpm} and  \textsc{MD-Patchy} mocks for the CMASS sample is related to their different biasing properties. We have found that the best-fitting bias parameters for the \textsc{MD-Patchy} mocks are: $b_1=1.97$ and $b_2=0.29$ (as it is later reported in Table~\ref{table:haloes}) while the \textsc{qpm} mocks have best-fitting values: $b_1=1.90$ and $b_2=-0.23$; the difference is especially notable for $b_2$, which flips from being positive for \textsc{MD-Patchy} to be negative for the  \textsc{qpm} mocks. In both cases the displayed bias parameters are the best-fitting to the mean of the 1000 and 2048 simulations, respectively,  at $k_{\rm max}=0.17\,h{\rm Mpc}^{-1}$. We do not quote the error-bar of the fit, as it is several order of magnitude smaller than the difference we want to stress.

We apply 3 halo mass cuts on the $N$-body dark matter haloes at 3.80, 5.75 and $8.36\times10^{12}\,M_\odot h^{-1}$, which select  haloes with more than 50, 75 and 110 particles, respectively. We name these 3 different haloes catalogues, {\it low-bias}, {\it mid-bias}  and {\it high-bias}, respectively. In all cases the redshift snapshot is at $z=0.55$.  
The properties of these halo catalogues are summarised in Table~\ref{table:haloes}. Note that since these 3 halo catalogues rely on the same underlying dark matter distribution their clustering is correlated. However, the potential relative change in their best-fitting parameters is mainly cosmic variance free, and can be used as a test of systematics of the model. For comparison, Table~\ref{table:haloes} displays as well the properties for the MD-\textsc{Patchy} mocks as well as for the data, in both CMASS and LOWZ samples.  
\begin{table}
\begin{center}
\begin{tabular}{ccccc}
Name &$M_{\rm min}^{\rm h}\, ( N_p )$ & $b_1$ & $b_2$ & $\bar{n}\times10^4$ \\
\hline
\hline
Low-bias $N$-body & 3.80 (50) & 1.75 & -0.26 & 6.75 \\
Mid-bias $N$-body & 5.75 (75) & 1.90 & 0.22 & 4.41  \\
High-bias $N$-body & 8.36 (110) & 2.07 & 0.49 & 2.90 \\ 
\hline
MD-\textsc{Patchy} (LOWZ) & - & $2.08$ & $0.43$ & $4.0$ \\
MD-\textsc{Patchy} (CMASS) & - & $1.97$ & $0.29$ & $4.5$ \\
 \hline
 Data (LOWZ) & - & $2.08$ & $0.92$ & $4.0$ \\
 Data (CMASS) & - & $2.01$ & $0.68$ & $4.5$ \\
\end{tabular}
\end{center}
\caption{Biasing properties of $N$-body haloes, MD-\textsc{Patchy} mocks and data. $N$-body dark matter halo catalogues are listed according to their mass cuts,  $M_{\rm min}^{\rm h}$: low-bias, mid-bias and high-bias. The columns show the bias properties for each of the catalogues. For haloes $b_1$ is estimated from the ratio between the real space power spectrum and the linear power spectrum at large scales. For mocks and data $b_1$ is estimated from the best-fitting model of the power spectrum and bispectrum up to $k_{\rm max}=0.17\,h{\rm Mpc}^{-1}$. For both $N$-body haloes, MD \textsc{Patchy} mocks and data, the quoted values of $b_2$ are estimated from the best-ft of the power spectrum and bispectrum up to $k_{\rm max}=0.17\,h{\rm Mpc}^{-1}$. When estimating the bias parameters the results are degenerate with $\sigma_8$, being $b_i\sigma_8$ the estimated quantity. We then divide $b_i\sigma_8$ by the true value of $\sigma_8$ in case of mocks and $N$-body haloes, and by the fiducial value for the data. The last column shows the number density of each catalogue in $({\rm Mpc}/h)^{-3}$ units. In the case of the mocks and data, the $n(z)$ is variable, and we only quote their maximum value. All masses are in units of $ [10^{12} M_\odot h^{-1}]$ and $M_{\rm min}^{\rm h} / m_p\equiv N_p$ is the minimum number of particles per halo. }
\label{table:haloes}
\end{table}%

\begin{figure*}
\centering
\includegraphics[scale=0.3]{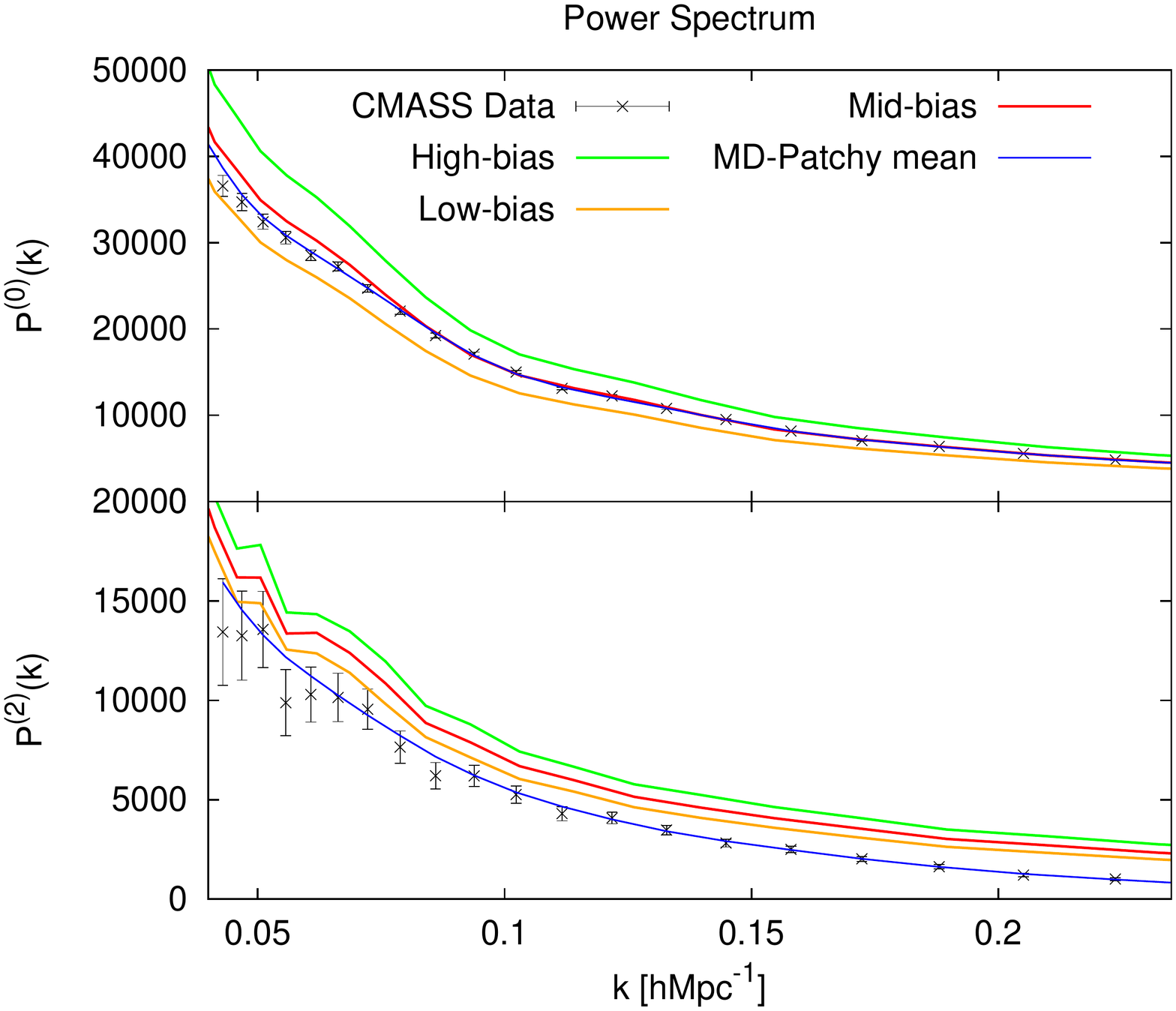}
\includegraphics[scale=0.3]{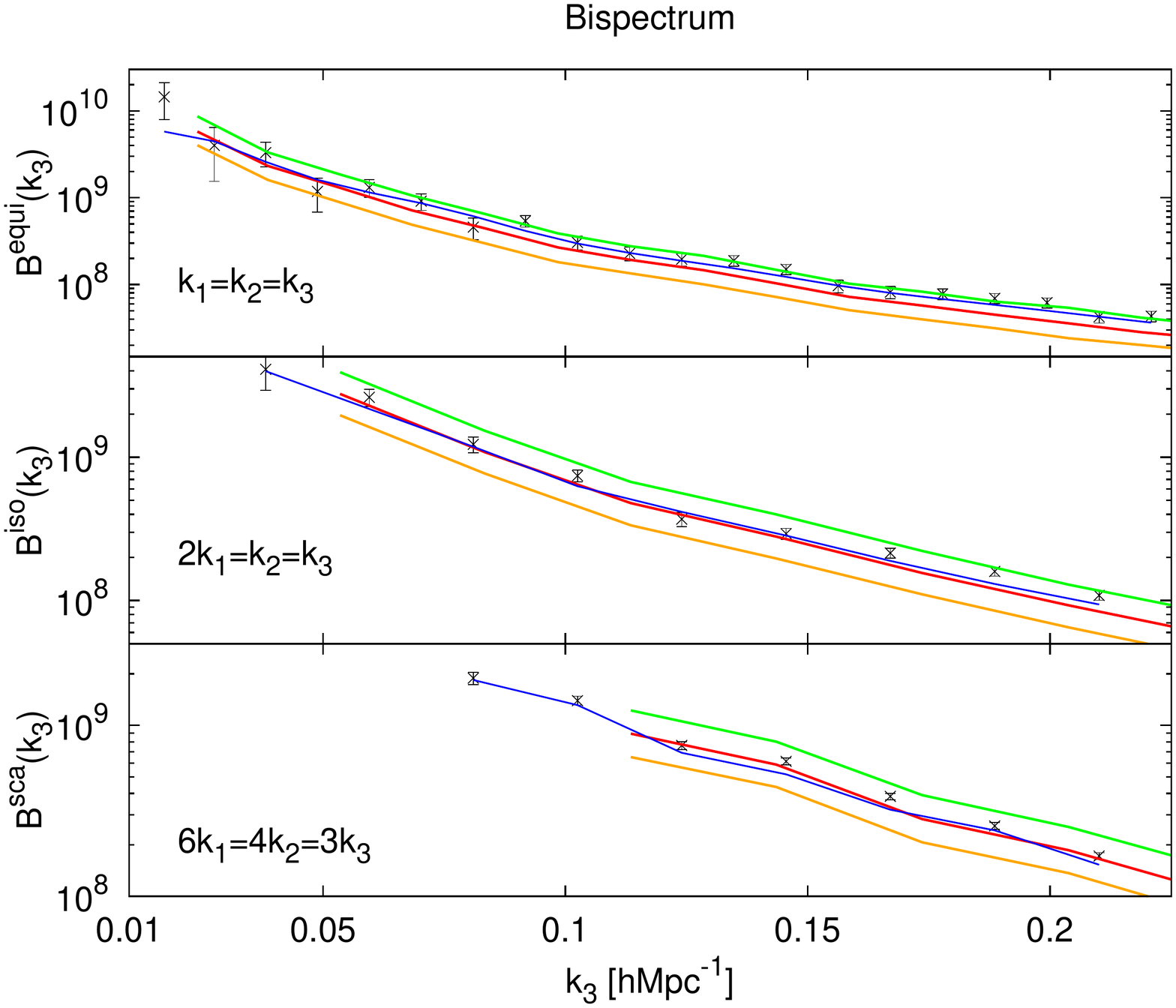}
\caption{The left panel presents the performance of the power spectrum monopole and quadrupole for the $N$-body dark matter halo catalogues, the MD-\textsc{Patchy} mocks, and the CMASS sample data as labeled; the right panel present the bispectrum for equilateral triangles (top sub-panel), isosceles $2k_1=k_2=k_3$ (middle sub-panel) and scalene $6k_1=4 k_2 =3 k_3$ (bottom sub-panel). The mass cuts of the different $N$-body halo catalogues are: $M_{\rm min}=3.80\times10^{12}M_\odot h^{-1}$ (low-bias haloes, in orange lines), $M_{\rm min}=5.75\times10^{12}M_\odot h^{-1}$ (mid-bias haloes, in red lines) and $M_{\rm min}=8.36\times10^{12}M_\odot h^{-1}$ (high-bias haloes in green lines). Details on the bias properties of these catalogues are presented in Table~\ref{table:haloes}.  The units of the power spectrum are $[({\rm Mpc}/h)^3]$ and of the bispectrum $[({\rm Mpc}/h)^6]$.}
\label{fig:data_nbody}
\end{figure*}

The measurements from the $N$-body haloes are compared with those from the data in Fig.~\ref{fig:data_nbody}, along with the CMASS data, and MD-\textsc{Patchy} mocks. The left panel presents the performance for the power spectrum monopole and quadrupoles, whereas the right panel for the bispectrum, for 3 particular shapes: equilateral (top sub-panel), isosceles $2k_1=k_2=k_3$ (middle sub-panel) and scalene $6k_1=4 k_2 =3 k_3$ (bottom sub-panel). The bias parameters of each of these halo catalogues are later reported in Table~\ref{table:haloes}. Note that in this case we do not present the bispectrum as a function of the triangle index as we have been doing so far. The reason is that the triangle-binning for the halo catalogues is different than that of mocks and data because the size of the periodic box for the halo catalogues is substantially smaller. Therefore, we choose to present the bispectrum as a function of the largest $k$, for 3 different triangle shapes. 

In terms of the power spectrum monopole, we see that the MD-\textsc{Patchy} mocks perfectly describe the clustering of the data up to $k\simeq0.24\,h{\rm Mpc}^{-1}$. The halo catalogue, corresponding to a mass cut of $M_{\rm min}=5.75\times10^{12}M_\odot h^{-1}$ (red lines), also shows an excellent agreement with the power spectrum monopole up to small scales. The differences on large scales $k\leq0.1\,h{\rm Mpc}^{-1}$, are caused by the sky selection function (which is not accounted in the $N$-body halo catalogues, but only in the data and mocks), which has the effect of reducing power at large scales.  The other two $N$-body catalogues, low-bias and high-bias. present lower and higher clustering, respectively, with respect to the data at all scales. At the level of the power spectrum monopole, both mid-bias haloes and MD-\textsc{Patchy} mocks shows an excellent agreement with the data. On the other hand the power spectrum quadrupole shows a significantly different behaviour. In order to create the MD-\textsc{Patchy} mocks, a large dispersion velocity had to be assigned to low mass objects considered satellite galaxies (see sections 2.2.4 of \citealt{Kitauraetal:2015} for further details) to match the observed data quadrupole. As the $N$-body halo catalogues have no satellite galaxies, they cannot be expected to account for such an effect, which explains the difference in the quadrupole signal. 

The performance of the bispectrum monopole for the MD-\textsc{Patchy} mocks, data and mid-halo $N$-body catalogue is similar for the isosceles and scalene shapes. However, we observe substantial difference for the equilateral triangles: at $k\geq0.15\,h{\rm Mpc}^{-1}$,  where the data is in better agreement with the bispectrum signal of the high-bias halo catalogue, rather than with the bispectrum signal of the mid-bias halo catalogue, as in the power spectrum monopole. This behaviour is also observed for the isosceles triangles at $k\geq 0.17\,h{\rm Mpc}^{-1}$. We see that MD-\textsc{Patchy} mocks are able to describe well the observed bispectrum from the CMASS sample data.  The $N$-body mid-bias halo catalogue also presents a good agreement with the CMASS sample data, except for the equilateral shapes at small scales, where the data is more in agreement with the $N$-body high-bias catalogue.

\section{Finding systematic errors for cosmological parameters}\label{sec:tests}
In this section we quantify the level of systematic errors on the measured cosmological parameters: $f$, $\sigma_8$, $\alpha_\parallel$ and $\alpha_\perp$, for a particular range of scales over which the power spectrum and bispectrum are modelled. In order to do so, we assume that the primary source of the systematic error lies in the mis-match between the theoretical model described in \S\ref{sec:model} and the true Universe, and furthermore that the \textsc{MD-Patchy} mocks and $N$-body dark matter haloes power spectra and bispectra measurements provide similar levels of disagreement as the data, and that these translate into similar inaccuracies in parameter measurement.

We start by taking the average of the power spectrum and bispectrum measurements of the 2048 realisations of the \textsc{MD-Patchy} mocks and that of the 20 realisations of the $N$-body dark matter halo simulations. For the galaxy mocks this reduces the statistical error by a factor of $\sqrt{2048}\simeq45$, so that any deviation between the measured signal on the mocks and the model must be caused by either systematics in the modelling or an inability of the theoretical model to fit the measured statistic.
For the 20 realisations of the dark matter haloes we estimate the covariance from 1000 of the redshift $z=0.505$ periodic cubic boxes from which the \textsc{MD-Patchy} mocks were drawn. 

\begin{figure*}
\centering
\includegraphics[scale=0.3]{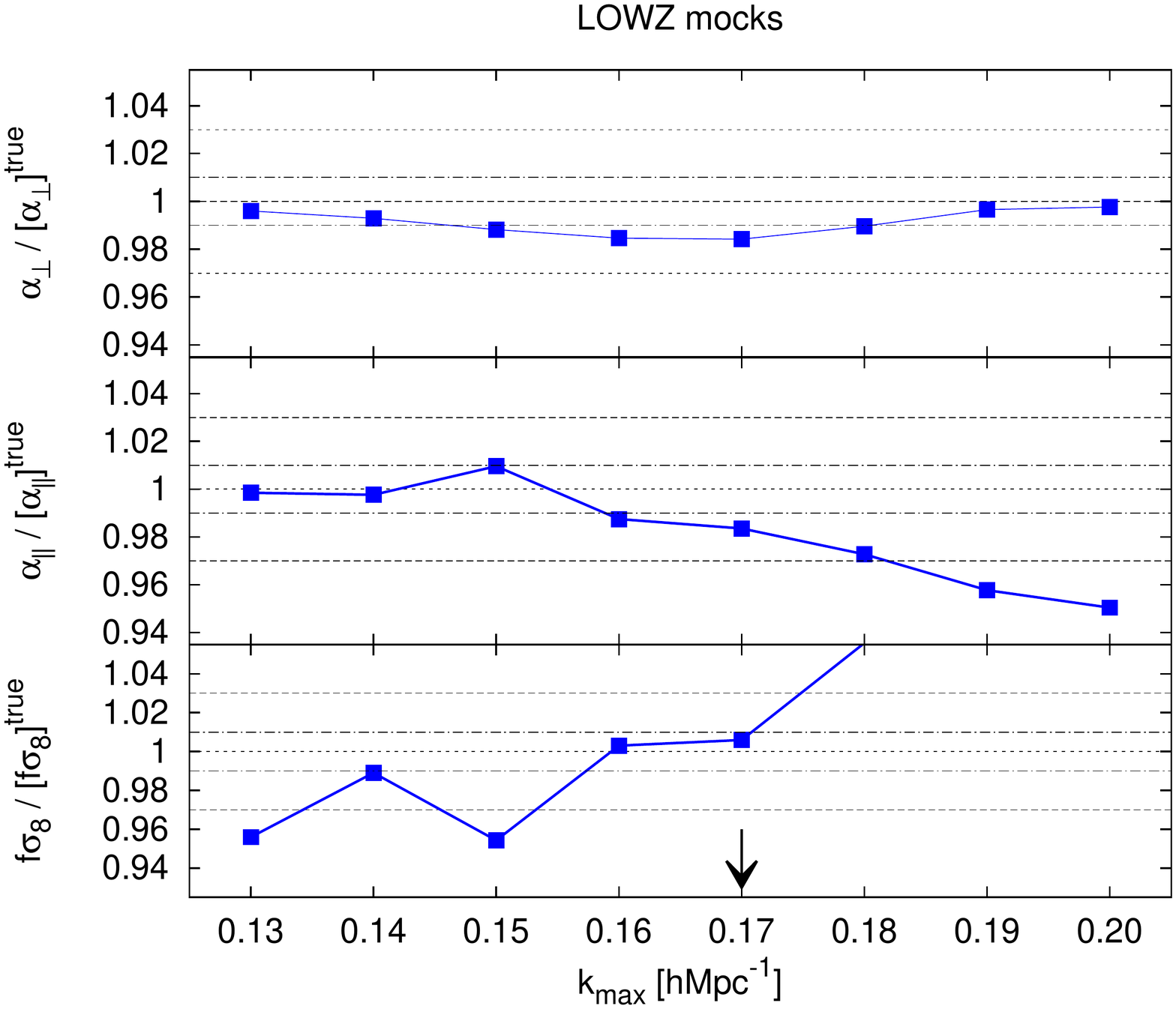}
\includegraphics[scale=0.3]{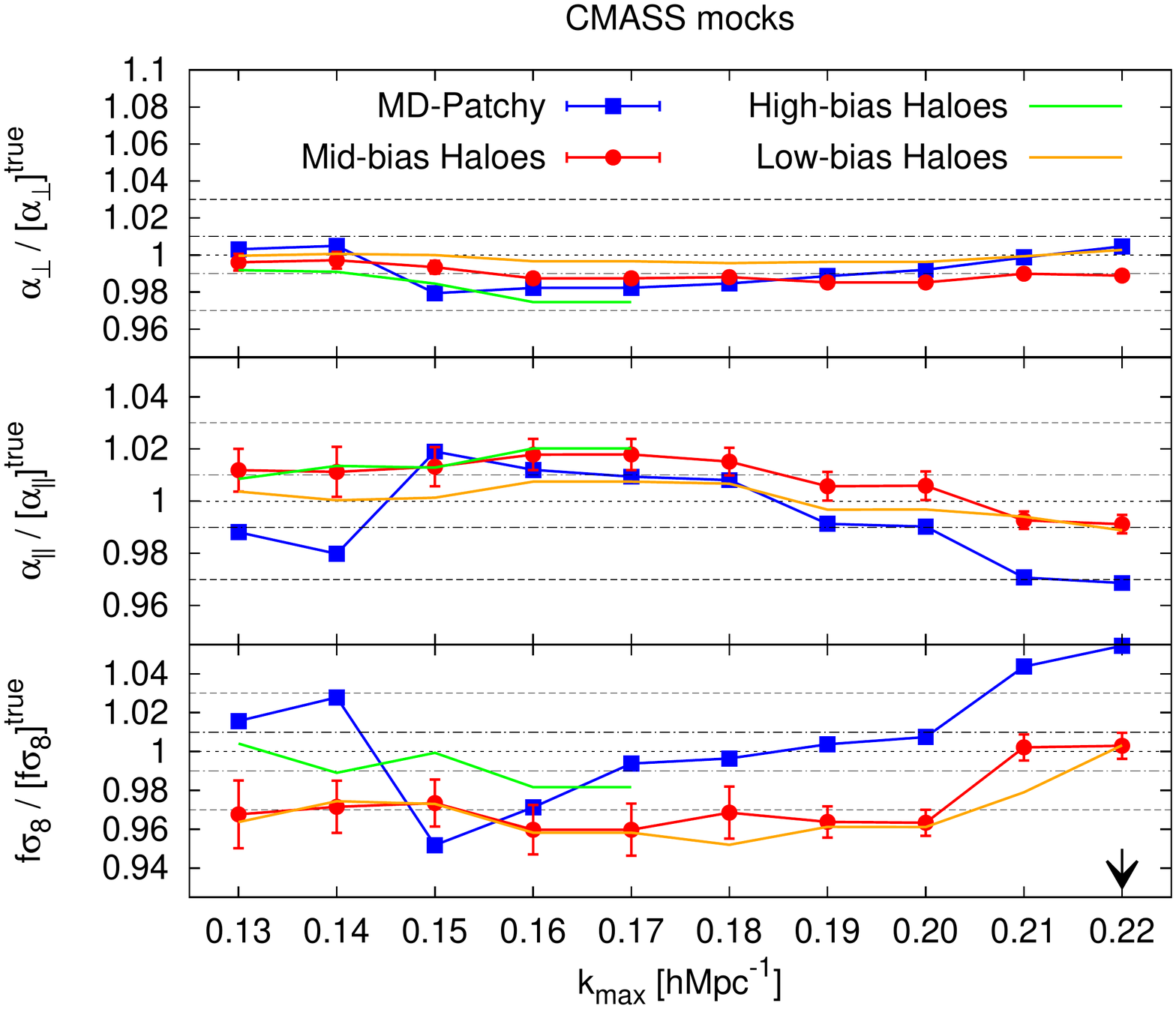}

\includegraphics[scale=0.3]{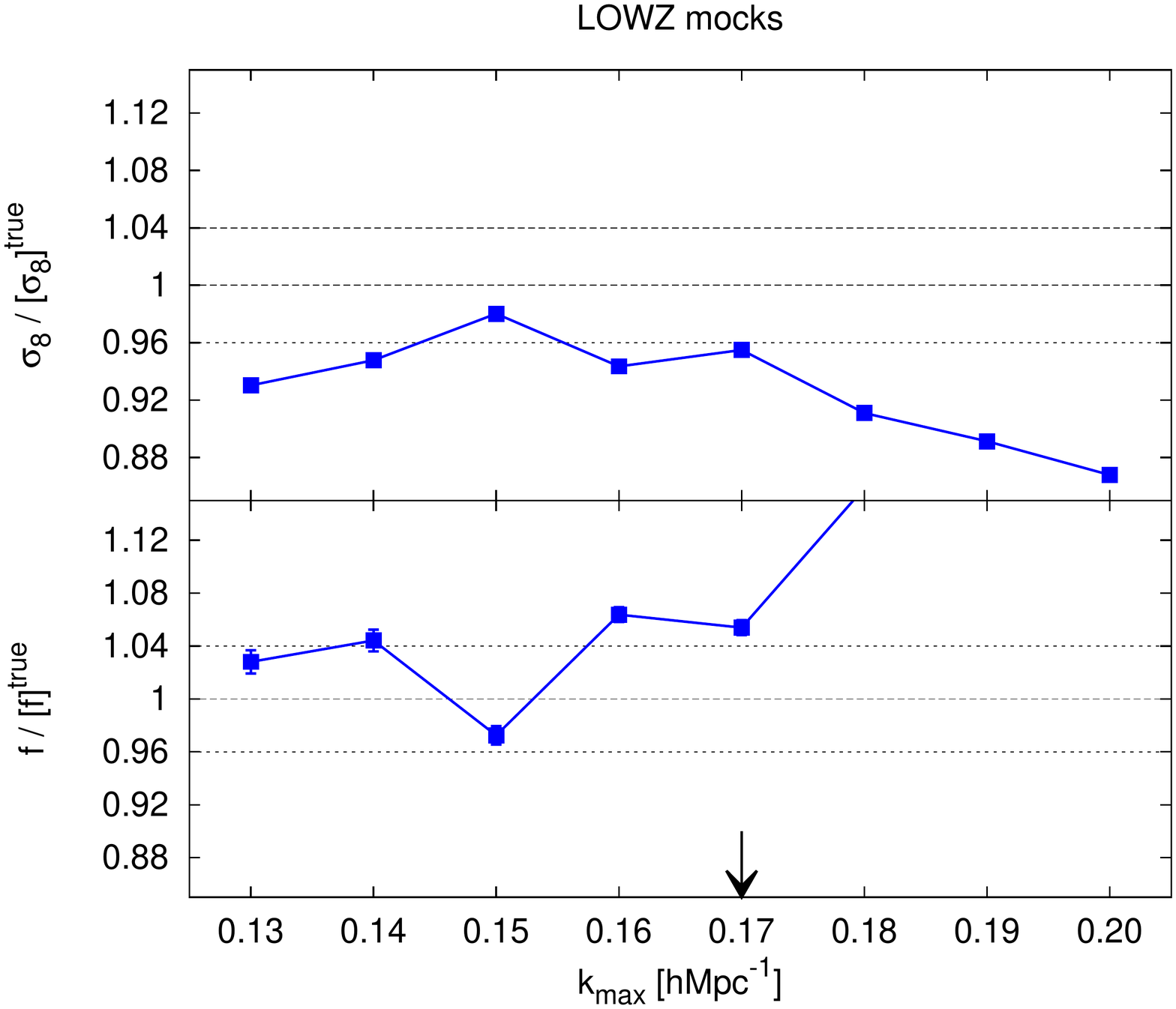}
\includegraphics[scale=0.3]{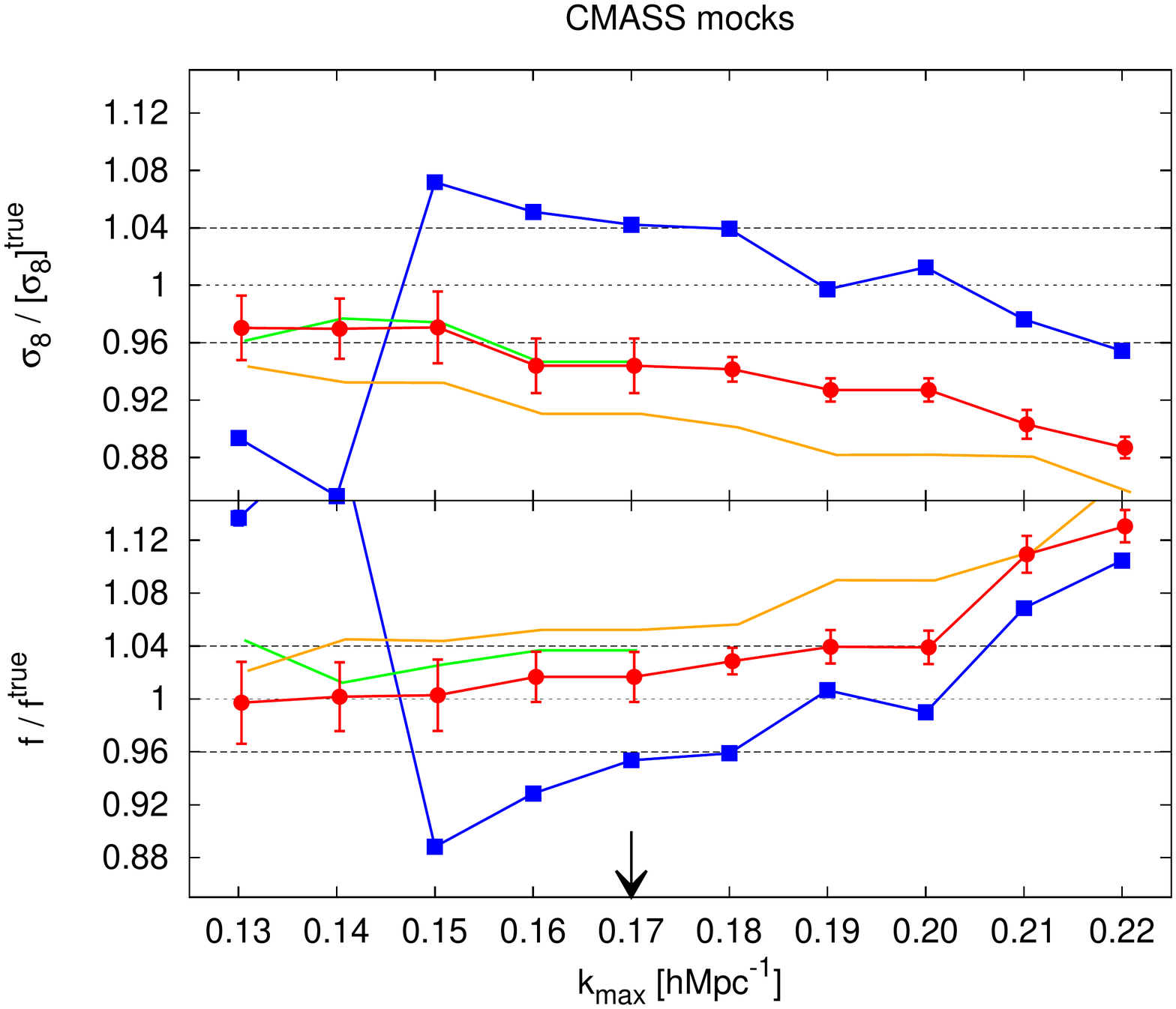}
\caption{The different colour lines and symbols display the best-fitting $\alpha_\parallel$, $\alpha_\perp$ and $f\sigma_8$ parameters (top panels) and $f$ and $\sigma_8$ (bottom panels) as a function of the minimum scale  used for fitting, $k_{\rm max}$ obtained by a combined fit to the power spectrum monopole, quadrupole and bispectrum monopole. Left panels display the performance for the average of 2048 realisations of the LOWZ MD-\textsc{Patchy} (blue symbols and lines). Right panels display the results for the average of 2048 realisations of the CMASS MD-\textsc{Patchy} mocks (blue symbols and lines) and the average of 20 realisations of  $N$-body dark matter haloes, in different colours according to the minimum mass cut applied (see Table~\ref{table:haloes} for details). On the top panels, horizontal dashed, dot-dashed and dotted lines show the $0\%$, $1\%$ and $3\%$ deviations, respectively, with respect to the corresponding true value for reference. On the bottom panels, horizontal dashed, dotted lines show the $0\%$ and $4\%$, respectively. The black arrows mark the truncation scale chosen for each set of variables.  }
\label{fig:kmax_mocks}
\end{figure*}

Fig.~\ref{fig:kmax_mocks} shows the best-fitting parameters as a function of the truncation scale, $k_{\rm max}$.  For the LOWZ sample we observe that, up to $k_{\rm max}=0.17\,h{\rm Mpc}^{-1}$, the AP parameters $\alpha_\parallel$ and $\alpha_\perp$ are recovered with a percentage accuracy of $\lesssim1.5\%$. For larger values of $k_{\rm max}$ we observe a significant underestimation on $\alpha_\parallel$, whereas $\alpha_\perp$ remains unbiased at the $\simeq1\%$ accuracy. The $f\sigma_8$ parameter is recovered at $\lesssim1\%$ accuracy for $0.16\leq k_{\rm max}\, [h{\rm Mpc}^{-1}]\leq 0.17$. For larger values of $k_{\rm max}$ $f\sigma_8$ is largely overestimated, whereas for smaller truncation $k$-values it oscillates between being slightly underestimated, $\simeq 4\%$, to just $\simeq1\%$. The recovered  $f$ and $\sigma_8$ present opposite behaviours. For $k_{\rm max}\leq0.17\,h{\rm Mpc}^{-1}$,  $f$ is mainly over-estimated by around $4\%$, whereas $\sigma_8$ is underestimated by around $\sim8\%$ in the worst case. For larger values of $k_{\rm max}$, the systematic errors grow rapidly in both parameters. From the performance of the theoretical model on the MD-\textsc{Patchy} mocks for the LOWZ sample we can establish that the truncation scale for the LOWZ sample data should be around $k_{\rm max}=0.17\,h{\rm Mpc}^{-1}$. The expected size of the statistical errors is larger than such corrections. We will see later in Table~\ref{table:results} that the statistical errors of the LOWZ sample of the data are: on $f\sigma_8$, 14~\%;  on $\alpha_\parallel$, 4.4~\%; and on $\alpha_\perp$, 2.9~\%; for a similar $k_{\rm max}$ cut. 

While the $N$-body simulations should describe gravitational clustering correctly down to non-linear scales, these dark matter haloes do not present the same clustering structure of the galaxies observed in CMASS, and therefore the systematic features observed for the haloes could be different than those obtained by applying the model to the actual data. On the other hand, the MD-\textsc{Patchy} mocks describe better the clustering observed in the data (see for example Fig.~\ref{fig:mocks} ), but the modelling of non-linear gravitational evolution cannot be as precise as a full $N$-body simulation. Because of this, we  try to find a compromise, in terms of systematic shifts, found by the MD-\textsc{Patchy} mocks and the $N$-body dark matter haloes. For the high-bias halo catalogue we only display results up to $k_{\rm max}=0.17\,h{\rm Mpc}^{-1}$. For higher values of $k$, the gravitational bispectrum signal (shot noise subtracted) becomes equal to its shot noise component, and therefore the information on those scales it is not very reliable, due to the low signal-to-noise ratio.

The systematic shifts on the AP parameters are similar for both dark matter haloes and  MD-\textsc{Patchy} mocks. For $\alpha_\perp$ we see that in all the $k_{\rm max}$-range studied the systematic shifts are of order $\lesssim1\%$, being higher for the high-bias halo catalogue, which reaches $\sim2\%$ at $k_{\rm max}=0.17\,h{\rm Mpc}^{-1}$.  For the $\alpha_\parallel$ we observe that up to $k_{\rm max}=0.18\,h{\rm Mpc}^{-1}$ the systematic shifts tend to over-estimate the parameter by $\sim1\%$ for both haloes and mocks. For higher $k_{\rm max}$ values the deviations in the low- and mid-bias halo catalogues stays below $1.5\%$. For the MD-\textsc{Patchy} mocks the systematics stays within $2\%$ for $k_{\rm max}\leq0.20\,h{\rm Mpc}^{-1}$ and reaches $3\%$ for higher values. 

The systematic shifts on the $f\sigma_8$ parameter for the $N$-body low- and mid-bias haloes is around $3\%$ for $k_{\rm max}\leq0.20\,h{\rm Mpc}^{-1}$, which reduces for higher $k_{\rm max}$ cuts. For the high-bias halo catalogue $f\sigma_8$ presents a lower systematic bias for $k_{\rm max}\leq0.17\,h{\rm Mpc}^{-1}$, staying below $2\%$. The systematics on the MD-\textsc{Patchy} mocks present a different behaviour: there is a moderate oscillatory systematic shifts up to $k_{\rm max}=0.16\,h{\rm Mpc}^{-1}$ and a stable $k_{\rm max}$-region of $0.17\leq k_{\rm max}\,[h{\rm Mpc}^{-1}] \leq 0.20$, where the systematic shifts stay below $1\%$. The expected size of the statistical errors are of the same order or larger than the observed systematics. We will see later in Table~\ref{table:results} that the for the statistical errors of the CMASS sample of the data are: on $f\sigma_8$, 5.7~\%;  on $\alpha_\parallel$, 1.7~\%; and on $\alpha_\perp$, 1.2~\%; for a similar $k_{\rm max}$ cut. 

When we analyse the  $f$ and $\sigma_8$ parameters  separately (bottom panels), we see that for the $N$-body dark matter haloes, there is a smooth transition in the range $0.13\leq k_{\rm max}\,[h{\rm Mpc}^{-1}]\leq0.17$, where $f$ presents systematic shifts of $\lesssim4\%$ and where $\sigma_8$ varies from $\lesssim4\%$ to $8\%$, being the mid-bias halo catalogue the less affected. At higher values of $k_{\rm max}$, for the low- and mid-bias halo catalogues, $f$ and $\sigma_8$ get individually under- and over-estimated, respectively. The behaviour on the $f$ and $\sigma_8$ best fit values on the MD-\textsc{Patchy} mocks is significantly different: there is an oscillatory behaviour at small values of $k_{\rm max}$. After that, there is a $k_{\rm max}$-range where $\sigma_8$ goes from being over-estimated by $\simeq 4\%$ at $k_{\rm max}=0.16\,h{\rm Mpc}^{-1}$ to being underestimated by $\simeq4\%$ at $k_{\rm max}=0.22\,h{\rm Mpc}^{-1}$; and $f$ goes from being under-estimated by $4\%$ at $k_{\rm max}=0.17\,h{\rm Mpc}^{-1}$ to being over-estimated by $\gtrsim4\%$ at $k_{\rm max}=0.21\,h{\rm Mpc}^{-1}$. 

The observed systematics on the different parameters of the $N$-body dark matter haloes and the MD-\textsc{Patchy} mocks is not always the same. As a general trend, the model used to describe the power spectrum and bispectrum presents a better agreement, in terms of the cosmological parameters $f$, $\sigma_8$, $\alpha_\parallel$ and $\alpha_\perp$,  with the $N$-body catalogues rather than with the MD-\textsc{Patchy} mocks. We know that the $N$-body haloes track very well the dark matter density field component, both in the power spectrum and in the bispectrum, up to non-linear scales. However, the power spectrum and bispectrum signal of the $N$-body haloes is slightly different with respect to the CMASS data, especially for the equilateral bispectrum and for the power spectrum quadrupole,  as shown in Fig.~\ref{fig:data_nbody}. The behaviour at the power spectrum level is well understood and is related to galaxy biasing and not to the underlying dark matter component: using the power spectrum monopole and quadrupole we can recover the $f\sigma_8$ parameter for both $N$-body haloes (shown in fig. 12 and 13 of \citealt{hector_bispectrum0}) and for MD-\textsc{Patchy} mocks (shown in fig. 6 of \citealt{gil-marin15_rsd}). Therefore, the observed differences in the $f\sigma_8$ parameter of Fig.~\ref{fig:kmax_mocks} must rely on differences in the bispectrum signal.  Since we know that for the $N$-body haloes the underlying dark matter component must be correct, the source of disagreement must come from the galaxy bias component, as the cosmology of the $N$-body haloes and {\it Planck15} are not that different. Tuning the minimum mass of the haloes is not sufficient for perfectly fitting the bispectrum signal of the data, although the agreement is reasonable good. Therefore, the bispectrum signal from the LRGs is dominated by the minimum mass of their host haloes, but for perfectly describing the full signal we would need to account for an additional bias parameter, which is likely related to the way the satellite galaxies are distributed inside the host haloes. The signal of the LRG satellite fraction, estimated around $10\%$,  is not captured by the $N$-body haloes and must have some contribution in the total bispectrum signal, yielding for (at least part of) the observed difference between $N$-body haloes and data. On the other hand the MD-\textsc{Patchy} galaxy mocks, account for this complex galaxy bias component of the observed LRGs through several internal free bias parameters (see section 3.3 from \citealt{Kitaura_gilmarinetal:2015} for details). Thus, the mocks should in principle be a better description of the LRG galaxy bias than the $N$-body haloes. However, we know that the mocks cannot keep track of full non-linear signal, as they rely on some approximation scheme (2LPT for the \textsc{qpm} mocks and ALPT for \textsc{Patchy}). The bispectrum signal is essentially a non-linear component of the density field (under the assumption of primordial Gaussian field). Therefore is not surprising that these schemes cannot fully track their behaviour at small scales. This point has a direct relation with the disagreement between the data and MD-\textsc{Patchy} mocks observed in Fig.~\ref{fig:mocks}. Thus, we decide to calibrate the truncation scale of our model, $k_{\rm max}$, according to their performance with the $N$-body haloes signal, in particular the mid-bias halo catalogue, which has the correct non-linear signature but a slightly different bias properties than those from the observed LRGs. We set the truncation scale for the data at around  $k_{\rm max}=0.22\,h{\rm Mpc}^{-1}$ for the CMASS sample, when estimating the AP parameters and $f\sigma_8$. In the case we are interested in recovering $f$ and $\sigma_8$ separately, the truncation scale we choose is  $k_{\rm max}=0.17\,h{\rm Mpc}^{-1}$.

\begin{table}
\begin{center}
\begin{tabular}{|c|c|c}
& LOWZ & CMASS \\
\hline
\hline
$f\sigma_8$ & $5.5\%$ & $5.5\%$ \\
$\sigma_8$ & $8\%$ & $8\%$ \\
$f$ & $4\%$ & $4\%$ \\
$\alpha_\parallel$ & $1.5\%$ & $1\%$ \\
$\alpha_\perp$ & $1.5\%$ & $1\%$ 
\end{tabular}
\end{center}
\caption{Systematic errors assumed on cosmological parameters caused by inaccuracies in the modelling for the LOWZ and CMASS samples. For LOWZ these systematics correspond to $k_{\rm max}=0.17\,h{\rm Mpc}^{-1}$. For CMASS these systematics correspond to a truncation scale of $k_{\rm max}=0.22\,h{\rm Mpc}^{-1}$ for $f\sigma_8$, $\alpha_\parallel$ and $\alpha_\perp$ and to a scale of $k_{\rm max}=0.17\,h{\rm Mpc}^{-1}$ for $f$ and $\sigma_8$.}
\label{table:systematics_model}
\end{table}

\subsection{Systematic test using the blind mock challenge results}
We participated in a blind mock challenge proposed by the BOSS galaxy clustering group \citep{Tinkeretal:2016}. This challenge was divided into two kind of analyses:  we had to recover the $f\sigma_8$ parameter with a similar truncation scale used in this paper for i) different periodic simulation boxes with different cosmologies and halo occupation distribution (HOD) parameters; ii) from a set of 84 N-body based mock catalogues with the same sky geometry than the CMASS sample.   We find that the deviation on the case i) is $\Delta f\sigma_8=0.0082$ when the power spectrum is used alone, and $\Delta f\sigma_8=0.024$ when we add the bispectrum signal. For the case ii) we find $\Delta f\sigma_8=0.016$ when the power spectrum is used alone, and $\Delta f\sigma_8=0.012$ when we add the bispectrum signal. Adding both contributions in quadrature, we establish a systematic error budget for the $f\sigma_8$ parameter: $\Delta f\sigma_8=0.018$ for the power spectrum analysis alone, which represent around $3.5\%$ of its value; and $\Delta f\sigma_8=0.027$ for power spectrum in combination with the bispectrum, which represents around $5.5\%$ of its value. The systematic error on $f\sigma_8$ estimated from the power spectrum and bispectrum analysis, $5.5\%$, is higher than the systematic error found by analysing the \textsc{MD-Patchy} mocks and N-body haloes, which was around 4\% and 3\%. In order to be conservative, we have decided to establish the higher systematic error budget found, $5.5\%$, as the error budget for $f\sigma_8$. For a more detailed discussion on the blind mock challenge, used cosmology and HODs we refer the reader to \cite{Tinkeretal:2016}.

Table~\ref{table:systematics_model} summarises the systematic error budget observed on the  LOWZ and CMASS sample in the different panels of Fig.~\ref{fig:kmax_mocks} at the chosen truncation scales. In section \S\ref{sec:results} we will add the total systematic error budget to the statistical errors of the data to obtain a total error budget on the final measurements. Since the shifts on the parameters of Fig.~\ref{fig:kmax_mocks} are sometimes of different magnitude among the different dark matter halo catalogues, we opt to not apply any systematic shift correction on the data, and we will only increase the size of the  errors in order to account for the systematic errors. 

\section{Results}\label{sec:results}
The main results of this paper are presented in Table  \ref{table:results}, which are the parameters of the model that better describes the power spectrum and bispectrum of the data. These values have been obtained by applying the large scales cuts $k_{\rm min}=0.02\,h{\rm Mpc}^{-1}$ for the power spectrum monopole, $k_{\rm min}=0.04\,h{\rm Mpc}^{-1}$ for the power spectrum quadrupole and $k_{\rm min}=0.03\,h{\rm Mpc}^{-1}$ for the bispectrum. For LOWZ and CMASS samples the small scale cut has been set to $k_{\rm max}=0.18\,h{\rm Mpc}^{-1}$ and $k_{\rm max}=0.22\,h{\rm Mpc}^{-1}$, respectively. Table~\ref{table:results} displays the results for both LOWZ and CMASS sample when the full set of 9 parameters are varied. Additionally, we also present the results when the AP variables, $\alpha_\parallel$ and $\alpha_\perp$ are kept fixed to their fiducial values. The first 3 rows correspond to the cosmological parameters of interest.
 The forth and fifth rows correspond to the AP-scale dilations, which are directly related to $Hr_s$ and $D_A/r_s$. The rest of the parameters correspond to either bias parameter or nuisance parameters.  In the last row the value of the $\chi^2$ of the fit is also presented. In all cases the values of $\chi^2$ are sufficiently close to 1 to be considered a good fit. The quoted error-bars correspond to the total error budget, which for the cosmological variables: $f\sigma_8$, $\alpha_\parallel$, $\alpha_\perp$ and derived quantities, include statistical errors derived from the analysis of the likelihood and the systematic errors of the models described in Table~\ref{table:systematics_model}. These 2 error-component have been added in quadrature in order to obtain a final total error budget. In parenthesis we show the contribution from the statistical errors only.  For the rest of variables, the quoted error-bar correspond to the statistical error. The impact of systematics caused by the close pairs has been shown very subdominant for the cosmological parameters compared to their statistical errors and their systematics of the model (see appendix \ref{appendixb} for details).

The model described by the parameters of the columns LOWZ and CMASS, i.e. when the full set of 9 parameters are varied, are plotted in Figs. \ref{fig:powerspectrum_data} and \ref{fig:bispectrum_data} in black lines along with the data points. 

\begin{figure*}
\includegraphics[scale=0.3]{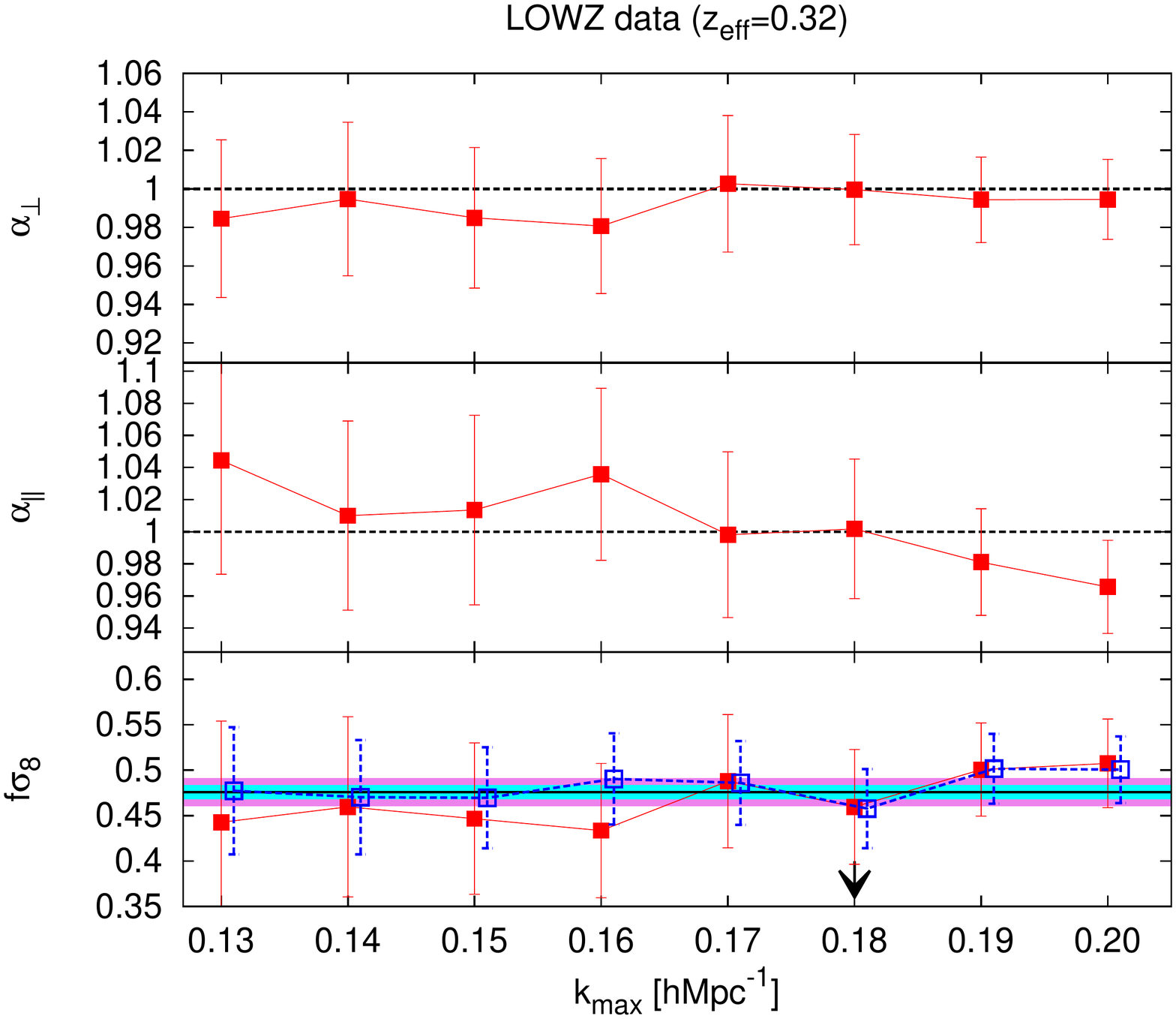}
\includegraphics[scale=0.3]{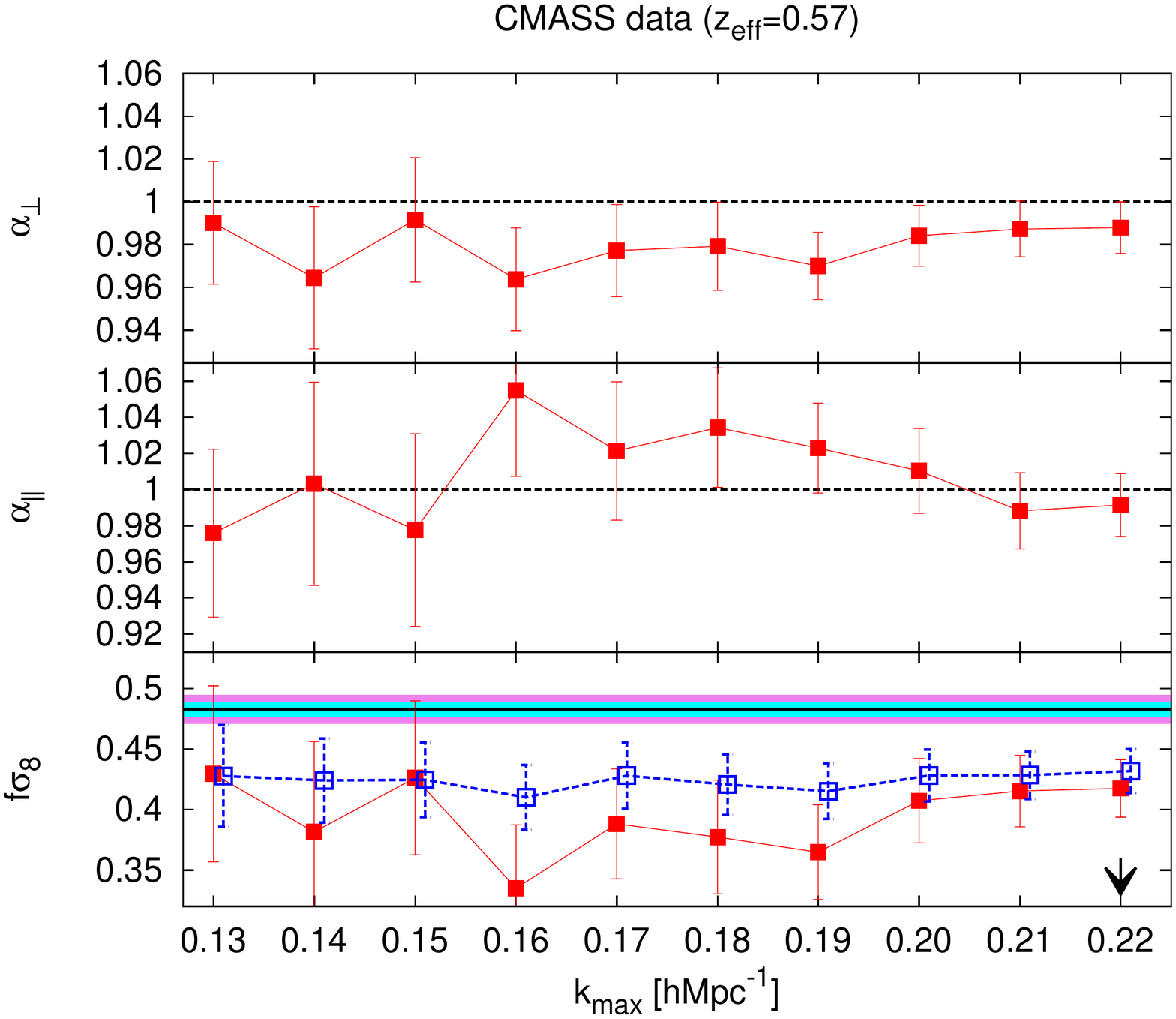}

\includegraphics[scale=0.3]{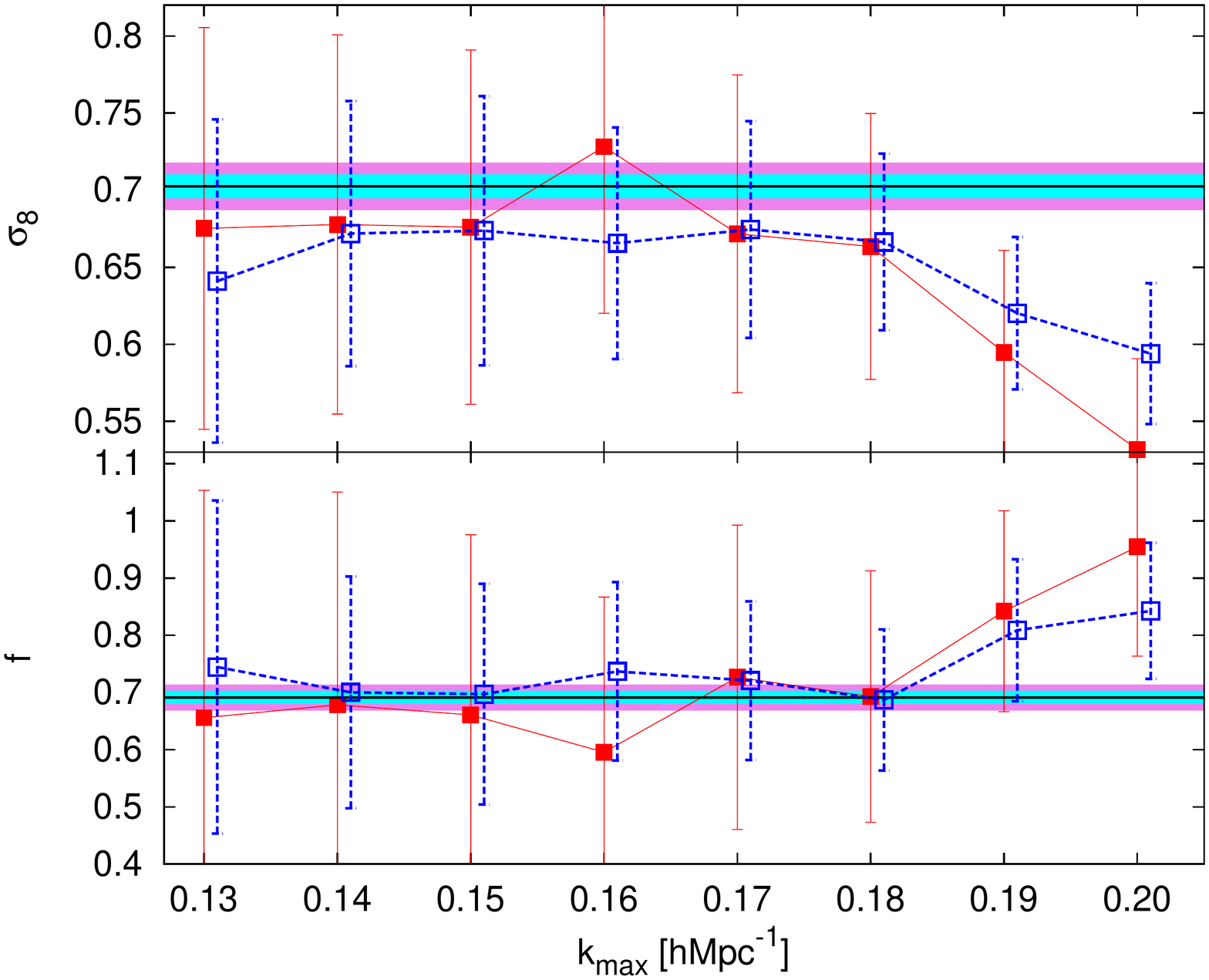}
\includegraphics[scale=0.3]{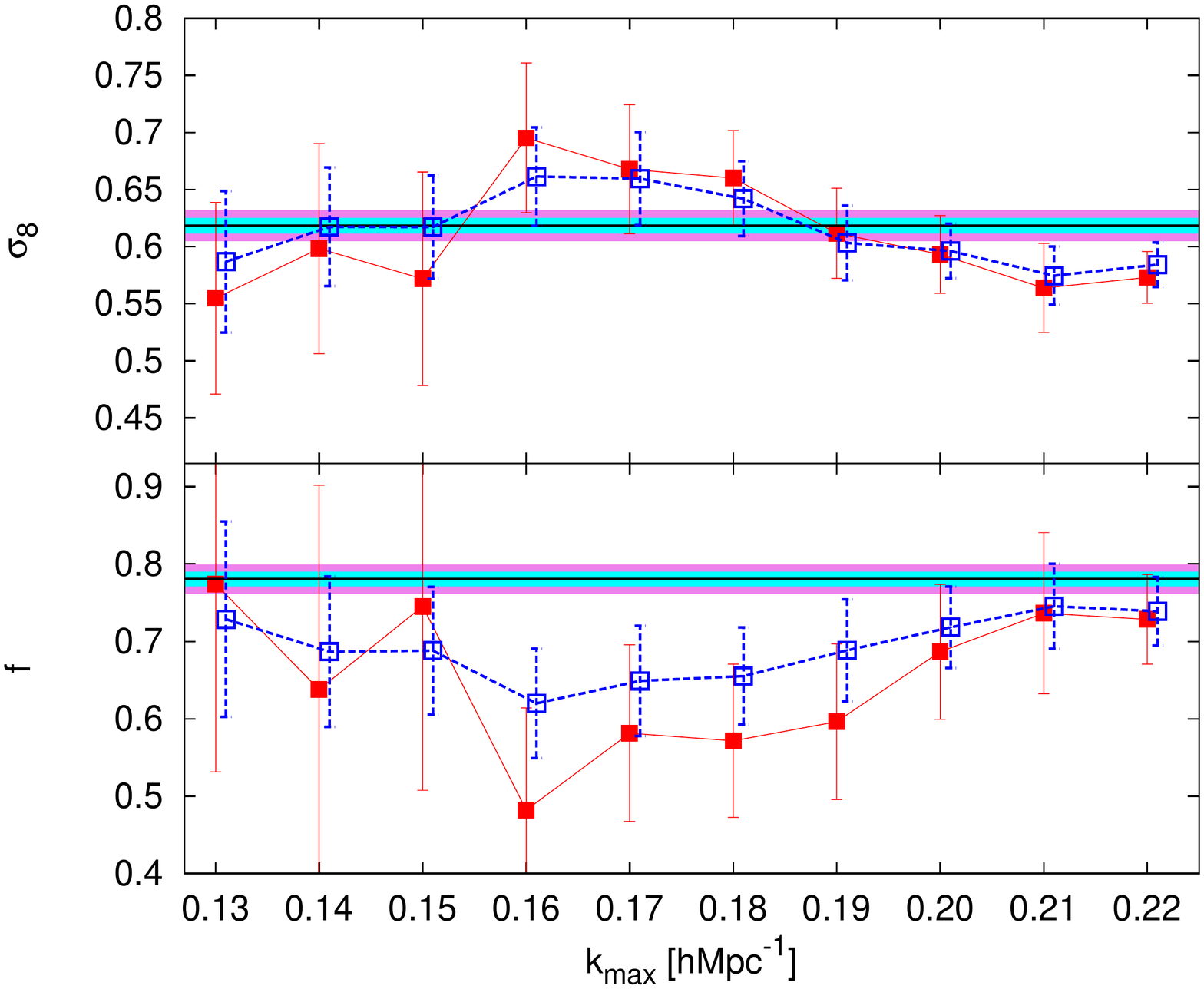}
\caption{Best-fitting cosmological parameters obtained from the combined fit of the BOSS DR12 power spectrum monopole, quadrupole and bispectrum as a function of the truncation scale of the fit, $k_{\rm max}$. Top panels display $\alpha_\parallel$, $\alpha_\perp$ and $f\sigma_8$, whereas bottom panels display $f$ and $\sigma_8$.  Left panels stands for LOWZ sample, whereas right panels for the CMASS sample. The red solid lines show the results when the full fit has been performed, whereas the blue dashed lines stands for the results when $\alpha_\parallel$ and $\alpha_\perp$ have been set to their fiducial value. In $f$, $\sigma_8$ and $f\sigma_8$ sub-panels the black solid line displays the {\it Planck15} prediction. The cyan and magenta bands represent the $1\sigma$ and $2\sigma$ error-bars, respectively, around the {\it Planck15} solution. The black arrow lines  marks the truncation scale which have been choose to take the parameters, according to the findings of \S\ref{sec:tests}. }
\label{fig:kmax_data}
\end{figure*}

\begin{table*}
\begin{center}
\begin{tabular}{ccccc}
\hline
 & LOWZ & LOWZ (no-AP) & CMASS &   CMASS (no-AP) \\
\hline
\hline
$f(z_{\rm eff})\sigma_8(z_{\rm eff})$ & $0.460 \pm 0.071\, (0.063)$  & $0.458 \pm 0.047\, (0.043)$ & $0.417 \pm 0.036\, (0.024)$  & $0.432 \pm 0.022\,(0.018)$   \\
$H(z_{\rm eff}) r_s(z_d)\,[10^3 {\rm km}s^{-1}]$ & $11.75\pm 0.55\,(0.52)$ & 11.773 & $13.78\pm 0.28\,(0.24)$  & 13.663  \\
$D_A(z_{\rm eff})/r_s(z_d)$ & $6.74\pm 0.22\,(0.19)$ & 6.7466 & $9.33\pm 0.15\,(0.11)$   & 9.4418 \\
\hline
 $\alpha_\parallel(z_{\rm eff})$ & $1.002 \pm 0.047\,(0.044)$ & 1  & $0.991 \pm 0.020\,(0.017)$   & 1 \\
  $\alpha_\perp(z_{\rm eff})$ & $1.000 \pm 0.032\,(0.029)$ & 1 & $0.988 \pm 0.016\,(0.012)$  & 1 \\
   $b_1(z_{\rm eff})\sigma_8(z_{\rm eff})$ & $ 1.323\pm0.033 $ & $ 1.322\pm 0.018 $  & $1.237 \pm 0.011$ & $1.2479 \pm 0.0072 $ \\
 $b_2(z_{\rm eff})\sigma_8(z_{\rm eff})$ & $0.60 \pm 0.26$ & $0.59 \pm 0.18$  & $0.606\pm 0.069 $ & $0.641 \pm 0.066$ \\
  $A_{\rm noise}(z_{\rm eff})$ &$0.20 \pm 0.11$  & $0.203\pm 0.082$ & $0.086 \pm 0.018$ & $0.104 \pm 0.015$  \\
  $\sigma_{\rm FoG}^P(z_{\rm eff})\,[{\rm Mpc}h^{-1}]$ & $3.85 \pm 0.37$ & $3.83 \pm 0.36$  & $ 3.45 \pm 0.14$ & $3.50 \pm 0.14$  \\
    $\sigma_{\rm FoG}^B(z_{\rm eff})\,[{\rm Mpc}h^{-1}]$ & $9.8 \pm 6.5$ & $9.8 \pm 7.2$ & $7.39 \pm 0.64$ & $7.54 \pm 0.70$  \\
\hline
  $\chi^2/({\rm d.o.f})$ & $147.84/(160-9)$ & $147.83/(160-7)$ & $746.24/(707-9)$ & $749.07/(707-7)$ \\
  \hline
\end{tabular}
\end{center}
\caption{Best-fitting parameters obtained from fitting the power spectrum monopole, quadrupole and bispectrum monopole of BOSS DR12 data, using the theoretical model described in \S\ref{sec:model}. The two first columns are the parameters obtained from fitting the LOWZ-DR12 data, whereas the third and the fourth  column are obtained from fitting the CMASS-DR12 data. The first and the third column corresponds to a full fit, when all the 9 parameters of the model are varied, whereas the second and fourth keep fixed the AP parameters, $\alpha_\parallel$ and $\alpha_\perp$ to their fiducial value ($\alpha_\parallel=\alpha_\perp=1$). The top 3 rows correspond to the parameters of cosmological interest. The 4th and 5th rows correspond to the AP-scale dilations, which are directly related to $Hr_s$ and $D_A/r_s$. The rest of the parameters correspond to either bias parameter or nuisance parameters.  The last row displays the value for the $\chi^2$ of the fit. For the LOWZ sample the minimum scale used for the fit is $k_{\rm max}=0.18\,h{\rm Mpc}^{-1}$, whereas for the CMASS sample is $k_{\rm max}=0.22\,h{\rm Mpc}^{-1}$. The displayed error-bars represent the $1\sigma$ deviation and in the case of $f\sigma_8$, $\alpha_\parallel$, $\alpha_\perp$, $Hr_s$ and $D_A/r_s$, they account for both a statistical and systematic contributions (added in quadrature). In parenthesis the statistical systematic contribution is also displayed for these parameters. For the rest of parameters, the quoted error-bars correspond to $1\sigma$ of the statistical error.  The correlations among the different cosmological parameters are presented in Eqs. \ref{cov1} - \ref{cov2}.    }
\label{table:results}
\end{table*}
Fig.~\ref{fig:kmax_data} shows the best-fitting cosmological parameters, $f$, $\sigma_8$, $\alpha_\parallel$, $\alpha_\perp$ and $f\sigma_8$, for the BOSS LOWZ and CMASS samples, as a function of truncation scale, $k_{\rm max}$, analogously to Fig.~\ref{fig:kmax_mocks}. The error-bars correspond to the statistical errors with no systematic correction. The red lines show the result when the full set of parameters are varied, and the blue lines when the AP parameters have been set to their fiducial values ($\alpha_\parallel=\alpha_\perp=1$). The black arrow marks the truncation scale which have been chosen to take the parameters, according to the findings of \S\ref{sec:tests}: the parameters displayed in Table~\ref{table:results} correspond to this scale. The magenta an cyan lines corresponds to the predictions by {\it Planck15} \citep{Planck_cosmology15}.

For the LOWZ sample we do not observe any large shifts in any of the parameters as we increase the truncation scale, for $k_{\rm max}\leq0.18\,h{\rm Mpc}^{-1}$, which suggests, along with the findings of \S\ref{sec:tests} (which advocated a slightly smaller value of $k_{\rm max}=0.17\,h{\rm Mpc}^{-1}$), that it is safe to truncate the fits at this scale. We neither observe any significant tension between the solutions when the AP parameters are varied or kept fixed. For larger values of $k_{\rm max}$ we observe that $f$ increases, whereas $\sigma_8$ tends to decrease as we move to larger values of $k_{\rm max}$. This is the same behaviour observed for the MD-\textsc{Patchy} mocks in Fig.~\ref{fig:kmax_mocks}, and it may be caused by a failure of the model. We also observe that $\alpha_\parallel$ starts changing significantly for $k_{\rm max}>0.18\,h{\rm Mpc}^{-1}$. 

For the CMASS sample we do not observe any large shifts in any of the parameters as we increase the truncation scale when the AP parameters are kept to their fiducial value. When the AP parameters are also varied  we see a $\sim1\sigma$ shift in all cosmological parameters when $k_{\rm max}$ increases from $k_{\rm max}=0.15\,h{\rm Mpc}^{-1}$ to $k_{\rm max}=0.16\,h{\rm Mpc}^{-1}$, which gets reduced to about $\sim0.5\sigma$ between $k_{\rm max}=0.15\,h{\rm Mpc}^{-1}$ and $k_{\rm max}=0.17\,h{\rm Mpc}^{-1}$, and even less to higher $k_{\rm max}$ values. For $k_{\rm max}=0.16\,h{\rm Mpc}^{-1}$ the $f\sigma_8$ results are about $1\sigma$ away between the full AP fit and the fit where the AP parameters are kept fixed. This difference is reduced when any other $k_{\rm max}$ truncation scale (higher or lower) is chosen. Given the observed features, we think that this $1\sigma$ fluctuation at $k_{\rm max}=0.16\,h{\rm Mpc}^{-1}$ is purely statistical. Besides from this point, we do not observe any other significant change in $f\sigma_8$, $\alpha_\parallel$ and $\alpha_\perp$ as we increase $k_{\rm max}$ up to $0.22\,h{\rm Mpc}^{-1}$, as we observed for the $N$-body haloes in \S\ref{sec:tests}. This supports the finding of \S\ref{sec:tests} of truncating the CMASS sample fits at $k=0.22\,h{\rm Mpc}^{-1}$ when estimating $f\sigma_8$ and the AP parameters. On the other hand, if we are interested in getting $f$ and $\sigma_8$ separately, the findings in \S\ref{sec:tests} suggested to use $k=0.17\,h{\rm Mpc}^{-1}$. We will come back to this point in \S\ref{sec:degeneracy} when we present the $f$ and $\sigma_8$ results separately. 

\subsection{Correlation of the measurements}\label{sec:81}
\begin{figure*}
\includegraphics[scale=0.3]{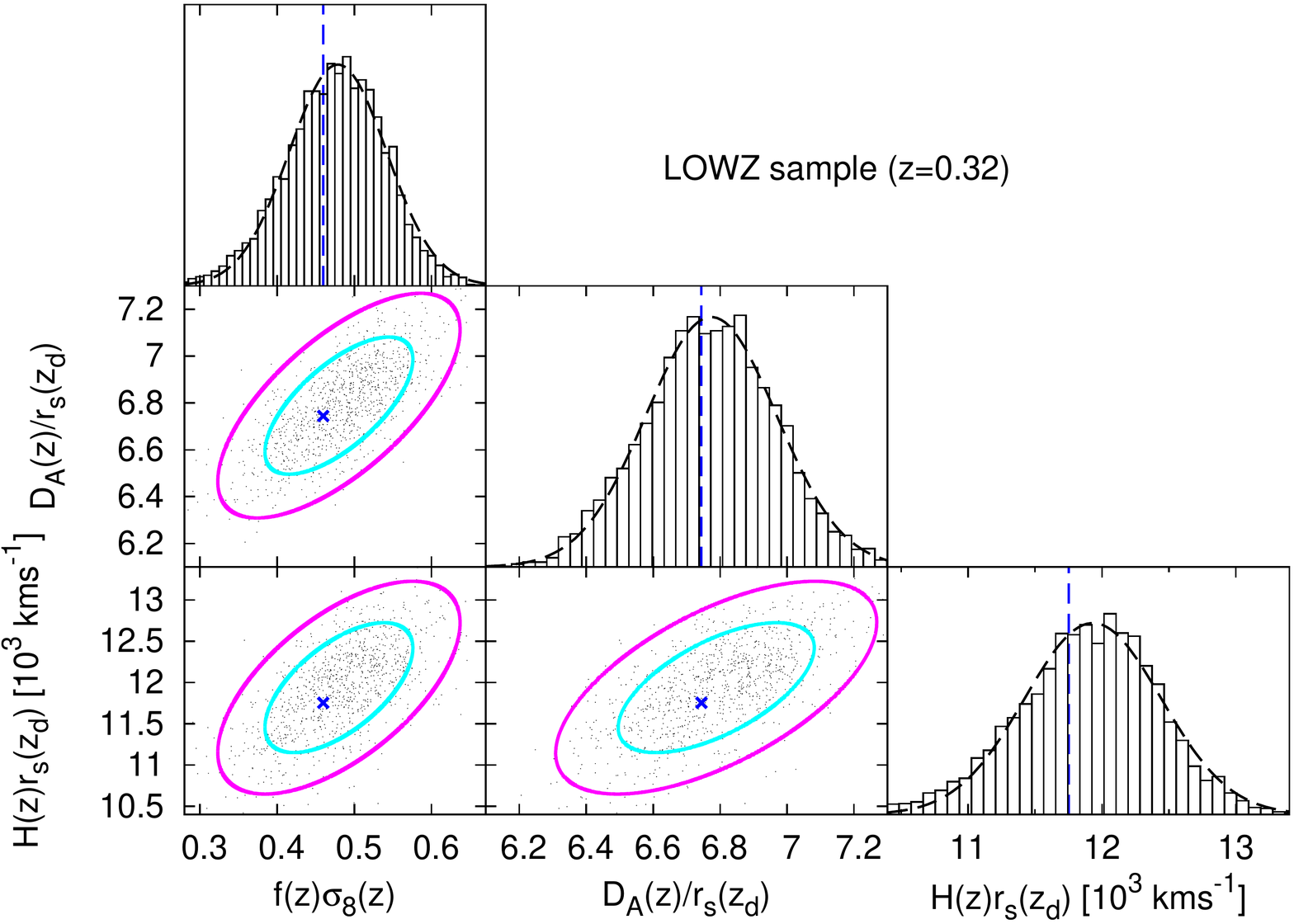}
\includegraphics[scale=0.3]{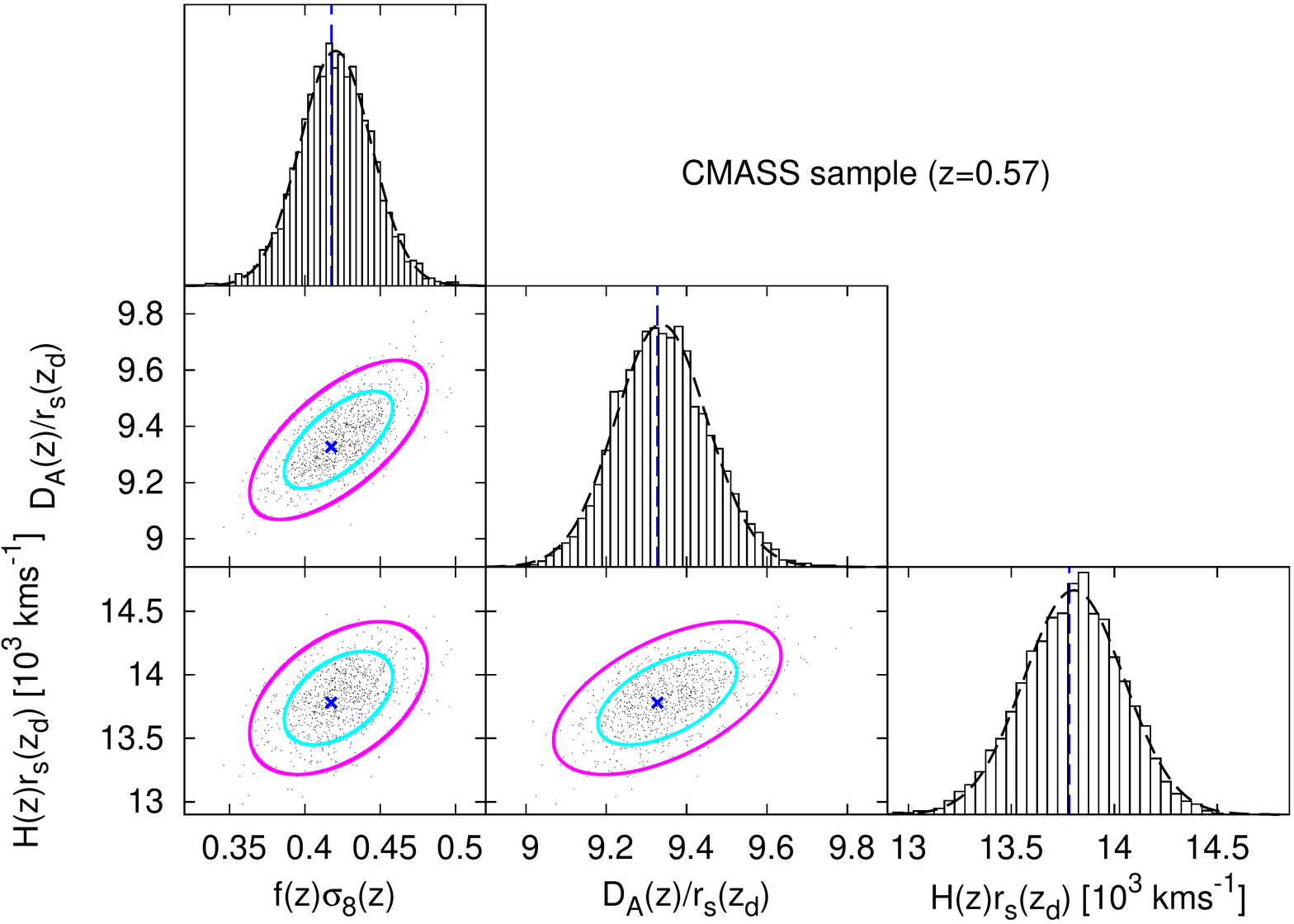}
\caption{Two dimensional likelihood distributions for the parameters $f(z)\sigma_8(z)$, $H(z)r_s(z_d)$ and $D_A(z)/r_s(z_d)$ for BOSS DR12 data sample. Left and right panels correspond to the LOWZ and CMASS samples, respectively. The black small dots correspond to the steps of the \textsc{mcmc}-chains, and the cyan and magenta contours correspond to the $1\sigma$ and $2\sigma$ confidence regions based on the Gaussian approximation obtained from the \textsc{mcmc}-chains. The blue crosses correspond to the best-fitting values of each parameter. The upper sub-panels show the histograms of the \textsc{mcmc}-chains, along with a Gaussian best-fitting values. In all the cases we see that the distribution of the \textsc{mcmc} is close to a Gaussian distribution.}
\label{cosmo_covariance2}
\end{figure*}

The measurements displayed in Table~\ref{table:results}, as well as the different parameters of Fig.~\ref{fig:kmax_data}, present important correlations among them that must be taken into account when performing cosmological constrains. Fig.~\ref{cosmo_covariance2} displays such correlation for the parameters $f(z)\sigma_8(z)$, $H(z)r_s(z_d)$ and $D_A(z)/r_s(z_d)$, where the black dots represent the down-sampled steps of the \textsc{mcmc}-chain corresponding to the data, and the cyan and magenta contours, correspond to the $1\sigma$ and $2\sigma$ confident regions, respectively, extracted from the \textsc{mcmc}-chains when the likelihood is assumed Gaussian.  The blue crosses mark the best-fitting solutions found by the minimisation algorithm. The correlation among the different variables can be directly inferred from the distribution of the \textsc{mcmc}-chains, as well as, from the size and orientation of the ellipses under the assumption of Gaussianity. The corresponding histograms complement the information by showing marginalised 1D posterior distributions. This distribution is, for all the variables, close to Gaussian, both in LOWZ and CMASS samples, suggesting that the statistical interpretation of the $1\sigma$ and $2\sigma$ contours as the $68.3\%$ and $95.5\%$ confident regions, respectively, is OK.

In order to present the correlation matrices we start by defining the data-vector containing the cosmology parameters of interest $f(z)\sigma_8(z)$, $H(z) r_s(z_d)$ (in $10^3 {\rm km}s^{-1}$ units) and $D_A(z)/r_s(z_d)$ as, 
\begin{equation}
 \label{data1}
D_{\rm data}(z) = 
 \begin{pmatrix}
  f(z)\sigma_8(z)  \\
  H(z) r_s(z_d)\, [10^3 {\rm km}s^{-1}] \\
  D_A(z)/r_s(z_d)
 \end{pmatrix}.
 \end{equation}
 The data-vectors for LOWZ and CMASS are formed from the results displayed in Table~\ref{table:results},
 \begin{equation}
 \label{data_lowz}
D_{\rm data}(z_{\rm LOWZ}) = 
 \begin{pmatrix}
   0.45960 \\
  11.753 \\
  6.7443
 \end{pmatrix},
 \end{equation}
 from the LOWZ sample at $k_{\rm max}=0.18\,h{\rm Mpc}^{-1}$, and
  \begin{equation}
 \label{data_cmass}
D_{\rm data}(z_{\rm CMASS}) = 
 \begin{pmatrix}
   0.41750 \\
  13.781 \\
  9.3276
 \end{pmatrix},
 \end{equation}
 from the CMASS sample at $k_{\rm max}=0.22\,h{\rm Mpc}^{-1}$.
 
The covariance matrices of these data-vectors are extracted from the Gaussian contours drawn in Fig.~\ref{cosmo_covariance2}. In addition, we add the systematic contribution of Table~\ref{table:systematics_model} to the diagonal elements. Thus, the following covariance matrices contain the total error budget, including systematic errors of the model. The covariance matrices read as, 
\begin{equation}
\label{cov1}
C^{{\rm LOWZ}} = 10^{-3}
 \begin{pmatrix}
 5.0837 & 23.818 & 10.490  \\
  &  300.30 & 73.448 \\
  &  & 47.493
 \end{pmatrix},
 \end{equation}
for the LOWZ sample at $k_{\rm max}=0.18\,h{\rm Mpc}^{-1}$, and, 
\begin{equation}
\label{cov2}
C^{{\rm CMASS}}= 10^{-3}
 \begin{pmatrix}
1.3046 & 4.6434 & 3.5329  \\
  & 77.713 & 22.773 \\
  &  & 21.700
 \end{pmatrix},
 \end{equation}
 for the CMASS sample at $k_{\rm max}=0.22\,h{\rm Mpc}^{-1}$. 
 
The corresponding likelihood of any cosmological model is given then by,
\begin{equation}
\label{likelihood}
\mathcal{L}\propto\exp\left[ -(D_{\rm data}-D_{\rm model})^T {C}^{-1} (D_{\rm data}-D_{\rm model}) \right],
\end{equation}
where $D_{\rm model}$ is the vector with the model prediction for the same cosmological parameters than $D_{\rm data}$ and $C^{-1}$ is the inverse of the covariance matrix.

\subsection{Breaking the $f$ and $\sigma_8$ degeneracy}\label{sec:degeneracy}
In the section above we presented the best-fitting value of the combined $f\sigma_8$ parameter. However, as we have mention before, in our fitting routine, as well as in the \textsc{mcmc}-chains, we allow $f$ and $\sigma_8$ to freely vary. The reason why we have presented $f\sigma_8$ as a single parameter in Table~\ref{table:results} is that $f$ and $\sigma_8$ present a strong degeneracy along the curve $f\sigma_8={\rm constant}$, and therefore are highly correlated. This is because most of the signal is coming from the power spectrum monopole and quadrupole at large scales, where these two parameters are perfectly degenerated (because of the Kaiser limit). However, as we explore smaller scales in the power spectrum multipoles, and especially when we add the bispectrum to the analysis, this degeneracy is partially broken and we can express as well the results in terms of $f$ and $\sigma_8$, along with their correlation parameter (see Appendix \ref{appendixc} for a further explanation). 

Table~\ref{table:results2} displays the results of the best-fitting for the CMASS sample in terms of $f$, $\sigma_8$, along with the rest of cosmological parameters of interest. We do not display the rest of the nuisance parameters as we did in Table~\ref{table:results} for simplicity. These results  correspond to a truncation scale of $k_{\rm max}=0.17\,h{\rm Mpc}^{-1}$, which is more conservative than the one used for presenting the results in terms of $f\sigma_8$ because, as we have seen in \S\ref{sec:tests}, the systematic errors in terms of $f$ and $\sigma_8$ are much more important than those on $f\sigma_8$. As in Table~\ref{table:results}, we show the results when the AP parameters are varied and when they are kept fixed to their fiducial value.  We do not attempt to present the results of the LOWZ sample in terms of $f$ and $\sigma_8$ as a separate variables because as the LOWZ effective volume is about 3 times smaller than the CMASS volume, the constrains on $f$ and $\sigma_8$ were very mild. 

\begin{table}
\begin{center}
\begin{tabular}{|c|c|c}
 & CMASS &   CMASS (no-AP) \\
 \hline
 \hline
 $f(z)$ & $0.58\pm0.12\,(0.11)$ & $0.649\pm0.076\,(0.071)$ \\
 $\sigma_8(z)$ & $0.668\pm0.078\,(0.056)$ & $0.660\pm0.067\,(0.041)$ \\ 
  $H(z)r_s(z_d)$ & $13.38\pm0.51\,(0.50)$ & $13.663$ \\
  $D_A(z)/r_s(z_d)$ & $9.227\pm0.223\,(0.203)$ & $9.4418$ \\
  \hline
  $\chi^2$ & 349.33/(370-9) & 350.95/(370-7) \\
  \end{tabular}
\end{center}
\caption{Best-fitting parameters obtained from fitting the power spectrum monopole, quadrupole and bispectrum monopole of BOSS DR12 data for the CMASS sample, using the theoretical model described in \S\ref{sec:model}. In contrast to Table~\ref{table:results} we express the results in terms of $f$ and $\sigma_8$ as a separate variables and for $k_{\rm max}=0.17\,h{\rm Mpc}^{-1}$. For this fit we only display the parameters of cosmological interest for simplicity. As in Table~\ref{table:results} the quoted error-bars contain the $1\sigma$  total error budget, this is the systematic and statistical contribution, both added in quadrature. In parenthesis we display the error contribution corresponding to the statistical contribution only. The correlations among the different cosmological parameters are presented in Eqs. \ref{cov3} - \ref{cov4}.  }
\label{table:results2}
\end{table}%

\begin{figure}
\includegraphics[scale=0.3]{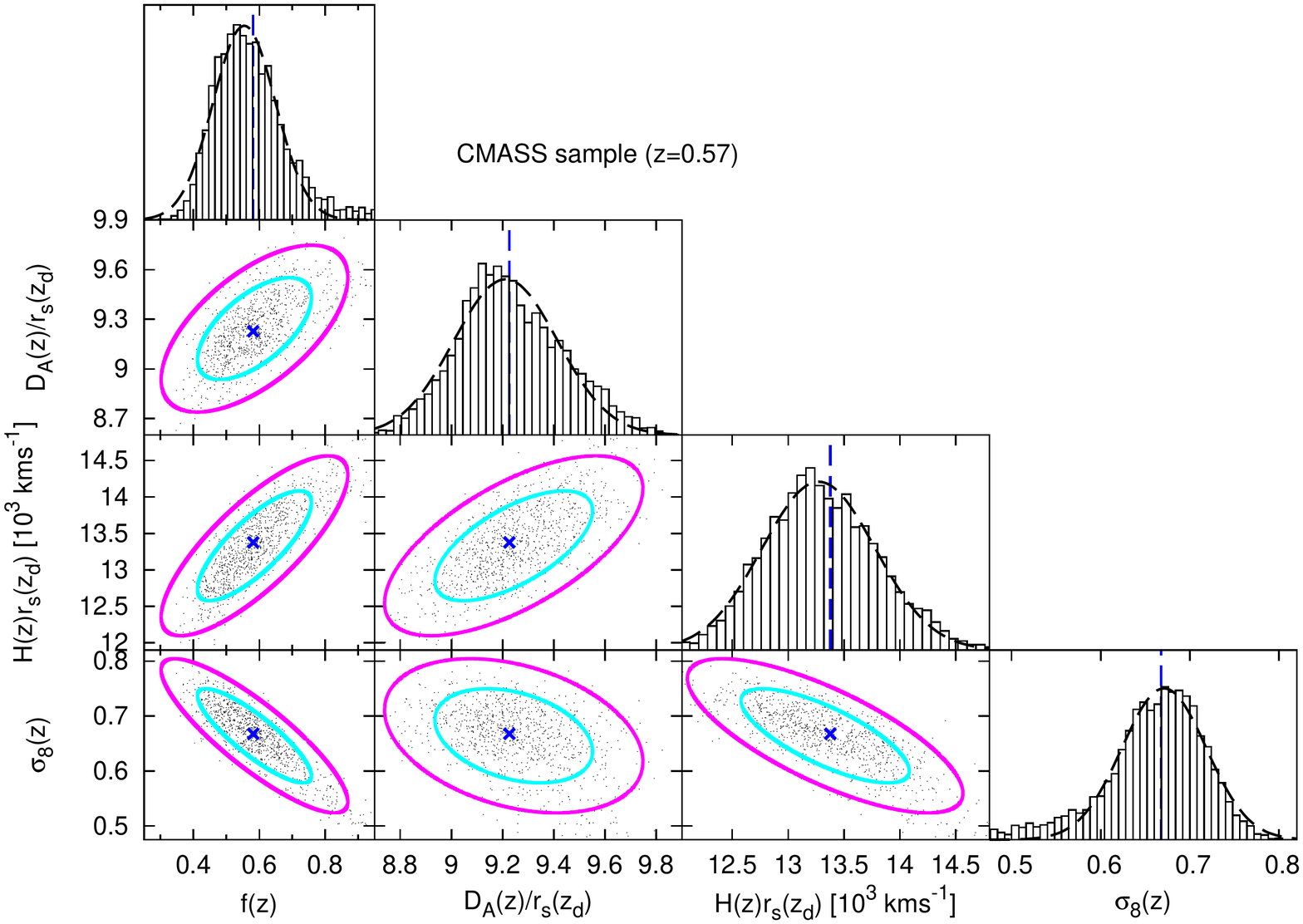}
\caption{Two dimensional posterior distributions (marginalised over all other parameters) for the parameters $f(z)$, $\sigma_8(z)$, $H(z)r_s(z_d)$ and $D_A(z)/r_s(z_d)$ for CMASS BOSS DR12 data sample corresponding the results presented in Table~\ref{table:results2} with a truncation scale of $k_{\rm max}=0.17\,h{\rm Mpc}^{-1}$. The black small dots correspond to the steps of the \textsc{mcmc}-chains, and the cyan and magenta contours correspond to the $1\sigma$ and $2\sigma$ confident regions based on the Gaussian approximation obtained from the \textsc{mcmc} chains. The blue crosses correspond to the best-fitting of each parameter to the data.  }
\label{cosmo_covariance}
\end{figure}

Fig.~\ref{cosmo_covariance} displays the correlation of the parameters presented in Table~\ref{table:results2}:  $f(z)$, $\sigma_8(z)$, $H(z)r_s(z_d)$ and $D_A(z)/r_s(z_d)$, using the same colour and symbol notation in Fig.~\ref{cosmo_covariance2}. We see that $f$ and $\sigma_8$ are highly correlation as expected, with a correlation coefficient of $r=-0.85$, similar to that found in the DR11 analysis \citep{hector_bispectrum2}.

Comparing Fig.~\ref{cosmo_covariance2} and \ref{cosmo_covariance} we see that the distribution of the \textsc{mcmc}-chains is closer to a Gaussian when it is displayed in terms of $f\sigma_8$, $Hr_s$, $D_A/r_s$ rather than the individual parameters, $f$ and $\sigma_8$. Certainly, this is also the case when only the power spectrum multipoles are used for constraining cosmological parameters: the degeneracy between $f$ and $\sigma_8$ is poorly broken because the overall signal is dominated by large scales, where the power spectrum Kaiser limit is insensitive to individual shifts of $f$ and $\sigma_8$, when $f\sigma_8$ is kept fixed (see Appendix \ref{appendixc} for further discussion). Adding the bispectrum monopole into the analysis helps significantly to break the $f\sigma_8$ degeneracy and  constraining $f$ and $\sigma_8$ individually, because in the large scale limit the bispectrum is no longer a function of $f\sigma_8$. As a consequence of the bispectrum signal, the $f-\sigma_8$ distribution in Fig.~\ref{cosmo_covariance} cannot take arbitrary low or large values of these variables. However, the efficiency on breaking the $f-\sigma_8$ degeneracy mainly depends on the quality of the bispectrum signal (for instance for the LOWZ sample the degeneracy is poorly broken and the resulting distribution is very non-Gaussian).
From the histograms of Fig.~\ref{cosmo_covariance}, the $f-\sigma_8$ distribution presents a behaviour which is close to be the Gaussian. However, we note that at $\sigma_8\lesssim0.57$ and $\sigma_8\gtrsim0.75$ for $\sigma_8$; and $f\lesssim0.4$ and $f\gtrsim0.7$ for $f$, there is a non-Gaussian feature, which present an excess and lack, respectively, compared to the Gaussian best-fitting values. However, at $1\sigma$ confident regions the histograms do not present any particular feature of non-Gaussianity. The non-Gaussian behaviour of the $f$ and $\sigma_8$ variables is mitigated when combining them into $f\sigma_8$ as in Fig.~\ref{cosmo_covariance2}, where the respective non-Gaussian behaviours are cancelled, resulting a Gaussian distribution. 

We present correlation matrices corresponding to the parameters presented in Table~\ref{table:results2} and in Fig.~\ref{cosmo_covariance}: $f(z)$, $\sigma_8(z)$, $H(z) r_s(z_d)$ (in $10^3 {\rm km}s^{-1}$ units) and $D_A(z)/r_s(z_d)$, corresponding to the best-fitting values whose truncation scale is $k_{\rm max}=0.17\,h{\rm Mpc}^{-1}$. We define the data-vector as, 
 \begin{equation}
 \label{data2}
D^{\rm data}(z_{\rm CMASS}) = 
 \begin{pmatrix}
  f(z)  \\
  \sigma_8(z) \\
  H(z) r_s(z_d)\, [10^3 {\rm km}s^{-1}] \\
  D_A(z)/r_s(z_d)
 \end{pmatrix}=
  \begin{pmatrix}
  0.58140  \\
  0.66778 \\
  13.377\\
  9.2265
 \end{pmatrix}.
 \end{equation}
 The corresponding covariance matrix, which incorporate both the statistical and systematic contribution, reads as,
  \begin{equation}
 \label{cov3}
C^{{\rm CMASS}} = 10^{-3}
 \begin{pmatrix}
13.581 & -7.7346 & 47.333 & 15.9145  \\
  & 6.01322 & -29.7585 & -5.7860\\
  &  & 264.27 & 73.165 \\
   &  &  & 49.794
 \end{pmatrix}.
 \end{equation}
 
In case the $H(z) r_s(z_d)$ and $D_A(z)/r_s(z_d)$ parameters are set to their fiducial value (second column of Table~\ref{table:results2}), the data-vector only depends on $f(z)$ and $\sigma_8(z)$, 
  \begin{equation}
   \label{data3}
D^{\rm data}(z_{\rm CMASS}) = 
 \begin{pmatrix}
  f(z)  \\
  \sigma_8(z) \\
 \end{pmatrix}=
  \begin{pmatrix}
  0.64893  \\
  0.65958 \\
 \end{pmatrix}.
 \end{equation}
 The corresponding covariance matrix reads as,  
  \begin{equation}
  \label{cov4}
C^{\rm CMASS} = 10^{-3}
 \begin{pmatrix}
5.3533 & -3.9574  \\
  & 4.32659 \\
 \end{pmatrix},
 \end{equation}
where, as before, it incorporates both the statistical and systematic contribution. As for the case of \S\ref{sec:81}, the corresponding likelihood for both cases corresponds to Eq. \ref{likelihood}.

\subsection{Comparison with other BOSS cosmological analyses}\label{sec:cosmo}

\begin{figure*}
\includegraphics[scale=0.3]{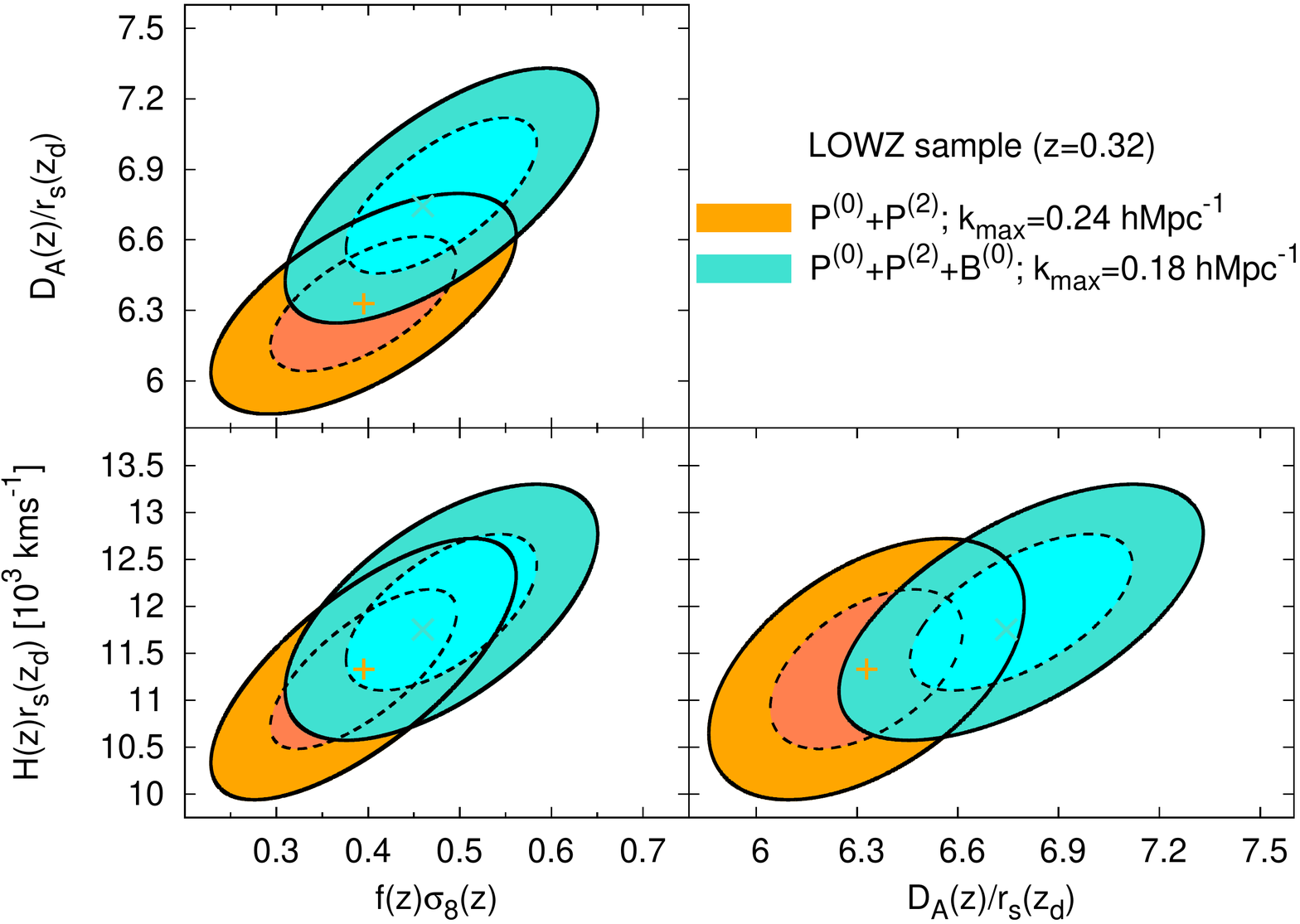}
\includegraphics[scale=0.3]{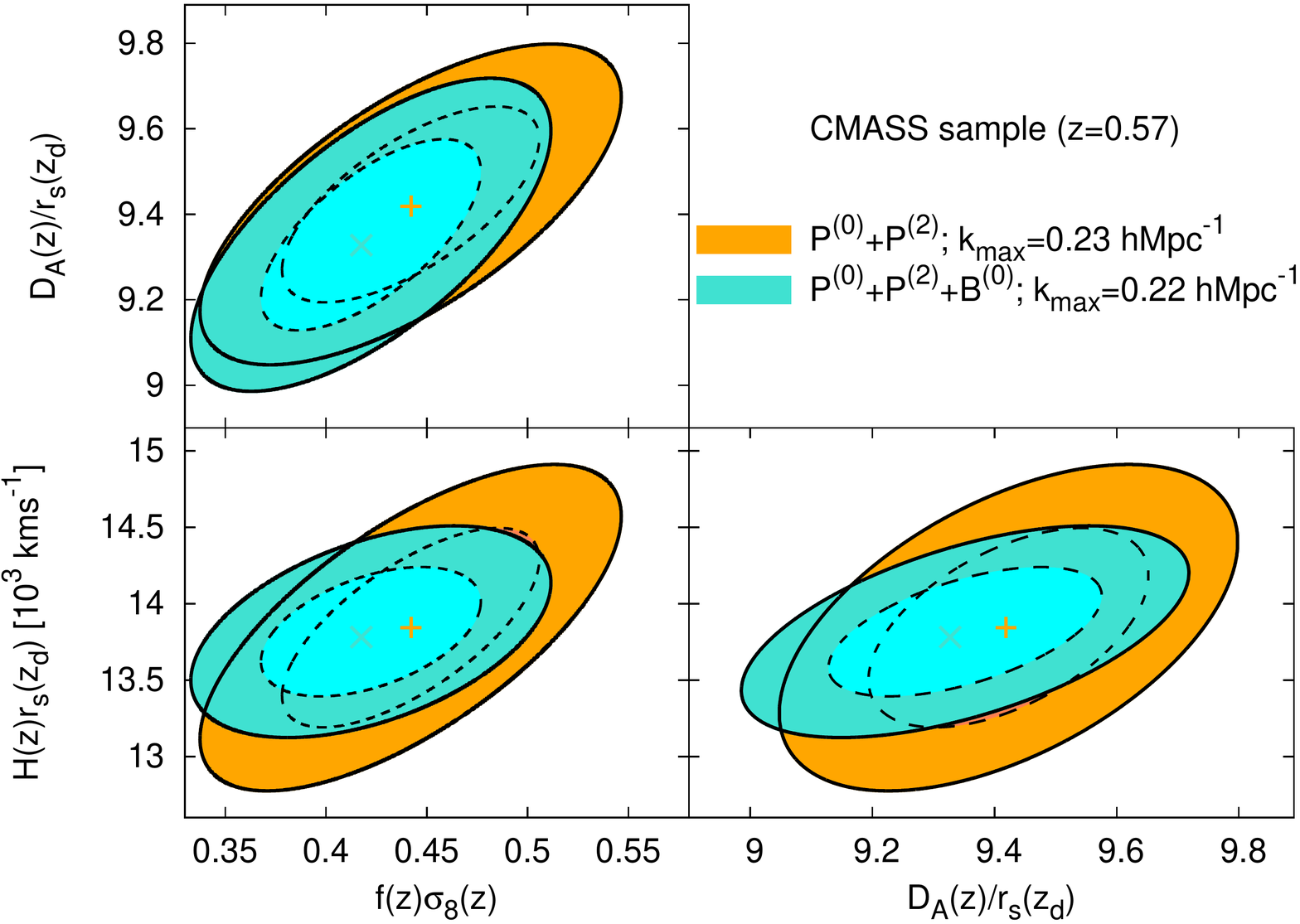}
\caption{Comparison of the $1\sigma$ (solid lines) and $2\sigma$ (dashed lines) confident regions of $f\sigma_8$, $D_A/r_s$ and $Hr_s$, corresponding to the LOWZ  and CMASS samples.  The orange contours correspond to the constrains obtained by analysing the power spectrum monopole and quadrupole up to $k_{\rm max}=0.24\,h{\rm Mpc}^{-1}$ for LOWZ and  $k_{\rm max}=0.23\,h{\rm Mpc}^{-1}$ for CMASS, using the covariance matrix obtained from the MD-\textsc{Patchy} mocks \citep{gil-marin15_rsd}. The turquoise contours are the $1\sigma$ and $2\sigma$ confident levels obtained from the analysis of the power spectrum multipoles in combination with the bispectrum up to $k_{\rm max}=0.18\,h{\rm Mpc}^{-1}$ for the LOWZ sample and $k_{\rm max}=0.22\,h{\rm Mpc}^{-1}$ for the CMASS sample, according to the covariance matrix of Eqs. \ref{cov1}-\ref{cov2}.}
\label{cosmo_covariance3}
\end{figure*}

In this section we compare our measurements with other studies of RSD based on DR11 and DR12 of BOSS LOWZ and CMASS samples. DR11 only contains about $10\%$ fewer galaxies than the final DR12 data set, so significant changes in measurements from DR11 and DR12 data are driven by changes in the methodology, rather than statistical errors. We start by comparing the differences between the present work and the results of \cite{gil-marin15_rsd}, based on the power spectrum monopole and quadrupole, which uses the same measurement technique, DR12 data and model presented here, but with a higher truncation $k_{\rm max}=0.24\,h{\rm Mpc}^{-1}$. 
Fig.~\ref{cosmo_covariance3}, compares constraints on $f\sigma_8$, $H r_s$ and $D_A/r_s$ for both the LOWZ and CMASS samples, and shows good agreement with both results, with shifts on the best-fitting results that are $\lesssim1\sigma$ for the CMASS sample and around between $1-1.5\sigma$ for the LOWZ sample. On the LOWZ sample, the effect of  adding the bispectrum to the power spectrum monopole and quadrupole analysis (and decreasing the  $k_{\rm max}$ truncation scale from $k_{\rm max}=0.24\,h{\rm Mpc}^{-1}$ to $k_{\rm max}=0.17\,h{\rm Mpc}^{-1}$) is to increase the $f\sigma_8$ best-fitting value,  along with the other two AP parameters. 
 On the other hand, on the CMASS sample, the change in the $f\sigma_8$ value is significantly smaller than $1\sigma$. 
 We note that there is a significant reduction on the error-bars for the CMASS sample, but not for the LOWZ sample. This is caused by the different values of $k_{\rm max}$ used for the LOWZ sample, when the bispectrum is added ($k_{\rm max}=0.18\,h{\rm Mpc}^{-1}$), and when only the power spectrum is used  ($k_{\rm max}=0.24\,h{\rm Mpc}^{-1}$).

\begin{figure}
\centering
\includegraphics[scale=0.3]{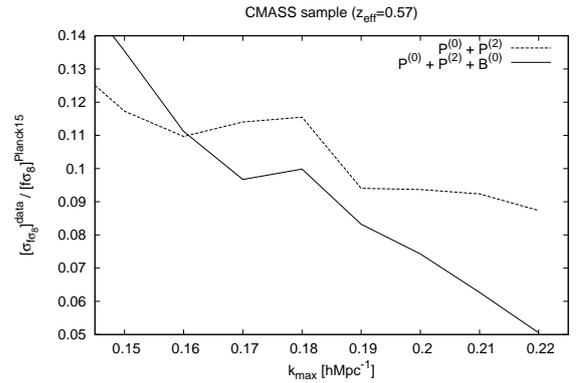}
\caption{Statistical errors on the $f\sigma_8$ parameters as a function of the truncation scale $k_{\rm max}$, extracted from the \textsc{mcmc}-chains of the CMASS-DR12 data. On dashed lines the prediction where the power spectrum monopole and quadrupole are used, and in solid lines when the bispectrum monopole is added to the analysis. At large scale we see that the effect of adding the bispectrum worsen the errors on $f\sigma_8$ (likely due to noise), but for $k_{\rm max}>0.16\,h{\rm Mpc}^{-1}$ the errors on $f\sigma_8$ are reduced by the effect of the bispectrum signal.  At $k_{\rm max}=0.22\,h{\rm Mpc}^{-1}$, the statistical error-bars are reduced by a factor of $\sim1.8$. }
\label{fig:errorsPB}
\end{figure}

Fig. \ref{fig:errorsPB} shows the role of the bispectrum in reducing the size of the error-bars of $f\sigma_8$ as a function of the truncation scale $k_{\rm max}$ for the CMASS sample. In this case  the relative error-bars have been extracted from the \textsc{mcmc}-chains of the data for the  $f\sigma_8$ parameter and are displayed as a function of the truncation scale, $k_{\rm max}$, for the case where the power spectrum monopole and quadrupole are used (dashed line), and where the bispectrum is added to these two statistics (solid line). At large scales we observe that adding the bispectrum worsens the constrains on $f\sigma_8$. This effect is probably due to noise in the data. As we explore smaller scale cuts, adding the bispectrum signal produces a reduction on the statistical error-bars for $f\sigma_8$ as expected. The effect starts to be important for $k_{\rm max}\geq0.19\,h{\rm Mpc}^{-1}$, and for $k_{\rm max}=0.22\,h{\rm Mpc}^{-1}$, the gain is a factor $\sim1.8$. This means that if we could model the bispectrum up to $k_{\rm max}\simeq0.22\,h{\rm Mpc}^{-1}$ with no systematic errors, we would improve by a factor of 1.8 the results obtained on $f\sigma_8$ by the classical analysis of the power spectrum multipoles. This suggests that the power spectrum and bispectrum are not very correlated in terms of the $f\sigma_8$ magnitude when the truncation scale, $k_{\rm max}$, is sufficiently large. We quantify the degree of correlation in \S\ref{sec:combining_boss}.

\begin{figure} 
\includegraphics[scale=0.3]{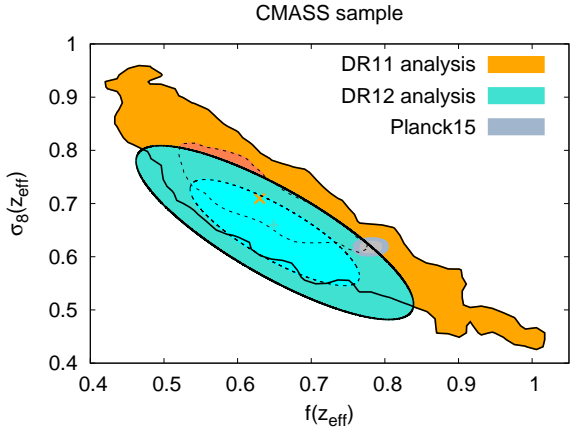}
\caption{Comparisons on the $f-\sigma_8$ constraints from the DR11-CMASS sample \citep{hector_bispectrum2} in orange contours, and DR12-CMASS sample (this work) in turquoise contours, when the AP parameters have been set to their fiducial value. The solid and dashed lines correspond to $1\sigma$ and $2\sigma$ contours, respectively. The improvement in the constraints between DR11 and DR12 CMASS samples is dominated by the inclusion of the increased bispectrum signal through including new triangles shapes. The blue contours show the predictions for the {\it Planck15} cosmology. }
\label{fig:dr11comparison}
\end{figure}

In Fig.~\ref{fig:dr11comparison} we compare the findings on the $f-\sigma_8$ plane drawn from the CMASS DR11 bispectrum analysis and from the current CMASS analysis, when the AP parameters have been set to their fiducial value \citep{hector_bispectrum2}. 
The measurements from the current analysis are drawn from the covariance matrix of Eq. \ref{cov4} and therefore include both statistical and systematic errors. The orange contours correspond to the constraints inferred from the measurements of the power spectrum monopole and bispectrum monopole presented in \cite{hector_bispectrum1}, in combination with those from the two-point correlation function monopole and quadrupole from \cite{Samushiaetal:2014}. We see how both DR12 and DR11 constraints are in good agreement. The reduction in the error-bars between DR11 and DR12 analyses corresponds to the inclusion of more triangular shapes in the DR12 respect to the DR11: for the DR11 bispectrum analysis, only those triangular shapes corresponding  to $k_2/k_1=1,\,2$ up to $k=0.20\,h{\rm Mpc}^{-1}$  were included in the analysis, whereas for the DR12 we have included all possible triangular shapes up to $k=0.22\,h{\rm Mpc}^{-1}$. In addition, the error-bars from the DR11 where drawn from the dispersion of 600 mocks, whose best-fitting was estimated taking only the diagonal errors of the covariance matrix. Although following this procedure does not bias the results, it does not provides a optimal estimator in terms of having a minimum variance estimator. On the other hand, for the DR12 analysis the errors are drawn from the posterior of the data, which has been computed taking into account the full covariance, which does guarantee a minimum-variance estimator. The correlation factor between $f$ and $\sigma_8$ is also very consistent. From DR11 analysis we obtained a correlation factor around $-0.90$, whereas for DR12 is around $-0.82$\footnote{Note that the $-0.82$ value for the correlation between $f$ and $\sigma_8$ is different from the case where the AP parameters are also varied, $-0.85$, previously described in Fig.~\ref{cosmo_covariance}.}.
This small change is either due to the inclusion of more triangles, or to the changing in the variance estimator or to the Gaussianization on the DR12 analysis. There is a slight tension within $2\sigma$ between DR12 and {\it Planck15} results. Also,  the $1.5\sigma$-low value of $f\sigma_8$ in comparison with the {\it Planck15} prediction observed in Fig.~\ref{fig:kmax_data} and \ref{cosmo_covariance3} is caused by the low value of $f$ and not the value of $\sigma_8$, which according to Fig.~\ref{fig:dr11comparison} is in good agreement with {\it Planck15} prediction. 
\begin{figure*}
\centering
\includegraphics[scale=0.3]{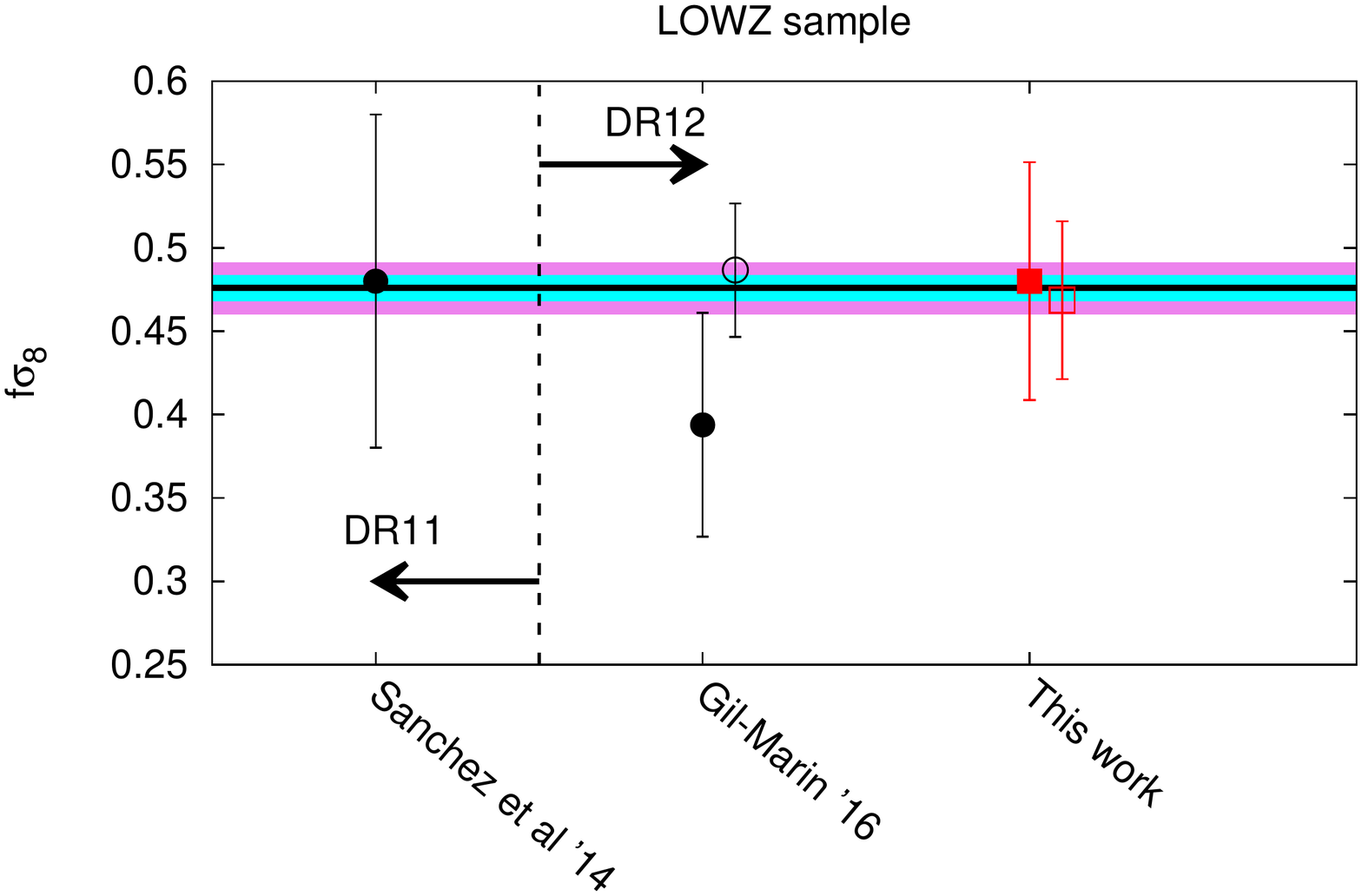}
\includegraphics[scale=0.3]{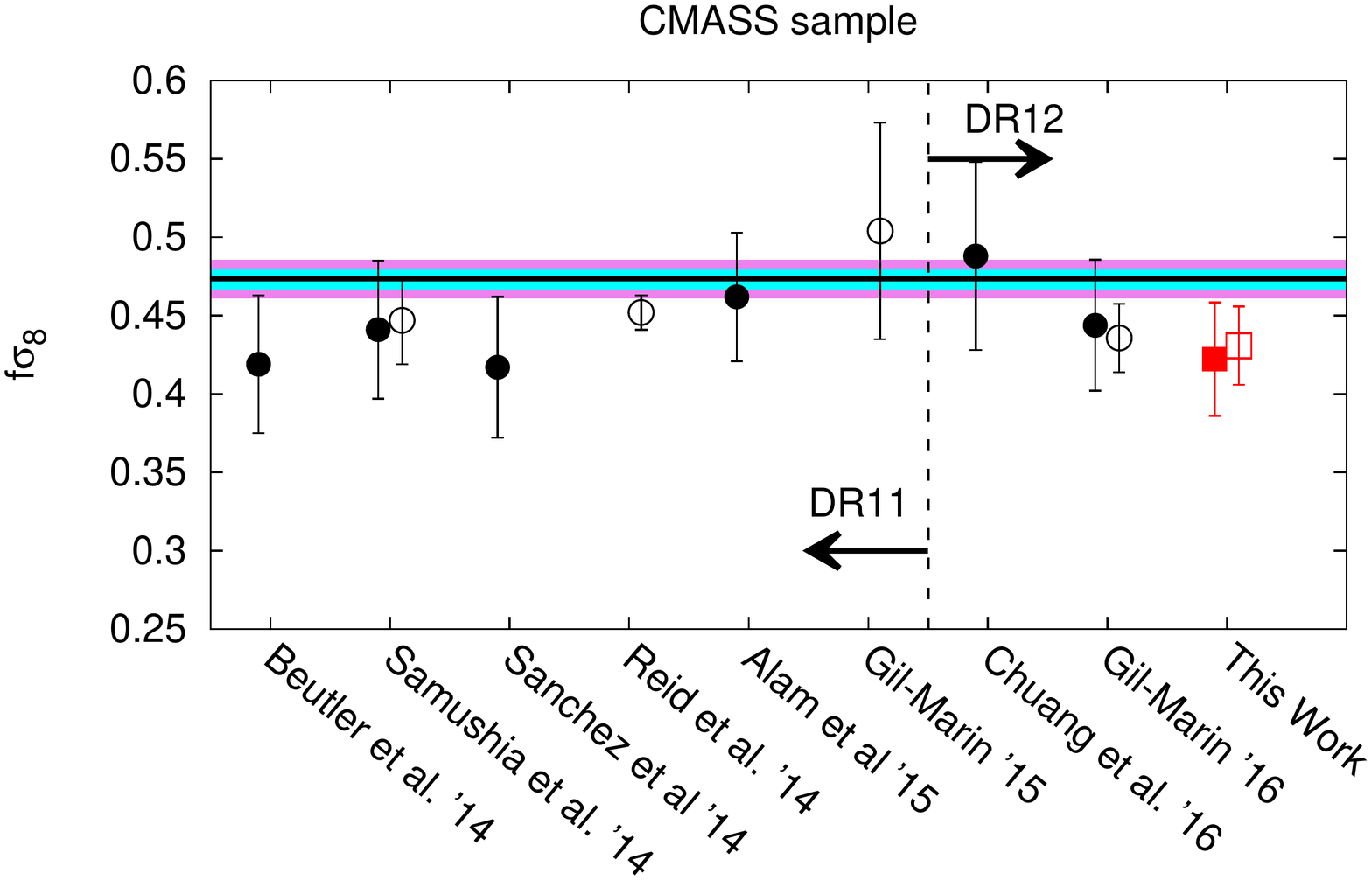}
\caption{Constrains on the $f\sigma_8$ parameter from the BOSS LOWZ- (left panel) and BOSS CMASS-sample (right panel) are displayed in black circles \citep{Chuangetal:2013, Beutleretal:2013, Samushiaetal:2014,Sanchezetal:2013,Reidetal:2013,Alametal:2015,hector_bispectrum1,gil-marin15_rsd}. Note that some of the measurements correspond to the DR11 release and others to the DR12, as indicated. In red squares the predictions of this work as they are listed in Table~\ref{table:results}. Empty symbols correspond to the analysis with no-AP test, whereas filled symbols to the analysis where the AP variables have been varied. The cyan and magenta bands display the $1\sigma$ and $2\sigma$ confident levels allowed by Planck TT+lowP+lensing in the base of a $\Lambda$CDM+GR model \citep{Planck_cosmology15}.}
\label{fig:boss_comparison}
\end{figure*}

Fig.~\ref{fig:boss_comparison} displays the $f\sigma_8$ measurements based on the following DR11 and DR12 works: \cite{Chuangetal:2013};\cite{Beutleretal:2013}; \cite{Samushiaetal:2014};\cite{Sanchezetal:2013};\cite{Reidetal:2013};\cite{Alametal:2015};\cite{hector_bispectrum1};\cite{gil-marin15_rsd}. 
A brief description on each of these works can be found in section 7.2 of \cite{gil-marin15_rsd}. The filled symbols correspond to analyses where the AP-variables have also been varied, whereas for the empty symbols they have been set to a certain fiducial value.  Both in LOWZ and CMASS samples, the observed differences between the DR11 and DR12 results are expected to be due to different systematics in the different models, scales and statistics considered. Note that the differences observed between the last two measurements are the same as those presented in Fig.~\ref{cosmo_covariance3}. For both LOWZ and CMASS samples our $f\sigma_8$ measurement is consistent within $1\sigma$ with the previous measurements. For the LOWZ sample, the $f\sigma_8$ results presented in this paper are in very good agreement with the {\it Planck15} prediction. For the CMASS sample we observe $\sim2\sigma$ tension with {\it Planck15}, being our $f\sigma_8$ measurement lower. This mild tension have been also reported in other analyses of the DR11 dataset such as the ones by \cite{Beutleretal:2013}; \cite{Samushiaetal:2014}; \cite{Sanchezetal:2013}  and is further discussed in \cite{Alametal:2016}.  In case we tune the AP parameters to the fiducial cosmology prediction, the $f\sigma_8$ best-fitting increases and the tension with {\it Planck15} is reduced to $\sim1.5\sigma$. 

\subsection{Combining  the cosmological parameters of RSD and BAO analyses of BOSS}\label{sec:combining_boss}

In this section we aim to combine the different cosmological parameters, $f\sigma_8$, $H(r)r_s(z_d)$ and $D_A(z)/r_s(z_d)$, obtained from different analyses of RSD and BAO from the LOWZ and CMASS samples of the DR12 BOSS survey. We focus on the RSD analysis of the power spectrum monopole and quadrupole presented in \cite{gil-marin15_rsd}, the RSD analysis of the power spectrum and bispectrum presented in this work, and the BAO post-reconstruction analysis of the power spectrum monopole and $\mu^2$-moment of \cite{gil-marin15_bao}. From the RSD analyses, both from the power spectrum multipoles only, and from the power spectrum in combination with the  bispectrum, we measure $f\sigma_8(z)$, $H(z)r_s(z_d)$, $D_A(z)/r_s(z_d)$; whereas from the post-reconstructed BAO analysis we measure $H(z)r_s(z_d)$, $D_A(z)/r_s(z_d)$. We denote the RSD measurements of the power spectrum multipoles with the superscript {\it RSD P}; the RSD measurements of the power spectrum combined with the bispectrum with the superscript {\it RSD P+B}; the BAO post-reconstructed measurements of the power spectrum moments with the superscript {\it BAO}. 

\begin{figure*}
\centering
\includegraphics[trim = 60mm 20mm 10mm 0mm, clip=false,scale=0.38]{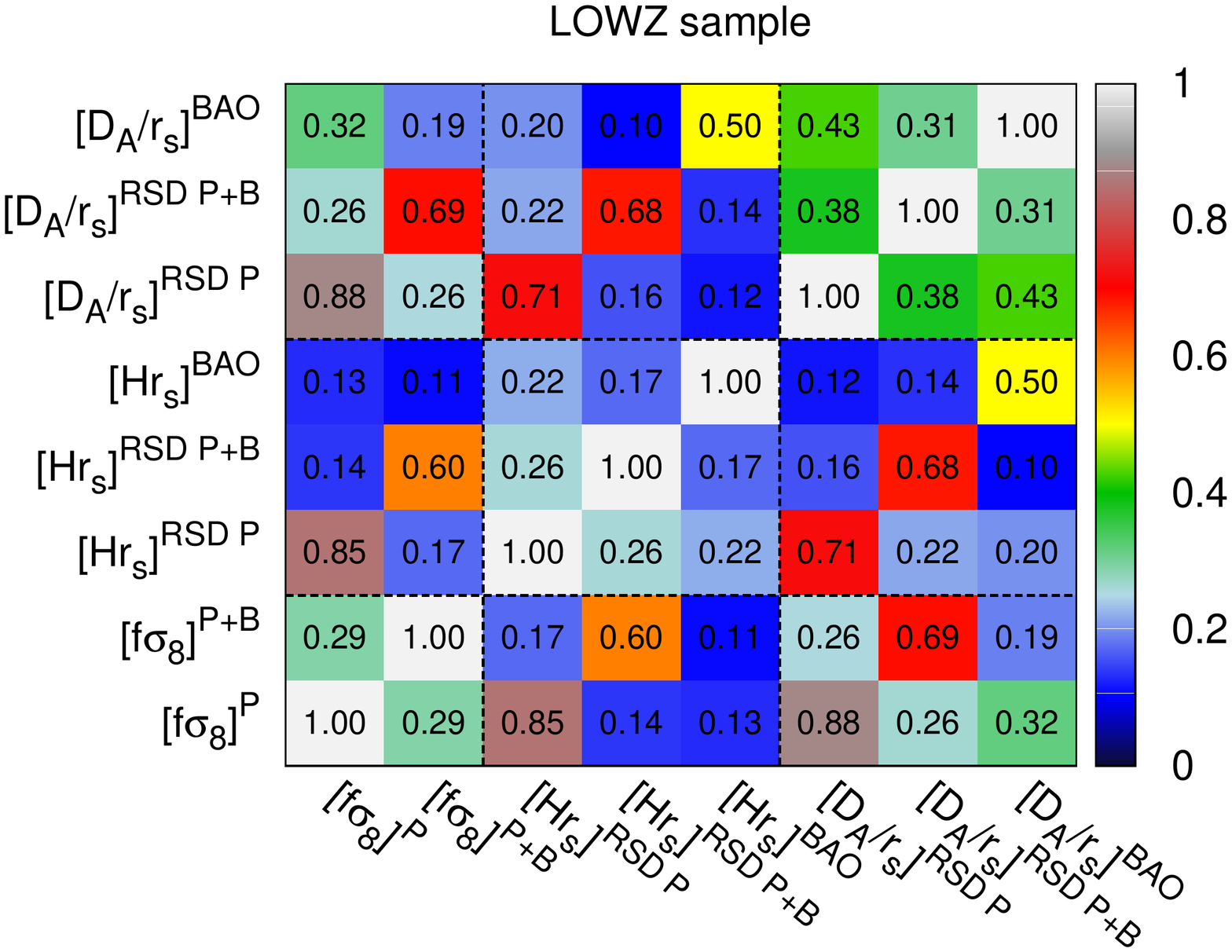}
\includegraphics[trim = 60mm 20mm 10mm 0mm, clip=false,scale=0.38]{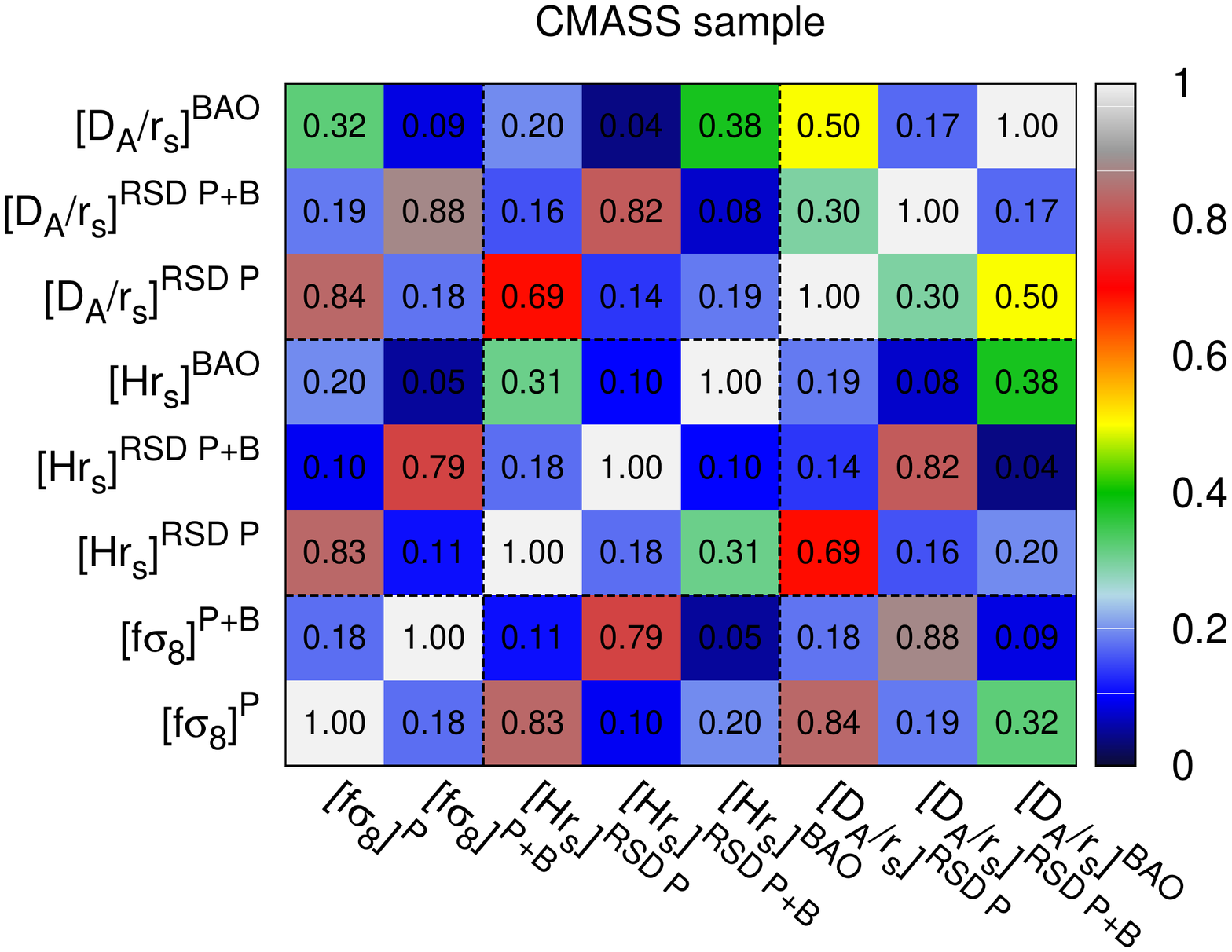}
\caption{Correlation coefficient  matrices for the LOWZ and CMASS samples, for the different cosmological parameters, $f\sigma_8(z)$, $H(z)r_s(z_d)$ and $D_A(z)/r_s(z_d)$, estimated from different probes: {\it RSD P} super-indices stand for the RSD analysis of the power spectrum monopole and quadrupole \citep{gil-marin15_rsd}; {\it RSD P+B} super-indices stand for the RSD analyses of the power spectrum and bispectrum (this work); {\it BAO} super-indices stand  for the BAO post-reconstruction analysis of the power spectrum multipoles \citep{gil-marin15_bao}. The colour scale stands for the degree of correlation among the displayed parameters. The numerical value of the correlation coefficient, $r_{ij}=C_{ij}/\sqrt{C_{ii}C_{jj}}$, is also displayed. The black dashed lines arrange the same physical cosmological parameters coming from the different probes.}
\label{fig:covariance_parameters}
\end{figure*}

The following data-vectors summarise the set of measurements of the different probes, 
   \begin{equation}
   \label{datatotlowz}
D^{\rm data}(z_{\rm LOWZ}) = 
 \begin{pmatrix}
  [f\sigma_8]^{{\rm RSD}\, P}  \\
  [f\sigma_8]^{{\rm RSD}\, P+B}  \\
  [Hr_s]^{{\rm RSD}\, P} \\
 [Hr_s]^{{\rm RSD}\, P+B} \\
[Hr_s]^{{\rm BAO}} \\
  [D_A/r_s]^{{\rm RSD}\, P} \\
 [D_A/r_s]^{{\rm RSD}\, P+B} \\
[D_A/r_s]^{{\rm BAO}} \\
 \end{pmatrix}=
  \begin{pmatrix}
  0.39529  \\
  0.45960 \\
  11.330 \\
  11.753 \\
  11.542\\
  6.3289 \\
  6.7443 \\
  6.6584 \\ 
 \end{pmatrix},
 \end{equation}
 for the LOWZ sample, and 
    \begin{equation}
   \label{datatotcmass}
D^{\rm data}(z_{\rm CMASS}) = 
 \begin{pmatrix}
  [f\sigma_8]^{{\rm RSD}\, P}  \\
  [f\sigma_8]^{{\rm RSD}\, P+B}  \\
  [Hr_s]^{{\rm RSD}\, P} \\
 [Hr_s]^{{\rm RSD}\, P+B} \\
[Hr_s]^{{\rm BAO}} \\
  [D_A/r_s]^{{\rm RSD} P} \\
 [D_A/r_s]^{{\rm RSD}v P+B} \\
[D_A/r_s]^{{\rm BAO}} \\
 \end{pmatrix}=
  \begin{pmatrix}
  0.44222  \\
  0.4175 \\
  13.844 \\
  13.781 \\
  14.552\\
  9.4187 \\
  9.3276 \\
  9.4220 \\ 
 \end{pmatrix},
 \end{equation}
 for the CMASS sample. In both cases the units of $H(z)r_s(z_d)$ are given in $[10^3 {\rm km}s^{-1}] $. We expect that the several parameters of the above data-vectors to be correlated, as they are coming from the same underlying dataset. We use the best-fitting values for each of the individual 2048 realisations of the MD-\textsc{Patchy} mocks for the RSD and BAO analyses in order to estimate the correlation coefficients of the data-vectors of Eq. \ref{datatotlowz}-\ref{datatotcmass}. These coefficients are displayed in Fig.~\ref{fig:covariance_parameters} for both the LOWZ and CMASS samples. The colour scale indicates the degree of correlation among parameters, and the overprinted value is the correlation coefficient obtained: $r_{ij}=C_{ij}/\sqrt{C_{ii}C_{jj}}$, where $C_{ij}$ is the $ij$-element of the covariance matrix and $C_{ii}$ the standard deviation of the parameter $i$. Using the variances of these parameters\footnote{As the variance here we use the diagonal elements of the individual covariances matrices of the RSD, and BAO analyses extracted from the likelihood of the data.} in combination with the correlation coefficients of Fig.~\ref{fig:covariance_parameters} we can generate a full $8\times8$ covariance matrix for the data-vectors of Eq. \ref{datatotlowz}-\ref{datatotcmass}. For clarity, this matrix is later displayed in Appendix \ref{appendix_matrix}.

In Fig.~\ref{fig:covariance_parameters}, the black dashed lines assemble the same cosmological parameters coming from different probes,
\begin{eqnarray}
\nonumber &&\{[f\sigma_8]^{{\rm RSD}\, P},\,[f\sigma_8]^{{\rm RSD}\, P+B}\}. \\
\nonumber &&\{[Hr_s]^{{\rm RSD}\, P},\,[Hr_s]^{{\rm RSD}\, P+B},\,[Hr_s]^{\rm BAO}\}, \\
\nonumber &&\{[D_A/r_s]^{{\rm RSD}\, P},\,[D_A/r_s]^{{\rm RSD}\, P+B},\,[D_A/r_s]^{\rm BAO}\}.
\end{eqnarray}

\begin{figure*}
\centering
\includegraphics[scale=0.3]{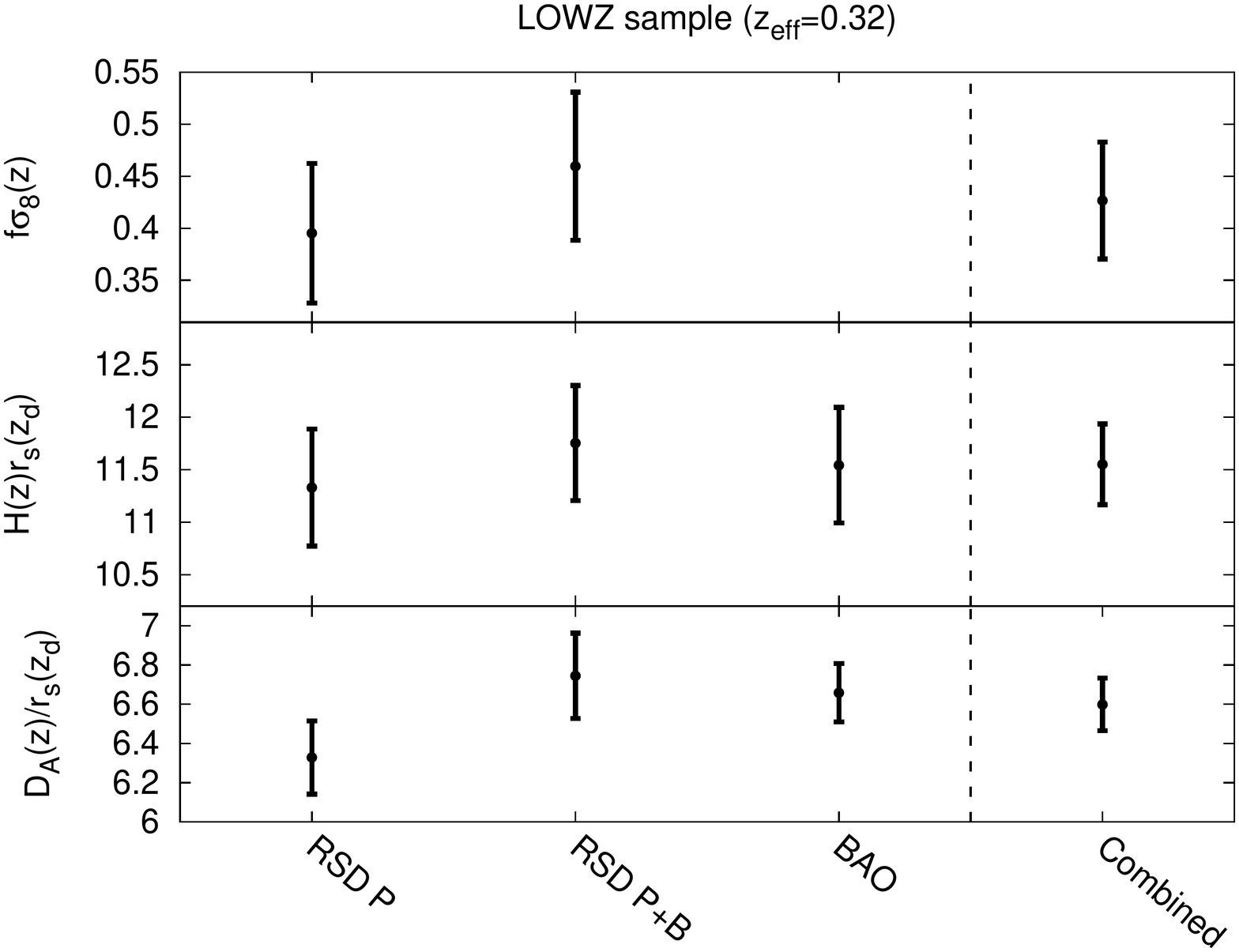}
\includegraphics[scale=0.3]{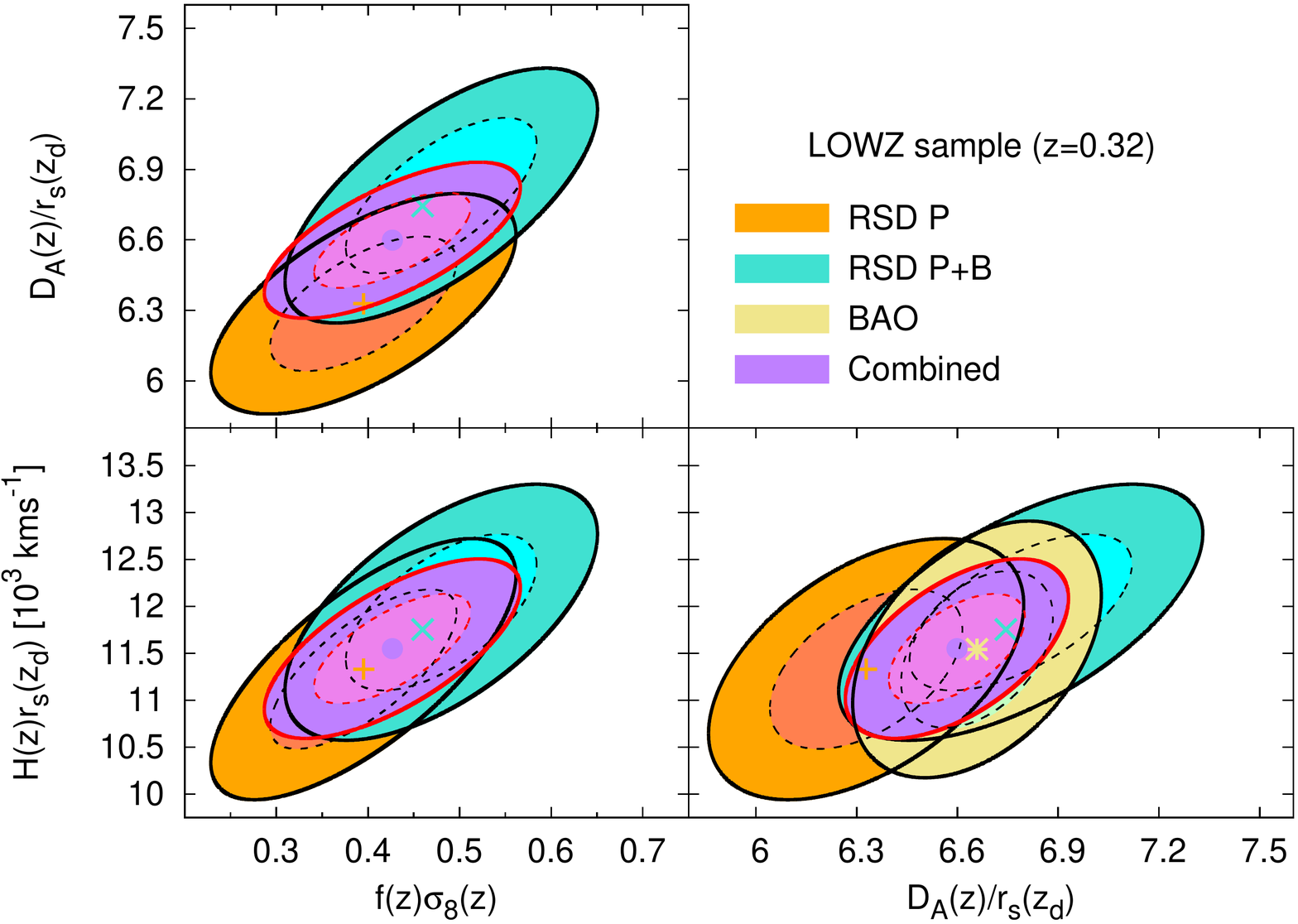}

\includegraphics[scale=0.3]{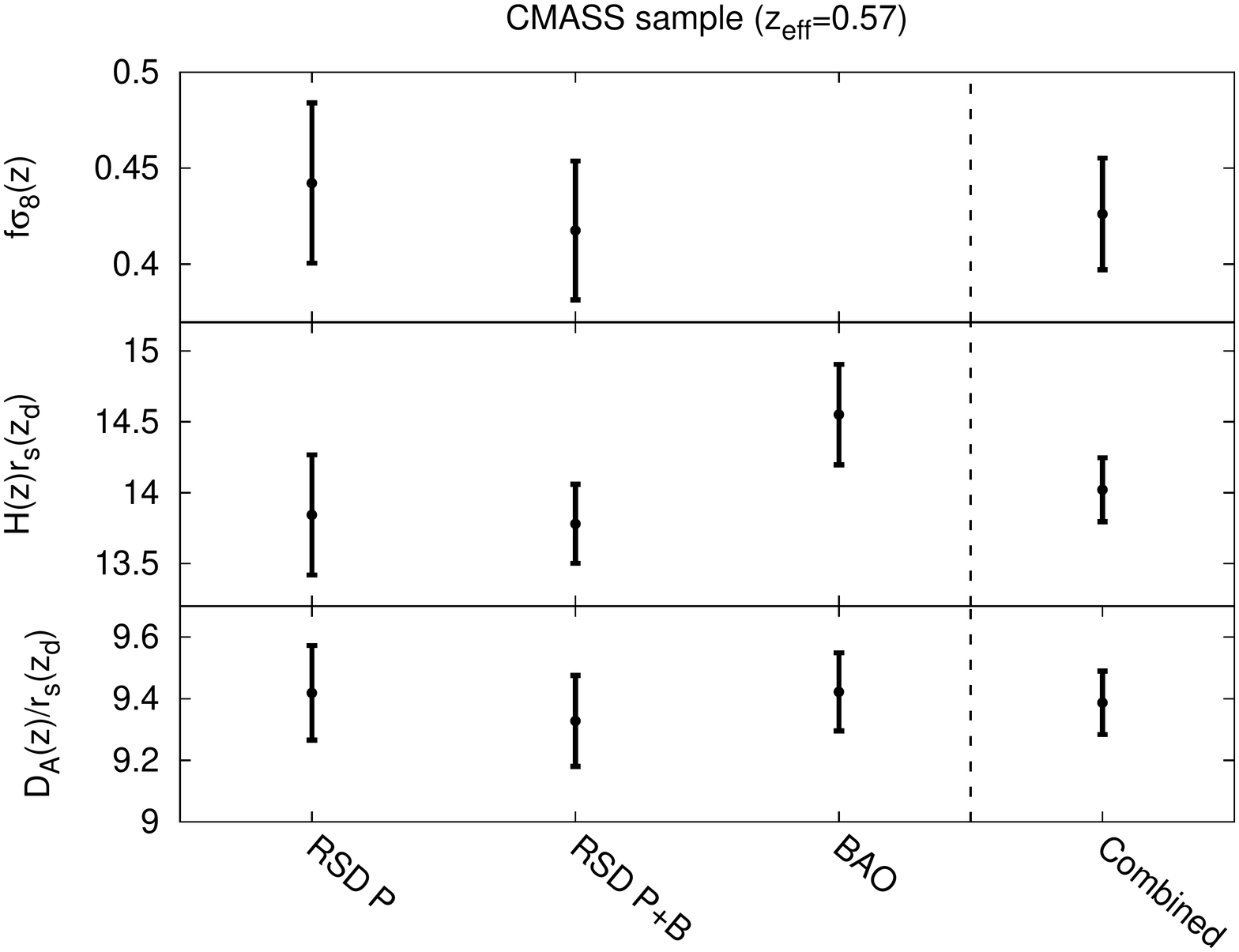}
\includegraphics[scale=0.3]{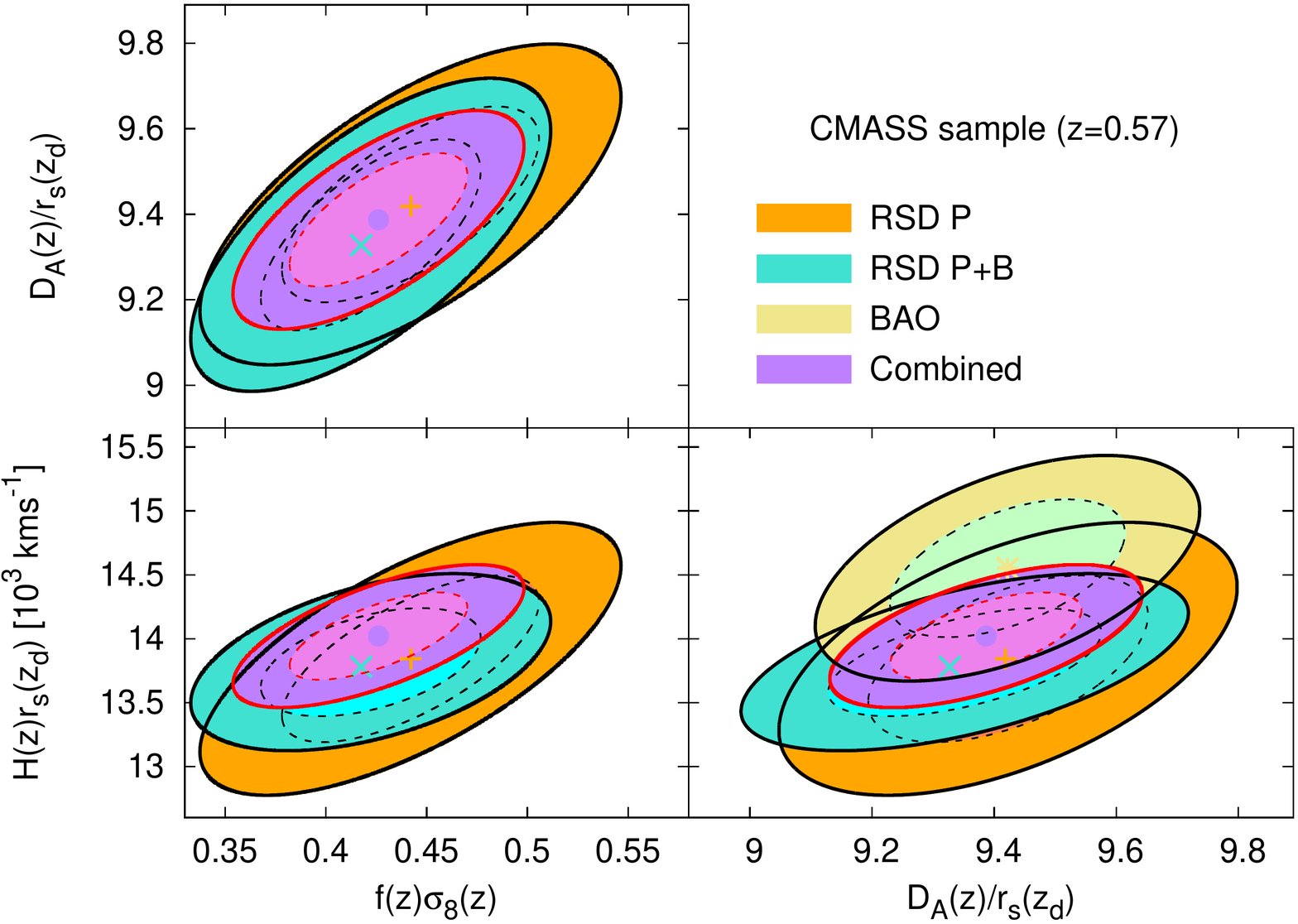}
\caption{The left panels show the individual measurements of $f\sigma_8$, $Hr_s$ and $D_A/r_s$ corresponding to the {\it RSD P} analysis \citep{gil-marin15_rsd}, {\it RSD P+B} analysis (this work), {\it BAO} analysis \citep{gil-marin15_bao}, as well as the combination of all of them according to Eq. \ref{eq:estimator}, using the correlation coefficients presented in Fig. \ref{fig:covariance_parameters}. The right panels show the same comparison in terms of the $1\sigma$ (solid lines) and $2\sigma$ (dashed lines) confident regions in the $f\sigma_8$-$Hr_s$-$D_A/r_s$ parameter space. Top panels refer to the LOWZ sample and bottom panels to the CMASS sample.}
\label{fig:boss_measurements}
\end{figure*}

These parameters can be recombined into a single cosmological parameter using the values the covariance matrices of Eq. \ref{covtotlowz}-\ref{covtotcmass}.
We aim to find a unique estimator for $f\sigma_8$, $Hr_s$ and $D_A$ by combining the measurements of the 3 probes. For simplicity we define the estimator to be a linear combination of the measurements of the individual probes, 
\begin{equation}\label{eq:estimator}
x^{\rm combined}=\sum_{i=1}^{N_{\rm probes}} w_i x_i.
\end{equation}
Here the variable $x$ stands either for $f\sigma_8$, $Hr_s$ or $D_A/r_s$; $N_{\rm probes}$ is 2 when $x$ stands for $f\sigma_8$ and 3 for the AP parameters; and finally $w_i$ are a weight given to each measurement that can only depend on the covariance matrix among the $x_i$ elements of the different probes. The variance of the estimator $x^{\rm combined}$ is given by, 
\begin{equation}\label{eq:variance}
\sigma^2_{x^{\rm combined}}=\sum_{ij} w_i w_j C_{ij},
\end{equation}   
where $C_{ij}$ is the $ij$-element of the covariance matrix. Minimising $\sigma^2_{x^{\rm combined}}$ subject to the condition of $\sum_i w_i=1$ we obtain, 
\begin{equation}\label{eq:weights}
w_i= \frac{\sum_k (C^{-1})_{ik}}{\sum_{jk}(C^{-1})_{jk}}.
\end{equation}
Therefore the linear combination of cosmological parameters of Eq. \ref{eq:estimator} with the weights defined by Eq. \ref{eq:weights} ensures the condition of minimum variance estimator.  When we apply the estimator of Eq. \ref{eq:estimator} to the data-vectors of Eq. \ref{datatotlowz} - \ref{datatotcmass} along with its covariance matrix of Eq. \ref{covtotlowz} - \ref{covtotcmass} we obtain that the combined parameters are given by the following data-vectors, 
   \begin{equation}
   \label{datalowzcombined}
D^{\rm data}(z_{\rm LOWZ}) = 
 \begin{pmatrix}
  [f\sigma_8]^{{\rm combined}}  \\
  [Hr_s]^{{\rm combined}} \\
[D_A/r_s]^{{\rm combined}} \\
 \end{pmatrix}=
  \begin{pmatrix}
  0.42660  \\
  11.549 \\
  6.5986 \\
 \end{pmatrix},
 \end{equation}
  for the LOWZ samples, and
   \begin{equation}
   \label{datacmasscombined}
D^{\rm data}(z_{\rm CMASS}) = 
 \begin{pmatrix}
  [f\sigma_8]^{{\rm combined}}  \\
  [Hr_s]^{{\rm combined}} \\
[D_A/r_s]^{{\rm combined}} \\
 \end{pmatrix}=
  \begin{pmatrix}
  0.42613  \\
  14.021 \\
  9.3869 \\
 \end{pmatrix},
 \end{equation}
for the CMASS sample; where the units of $H(z)r_s(z_d)$ are $[10^3 {\rm km}s^{-1}]$. The variance elements of these data-vectors are given by Eq. \ref{eq:variance}, whereas the correlation coefficients can be estimated them from the MD-\textsc{Patchy} mocks following the same procedure applied to the data, i.e., for each individual mock apply Eq. \ref{eq:estimator} and from those compute the correlation of the $x^{\rm combined}$ parameters. By doing this we obtain the following covariance matrices, 
  \begin{equation} 
\label{covlowzcombined}
C^{{\rm LOWZ}} = 10^{-3}
 \begin{pmatrix}
3.1667 & 14.726 & 5.0871 \\
& 148.099 & 28.929 \\
& &  17.883 \\
 \end{pmatrix},
\end{equation}
for the LOWZ sample, and 
 \begin{equation} 
\label{covcmasscombined}
C^{{\rm CMASS}} = 10^{-3}
 \begin{pmatrix}
0.84506 & 4.3722 & 2.0151 \\
& 50.717 &  13.827 \\
& &  10.613 \\
 \end{pmatrix}.
\end{equation}
for the CMASS sample.

This correspond to measurements of $f\sigma_8(z_{\rm LOWZ})=0.427\pm 0.056$, $H(z_{\rm LOWZ})r_s(z_d)=(11.55\pm 0.38)\cdot [10^3 {\rm km}s^{-1}]$ and $D_A(z_{\rm LOWZ})/r_s(z_d)=6.60\pm 0.13$ for the LOWZ sample and $f\sigma_8(z_{\rm CMASS})=0.426\pm 0.029$, $H(z_{\rm CMASS})r_s(z_d)=(14.02\pm 0.22)\cdot [10^3 {\rm km}s^{-1}]$ and $D_A(z_{\rm CMASS})/r_s(z_d)=9.39\pm 0.10$ for the CMASS sample, as it is displayed in Table \ref{table:combined_results}.
\begin{table}
\begin{center}
\begin{tabular}{|c|c|c}
 & LOWZ & CMASS \\
 \hline
 $f\sigma_8(z_{\rm eff})$ & $0.427\pm 0.056$ & $0.426\pm 0.029$  \\
 $H(z_{\rm eff})r_s(z_d)\,[10^3 {\rm km}s^{-1}]$ & $11.55\pm 0.38$ & $14.02\pm 0.22$ \\
 $D_A(z_{\rm eff})/r_s(z_d)$ & $6.60\pm 0.13$ & $9.39\pm 0.10$ \\
\end{tabular}
\end{center}
\caption{Cosmological parameters obtained by combining the individual measurements corresponding to the {\it RSD P} analysis, {\it RSD P+B} analysis, {\it BAO} analysis,  according to Eq. \ref{eq:estimator}. The correlation among these parameters is given by the covariance matrices of Eqs. \ref{covlowzcombined}-\ref{covcmasscombined}. These measurements correspond to the results labeled as ``combined" and shown in the left panels of Fig. \ref{fig:boss_measurements}. }
\label{table:combined_results}
\end{table}%
Comparing the error-bars of the parameters $f\sigma_8$ and $Hr_s$ coming from combining the 3 probes to those error-bars obtained in Table~\ref{table:results} from the {\it RSD P+B} analysis we see an improvement of $\sim25\%$ for the LOWZ sample and $\sim20\%$ for the CMASS sample. The gain in  the $D_A/r_s$ parameter is much higher because most of the signal in this parameter comes from the post-reconstruction BAO analysis, as can be inferred from the diagonal elements of the matrices of Eq. \ref{covtotlowz} and \ref{covtotcmass}. The different panels of Fig.~\ref{fig:boss_measurements} present the comparison between the combination of measurements (in purple contours) with the different individual probes, {\it RSD P} (orange contours), {\it RSD P+B} (turquoise contours) and {\it BAO} (yellow contours). 
The top and bottom panels show the results for the LOWZ and CMASS samples, respectively. The left panels present the measurements along with their standard deviation error-bars of the individual $f\sigma_8$, $Hr_s$ and $D_A/r_s$ parameters; whereas the right panels display the correlation ellipses of the same parameters. These results are as well displayed for clarity in a Table \ref{table_e}.

\begin{table*}
\begin{center}
\begin{tabular}{|c|c|c|c|c|}
Sample & Analysis & $f\sigma_8(z_{\rm eff})$ & $H(z_{\rm eff})r_s(z_{d})$ & $D_A(z_{\rm eff})/r_s(z_d)$ \\
\hline
\hline 

\multirow{4}{*}{LOWZ} & {\it RSD P}  & $0.394\pm0.064$  & $11.41\pm0.56$ & $6.35\pm0.19$ \\ &   {\it BAO} & $-$ &  $11.60\pm0.60$ & $6.66\pm0.16$ \\  &   {\it RSD P+B}  & $0.460\pm0.071$ & $11.75\pm0.55$ & $6.74\pm0.22$ \\ & Combined & $0.427\pm0.056$ & $11.55\pm0.38$ & $6.60\pm0.13$\\

\hline
\multirow{4}{*}{CMASS} &{\it RSD P} & $0.444\pm0.042$  & $13.92\pm0.44$ & $9.42\pm0.15$ \\ &   {\it BAO} & $-$ &  $14.56\pm0.37$ & $9.42\pm0.13$ \\  & {\it RSD P+B}  & $0.417\pm0.036$ & $13.78\pm0.28$ & $9.33\pm0.15$\\  & Combined  & $0.426\pm0.029$ & $14.02\pm0.22$ & $9.39\pm 0.10$\\
\end{tabular}
\caption{Best-fitting cosmological parameters, $f\sigma_8(z)$, $H(z)r_s(z_d)$ and $D_A(z)/r_s(z_d)$ inferred from a pre-reconstruction RSD analysis of the power spectrum monopole and quadrupole \citep{gil-marin15_rsd}, post-reconstruction BAO analysis of the power spectrum monopole and quadrupole \citep{gil-marin15_bao}, pre-reconstruction RSD analysis of the power spectrum monopole, quadrupole and bispectrum monopole (this work, Table \ref{table:results}) and the combination of these 3 measurements (this work, Table \ref{table:combined_results}). These results correspond to the plots of Fig. \ref{fig:boss_measurements}.}
\label{table_e}
\end{center}
\end{table*}%

From the panels of Fig.~\ref{fig:boss_measurements} we observe a $\lesssim1\sigma$ agreement among most of the parameters coming from different analysis techniques. The unique case where the tension reaches $\sim2\sigma$ tension is for the $Hr_s$ parameter for the CMASS sample, where the prediction from the power spectrum BAO analysis is about $\lesssim2\sigma$ higher than those predictions from both RSD analyses. This tension was already reported in \cite{gil-marin15_bao} when comparing the pre-recon with the post-recon best-fitting values (see $\alpha_\parallel$ values of table 3 in \citealt{gil-marin15_bao}). In particular this mild tension is related to the shift in the BAO peak position in the $\mu^2$-moment of the pre-reconstructed and post-reconstructed data catalogue. If we were plotting the pre-reconstruction prediction (which would be coming from the exact same data-set as the RSD analysis) the tension between the RSD analysis and BAO for the $Hr_s$ parameter would be reduced to $\leq1\sigma$, as the $Hr_s$ best-fitting value form the pre-recon data-set is lower than $Hr_s$ best-fitting value from the post-recon data-set. Therefore, this discrepancy has its origin in the effect of the reconstruction process in the anisotropic signal of the data and is likely to be just statistical. We believe that such large effect is not caused by systematic effects in the reconstruction process. Such potential systematics were quantified in \citep{gil-marin15_bao}, and resulted negligible compared to the statistical budget.

\subsection{Comparison with other galaxy surveys}

In this section we compare our measurements on $f\sigma_8$ for the LOWZ and CMASS with the $f\sigma_8$ values reported by other surveys at redshifts, along with {\it Planck15} predictions.

\begin{figure}
\centering
\includegraphics[scale=0.31]{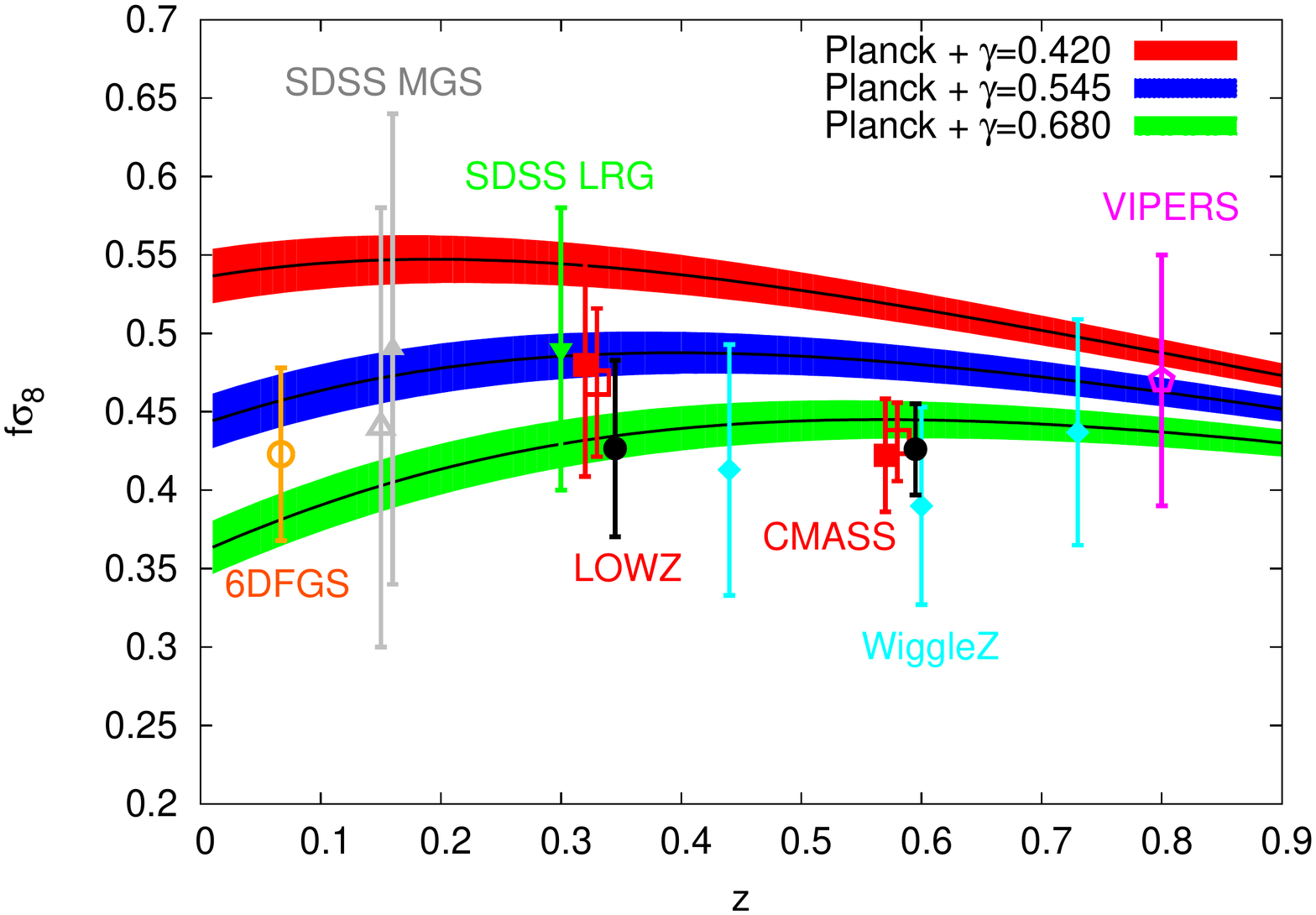}
\caption{Constrains on $f\sigma_8$ from several galaxy redshift surveys in the base of a $\Lambda$CDM model with $f(z)=\Omega_m^\gamma$: orange circles (6dFGS) by \citet{Beutleretal:2012}; grey triangle (SDSS Main Galaxy Sample) by \citet{Howlettetal:2015}; green inverse triangles (SDSS Luminous Red Galaxies) by \citet{Okaetal:2014}; cyan diamonds (WiggleZ) by \citet{Blakeetal:2012}; and purple pentagon (VIPERS) by \citet{Vipers}. In red squares the results from BOSS-DR12 according to Table~\ref{table:results}. In black circles the combined measurement from the RSD and BAO CMASS and LOWZ DR12 BOSS analyses, derived in \S\ref{sec:combining_boss}. Filled symbols represent the $f\sigma_8$ measurements when both the RSD and AP parameters have been constrained and filled symbols then only the RSD parameters are constrained. For the empty symbols, as well as for the combined measurement, the redshift position have been slightly displaced for clarity. The red, blue and green bands show the $1\sigma$ confident levels allowed by Planck TT+lowP in the base of a $\Lambda$CDM model \citep{Planck_cosmology15} when $\gamma=0.420$, $\gamma=0.55$ (GR) and $\gamma=0.680$.}
\label{fig:survey_comparison}
\end{figure}

Fig.~\ref{fig:survey_comparison} compares our measurements of $f\sigma_8$ (red symbols), with those from the 6dFGS by \citet{Beutleretal:2012}, SDSS Main Galaxy Sample by \citet{Howlettetal:2015}, SDSS Luminous Red Galaxies by \citet{Okaetal:2014}, WiggleZ by \citet{Blakeetal:2012}; and VIPERS by \citet{Vipers}. A brief description of each of these measurements was presented in section 7.3 of \cite{gil-marin15_rsd}, and we do not repeat it here. As in Fig.~\ref{fig:boss_comparison}, full symbols correspond to the results whose analyses also fit for the AP variables, whereas empty symbols keep them fixed to a fiducial cosmology. We also include the combined $f\sigma_8$ measurement (black symbols) by using {\it RSD P}, {\it RSD P+B} and {\it BAO}, derived in \S~\ref{sec:combining_boss}.
The coloured bands present the model prediction for {\it Planck15} best-fitting $\Omega_m=0.308$, when the different theories models for the theory of gravity are adopted. Under the assumption of $f(z)=\Omega_m^\gamma(z)$, we display the results for the GR prediction $\gamma=0.55$ (blue bands, $1\sigma$ confident levels), and as well two extra values of $\gamma$, $\gamma=0.420$ and $\gamma=0.680$, in green and red bands, respectively. In general all the results are in agreement with {\it Planck15}+GR prediction within $1\sigma$ and $2\sigma$ confident levels. Lower-than-GR values for $\gamma$ are disfavoured by the measurements, whereas higher-than-GR values for $\gamma$ are slightly favoured as it was noted in \cite{Macaulayetal:2013}.

\subsection{Deviations from GR predictions}
In the $\Lambda$CDM scenario the growth rate of structure, $f$, can be expressed as a function of the matter density of the Universe $\Omega_m(z)$ through \citep{Linder2005}, $f(z)=\Omega_m^\gamma(z)$; where $\gamma$ is the so called growth index, which under the assumption of GR takes the value $\gamma_{\rm GR}\simeq0.55$. The growth rate of structure is then related, not only to the amount of matter of the Universe, but also to the theory of gravity that rules the matter component (both baryonic and dark) of the Universe. Therefore, determining $\Omega_m(z)$ and $f(z)$ independently allow us to  perform a consistency test on the $\gamma$ parameter, which could potentially deviate from the GR prediction. In particular, we  measure the value of $f(z)$ through RSD and the value of $\Omega_m(z)$ through the AP-parameters in addition to CMB data.

In order to obtain the constrain on $\gamma$ we use the {\it Planck15} results, TT+lowP\footnote{This corresponds to the first column of table 3 in \cite{Planck_cosmology15}}, and combine them with the BOSS measurements from the LOWZ and CMASS samples presented in this paper. We build the total likelihood to be the product of the individual likelihoods of {\it Planck15}, CMASS and LOWZ: $\mathcal{L}=\mathcal{L}_{\rm Planck15}\times\mathcal{L}_{\rm CMASS}\times\mathcal{L}_{\rm LOWZ}$, assuming they are independent. For simplicity we do not exploit the Integrated Sachs-Wolfe (ISW) effect here, and therefore, we do not use the CMB data to directly put constrains on $\gamma$. In this case the CMB data is only used to provide tight constrains on ${\Omega_m}_0$ and on ${\sigma_8}_0$

For each \textsc{mcmc} chain element we randomly choose a value for $\{{\Omega_m}_0,\, H_0,\, {\sigma_8}_0,\, \gamma\}$, where the $0$-sub index stands for the quantities at $z=0$. The linear growth factor $D$, is then computed as the integration of the logarithmic growth rate: $d{\ln D(a,\gamma)}/d{\ln a}\equiv f(a,\gamma) = \Omega_m(z)^{\gamma}$; where $a$ is the scale factor, $a\equiv1/(1+z)$. In order to obtain the $\sigma_8$ value at a given redshift we then propagate it as, $\sigma_8(z_{\rm eff},\gamma)=D(z_{\rm eff},\gamma){\sigma_8}_0$. One has to bear in mind that the ${\sigma_8}_0$ value that {\it Planck15} provides has been obtained by propagating its linear evolution from the CMB epoch until today using a linear growth factor which intrinsically assumes $\gamma=\gamma_{\rm GR}$. In order to undo this assumption we take the value of the $\sigma_8$ at $z=0$ as ${\sigma_8}_0(\gamma)={\sigma_8}_0|_{\rm Planck15} D(z_{\rm CMB}, \gamma_{\rm GR}) / D(z_{\rm CMB}, \gamma)$; where ${\sigma_8}_0|_{\rm Planck15}$ is the $\sigma_8$ value at $z=0$ provided by {\it Planck15} and $z_{\rm CMB}\simeq 1100$. Using all these ingredients we build $f(z_{\rm eff})\sigma_8(z_{\rm eff})$ form each \textsc{mcmc} chain element. In addition we also constrain the BOSS measured quantities, $H(z_{\rm eff})r_s(z_d)$ and $D_A(z_{\rm eff})/r_s(z_d)$ using the relations,
\begin{eqnarray} 
H(z_{\rm eff})&=&H_0\sqrt{{\Omega_m}_0(1+z_{\rm eff})^3+1-{\Omega_m}_0} \\ 
D_A(z_{\rm eff})&=& \frac{1}{1+z_{\rm eff}}\int_0^{z_{\rm eff}} \frac{c\,dz}{H(z)}
\end{eqnarray}

In order to  perform the constrains on $\gamma$ we consider following 3 BOSS datasets,
\begin{enumerate}
\item $\{ f\sigma_8,\, D_A/r_s,\, Hr_s\}$ parameters from the LOWZ and CMASS samples displayed by Eq. \ref{data_lowz}-\ref{data_cmass}. We refer to this case as $[f\sigma_8,\, D_A,\,H]^{\rm P+B}$.
\item $\{ f\sigma_8,\, D_A/r_s,\, Hr_s\}$ combined parameters presented in \S{\ref{sec:combining_boss}}, corresponding to the LOWZ and CMASS samples and described by Eq. \ref{datalowzcombined} - \ref{datacmasscombined}.  We refer to this case as $[f\sigma_8,\, D_A,\,H]^{\rm combined}$.
\item $\{ f,\, D_A/r_s,\, Hr_s\}$ parameters from the CMASS sample only, displayed by Eq. \ref{data2}. We refer to this case as $[f,\, D_A,\, H]^{\rm P+B}$. Note that no constrains on $\sigma_8$ are assumed, and therefore the constrains on $\gamma$ come only from $f(z,\gamma)=\Omega_m(z)^\gamma$.
\end{enumerate}

\begin{table}
\begin{center}
\begin{tabular}{|c|c|c}
Assumptions & $\gamma$ & ${\Omega_m}_0$ \\
\hline
\hline
$[f,\, D_A,\, H]^{\rm P+B}$ & $0.80^{+0.31}_{-0.23}$ & $0.20_{-0.10}^{+0.14}$ \\
$[f\sigma_8,\, D_A,\, H]^{\rm P+B}$ & $0.33^{+0.41}_{-0.47}$ & $0.330^{+0.058}_{-0.064}$ \\
$[f\sigma_8,\, D_A,\, H]^{\rm combined}$ & $0.18^{+0.29}_{-0.34}$ & $0.341^{+0.045}_{-0.046}$ \\
\hline
$[f,\, D_A,\, H]^{\rm P+B}$+{\it Planck15} & $0.95^{+0.26}_{-0.23}$ & $0.315\pm 0.012$ \\
$[f\sigma_8,\, D_A,\, H]^{\rm P+B}$+{\it Planck15} & $0.701^{+0.088}_{-0.093}$ & $0.318^{+0.012}_{-0.011}$ \\
$[f\sigma_8,\, D_A,\, H]^{\rm combined}$+{\it Planck15} & $0.733^{+0.068}_{-0.069}$ & $0.320^{+0.011}_{-0.012}$  
\end{tabular}
\end{center}
\caption{Constrains on $\gamma$ and ${\Omega_m}_0$ parameters obtained from the BOSS datasets, $[f\sigma_8,\, D_A,\,H]^{\rm P+B}$, $[f\sigma_8,\, D_A,\,H]^{\rm combined}$ and $[f,\, D_A,\, H]^{\rm P+B}$ (see text), when they are used alone (first 3 rows) and when they are combined with CMB data from {\it Planck15} (last 3 rows). The quoted error-bars correspond to $1\sigma$ confident regions. Fig. \ref{fig:gamma} displays the likelihoods of these models when they are combined with {\it Planck15} data. The constrains on $\gamma$ comes uniquely from BOSS data, as the ISW from the CMB has not been exploited for simplicity.}
\label{table:gamma}
\end{table}%

Table \ref{table:gamma} displays the constrains on $\gamma$ and ${\Omega_m}_0$ obtained from the datasets described above, when they are used alone (first 3 rows), and when they are combined with {\it Planck15} results (last 3 rows), as previously described. The constrains on $\gamma$ are very mild when the BOSS dataset is used alone, whereas when the {\it Planck15} data is added we obtain much better constrains: when  the information on $f\sigma_8$ along with the AP parameters is used we obtain $\gamma=0.701^{+0.088}_{-0.093}$ (13\% precision) for the power spectrum and bispectrum measurements  (P+B model); and $\gamma=0.733^{+0.068}_{-0.069}$ (10\% precision) for the combined model. As the ISW information has not been exploited, the improvement in the determination of $\gamma$ arises only from a better determination of ${\Omega_m}_0$  and not through a direct constrain of $\gamma$ using the CMB data. These predictions are slightly better than those found by \cite{Beutleretal:2013}; \cite{Samushiaetal:2014};\cite{Sanchezetal:2013} using the power spectrum and correlation function multipoles of the BOSS DR11 datasets. On the other hand, if we only use $f$ to constrain $\gamma$ we obtain $\gamma=0.95^{+0.26}_{-0.23}$, which has a significantly larger error-bars, but is independent of the $\sigma_8$ power spectrum normalisation.

\begin{figure}
\centering
\includegraphics[scale=0.31]{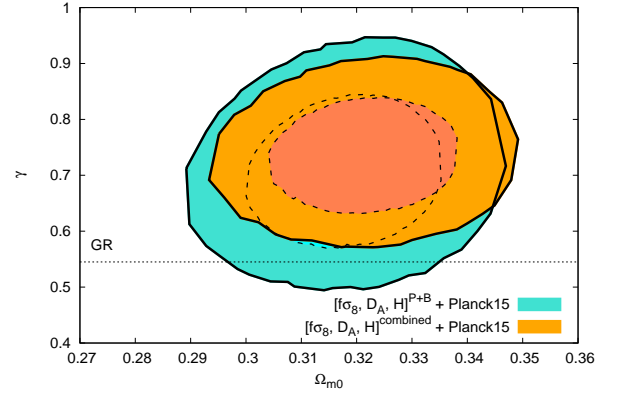}
\caption{Two dimensional posterior distribution for $\gamma$ and ${\Omega_m}_0$ from different BOSS datasets in combination with CMB data, in different colour lines (see legend and text). The solid and dashed lines correspond to the 68 and 95.4\% confidence levels, respectively. The black dotted horizontal line display the GR prediction for $\gamma$, $\gamma_{\rm GR}=0.55$.  As the ISW information has not been exploited, the constrains on $\gamma$ are directly related to the constrains on $f$ and $f\sigma_8$ from BOSS data, whereas the CMB data is only used to set constrains on ${\Omega_m}_0$ and ${\sigma_8}_0$.}
\label{fig:gamma}
\end{figure}

Fig. \ref{fig:gamma} displays the two dimensional posterior distribution for  $\gamma$ and ${\Omega_m}_0$ corresponding to the results from Table \ref{table:gamma}. For clarity only the results where the CMB data has been combined are shown:  the turquoise contours for $[f\sigma_8\,,D_A,\,H]^{\rm P+B}$ and the orange contours for $[f\sigma_8\,,D_A,\,H]^{\rm combined}$. In all cases the solid and dashed lines show the $1\sigma$ and $2\sigma$ confident levels, respectively. The horizontal black dotted line show the GR prediction for $\gamma$, $\gamma_{\rm GR}=0.55$. 

The measurements of $\gamma$ presented in Fig. \ref{fig:gamma} and Table \ref{table:gamma} are in mild or no-tension with the GR prediction: $1.7\sigma$ for $[f,\,D_A,\,H]^{\rm P+B}$; $1.6\sigma$ for $[f\sigma_8,\,D_A,\,H]^{\rm P+B}$ and $2.7\sigma$ for $[f\sigma_8,\,D_A,\,H]^{\rm combined}$; where in all cases a weaker-than-GR model (higher $\gamma$, lower $f\sigma_8$) is favoured. Consistent values of $\gamma$ were found using the DR11 dataset \citep{Beutleretal:2013};  \cite{Samushiaetal:2014};\cite{Sanchezetal:2013}. The observed $2.7\sigma$ tension for the combined case could be due to either a statistical fluctuation on the $f\sigma_8$ measured values, or a unaccounted systematic uncertainties in BOSS and/or {\it Planck15} data; or an indication of a failure of the $\Lambda$CDM + GR gravity. Similar tension between {\it Planck15} and RSD-analyses of different galaxy surveys was reported by \cite{Macaulayetal:2013}, where a similar $2.5\sigma$ tension was found. Further investigation into the preference of BOSS data for $>\gamma_{\rm GR}$ when a $\Lambda$CDM model is assumed is left for future work. These additional analyses may include Bayesian selection comparing the evidence ratio of a GR+$\Lambda$CDM model with phenomenological non-GR models with free $\gamma$.

\section{Conclusions}\label{sec:conclusions}

In this paper we have presented the measurement of the isotropic bispectrum of the LOWZ and CMASS DR12 galaxy samples of the Baryon Oscillation Spectroscopic Survey of the Sloan Digital Sky Survey III. We report a detection of the bispectrum monopole at  high statistical significance, which enables to use it to measure cosmological parameters of interest.  

This paper extends, improves and complements the previous bispectrum analysis of the BOSS DR11 galaxy sample presented in \cite{hector_bispectrum1,hector_bispectrum2}: i) we use an improved bispectrum estimator that enables a fast measurement of all triangular shapes, which boosts the statistical gain when constraining cosmological parameters; ii) we use a full covariance matrix of the power spectrum and bispectrum estimated from 2048 realisations of MD-\textsc{Patchy} mocks, which allows the use of a minimal error estimator for the inferred cosmological parameters; iii) the geometrical Alcock-Paczynski effect has been included on the galaxy bispectrum, which allows us to set constrains on the angular diameter distance parameter, $D_A(z_{\rm eff})/r_s(z_d)$ and the Hubble parameter $H(z_{\rm eff})r_s(z_d)$; iv)  we perform our analysis on both the  LOWZ and CMASS samples, and therefore we provide measurements on two redshift bins. 
 
We have analysed the effects of the redshift space distortions in the bispectrum monopole, in combination with the power spectrum monopole and quadrupole to constrain on the growth factor times the amplitude of the linear power spectrum, $f\sigma_8$. In order to extract cosmological information from the galaxy bispectrum measurements in combination with the power spectrum multipoles we have used a non-local and non-linear bias model \citep{McDonaldRoy:2009}. After imposing the condition of locality in Lagrangian space only two free parameters are left to marginalise over, $b_1$ and $b_2$.  The RSD in the bispectrum are described through the phenomenological model presented in \citep{hector_bispectrum0}, which has been used in the previous bispectrum analysis of DR11. The RSD model depends on the logarithmic rate of structure growth, $f$,  on two FoG damping parameters, $\sigma_{\rm FoG}^P$ and $\sigma_{\rm FoG}^B$,  one for the power spectrum and another for the bispectrum, respectively; and on the amplitude of the shot noise relative to the Poisson prediction. Although $f$ is directly related to the fiducial cosmological model when GR is assumed as a theory of gravity, we have kept $f$ free in order to test possible deviations from GR. In addition to the RSD,  we have also included the geometrical AP effect, through the dilation parameters $\alpha_\parallel$ and $\alpha_\perp$, which modifies the wave modes parallel and perpendicular to the LOS, respectively. These parameters are related to the angular diameter distance and the Hubble parameter, which we are also able to constrain. In our analysis we have fixed the shape of the linear power spectrum using the fiducial cosmology ${\boldsymbol \Omega}^{\rm fid}$, but we have marginalised over the amplitude $\sigma_8$. In total, our galaxy redshift space power spectrum and bispectrum model has 9 free parameters we marginalise over, ${\bf \Psi}=\{b_1,b_2,A_{\rm noise}, \sigma_{\rm FoG}^P, \sigma_{\rm FoG}^B, f,\sigma_8, \alpha_\parallel,\alpha_\perp  \}$.

We have computed the full covariance matrix of the power spectrum monopole, quadrupole and bispectrum monopole using  2048 realisations of \textsc{MD-Patchy} mocks. We have observed that there is a strong correlation among similar triangle shapes of the bispectrum, as well as between the power spectrum monopole and bispectrum monopole, for the triangles that share at least one $k$-vector. The correlation between the power spectrum quadrupole and bispectrum monopole has been observed to be consistent with 0. 

We have tested possible systematics of our bispectrum model using the \textsc{MD-Patchy} mocks and dark matter halo $N$-body simulations. We have found significant disagreement between the behaviour of the model when it was applied to the mocks and to $N$-body, especially at small scales, where the dark matter halo bispectrum is in better agreement with the model prediction than the \textsc{MD-Patchy} mocks. Using these resources we have estimated the truncation scale to be applied to the data, and as well the systematics of our model, that have been added in quadrature to the statistical errors in the final measurement from the data. 

When analysing the data, we find that for the DR12 LOWZ sample $f(z_{\rm LOWZ})\sigma_8(z_{\rm LOWZ})=0.460\pm 0.071$, $D_A(z_{\rm LOWZ})/r_s(z_d)=6.74\pm0.22$, $H(z_{\rm LOWZ})r_s(z_d)=(11.75\pm 0.55)\,{10^3\rm km}s^{-1}$, where $z_{\rm LOWZ}=0.32$. For DR12 CMASS we find $f(z_{\rm CMASS})\sigma_8(z_{\rm CMASS})=0.417\pm 0.036$, $D_A(z_{\rm CMASS})/r_s(z_d)=0.33\pm0.15$, $H(z_{\rm CMASS})r_s(z_d)=(13.78 \pm 0.28)\, {10^3\rm km}s^{-1}$, where $z_{\rm CMASS}=0.57$. All the quoted error-bars include the statistic and systematic error budget, both added in quadrature.  The correlation among these parameters have been also presented in the covariance matrices of Eq. \ref{cov1} and \ref{cov2}, for LOWZ and CMASS samples, respectively. These are the main results of this paper and are in general agreement with previous BOSS DR11 and DR12 measurements. 

Adding the bispectrum to the traditional power spectrum multipole analysis have enable us to measure separately $f$ and $\sigma_8$ for the CMASS sample, along with their correlation factor. We have found that when the AP parameters are set to their fiducial value, $f(z_{\rm CMASS})=0.649\pm0.076$ and $\sigma_8(z_{\rm CMASS})=0.660\pm0.067$, with a correlation factor of $-0.82$. When the AP parameters are also varied we find that  $f(z_{\rm CMASS})=0.58\pm0.12$ and $\sigma_8(z_{\rm CMASS})=0.668\pm0.076$ with the correlation matrix given by Eq. \ref{cov3}. 

When we combine the LOWZ and CMASS BOSS data coming from the RSD and BAO analyses of \cite{gil-marin15_rsd,gil-marin15_bao} along with the results presented in this work we are able to improve significantly the constrains on the cosmological parameters: $f\sigma_8(z_{\rm LOWZ})=0.427\pm 0.056$, $H(z_{\rm LOWZ})r_s(z_d)=(11.55\pm 0.38)\cdot [10^3 {\rm km}s^{-1}]$ and $D_A(z_{\rm LOWZ})/r_s(z_d)=6.60\pm 0.13$ for the LOWZ sample and $f\sigma_8(z_{\rm CMASS})=0.426\pm 0.029$, $H(z_{\rm CMASS})r_s(z_d)=(14.02\pm 0.22)\cdot [10^3 {\rm km}s^{-1}]$ and $D_A(z_{\rm CMASS})/r_s(z_d)=9.39\pm 0.10$ for the CMASS sample.

We have performed a $\Lambda$CDM-GR consistency check using the $f\sigma_8$ along with the AP parameters measured from BOSS data, in combination with ${\Omega_m}_0$, $H_0$ and ${\sigma_8}_0$ constrains from CMB using ${\it Planck15}$ data. We measure the growth index $\gamma=0.701^{+0.088}_{-0.093}$, using the power spectrum and bispectrum measured quantities. When we use the combined BOSS measurements coming from RSD and BAO analyses along with {\it Planck15} measurements we find $\gamma=0.733^{+0.068}_{-0.069}$.  We find that this result is in $2.7\sigma$ tension with the predictions from GR, $\gamma_{\rm GR}\simeq0.55$.  This tension could be due to i)  a statistical fluctuation on the $f\sigma_8$ measured values; ii) an unaccounted systematic uncertainties in BOSS or {\it Planck15} data; iii) an indication of a failure of the $\Lambda$CDM or the GR gravity model. Future galaxy surveys using more redshift bins and more accurate data may shed light on this tension revealing the origin of this discrepancy.  

The constraints on $f(z_{\rm eff})\sigma_8z_{\rm eff}$, along with $H(z_{\rm eff})r_d(z)$ and $D_a(z_{\rm eff})r_d(z_{\rm eff})$, will be useful in a joint analysis with other cosmological data sets (in particular CMB data) for setting stringent constraints on neutrino mass, dark energy, gravity, curvature as well as number of neutrino species.

\section*{Acknowledgements}
We thank Beth Reid for providing the $N$-body halo catalogues used in this paper. We thanks Julien Guy, Rom\'an Scoccimarro and Chang Hoon Hahn for useful comments on the final draft. 

This work has been done within the Labex ILP (reference ANR-10-LABX-63) part of the Idex SUPER, and received financial state aid managed by the Agence Nationale de la Recherche, as part of the programme Investissements d'avenir under the reference ANR-11-IDEX-0004-02.

WJP is grateful for support from the UK Science and Technology Facilities Research Council through grants ST/M001709/1 and ST/N000668/1. WJP is also grateful for support from the European Research Council through grant 614030 Darksurvey, and support from the UK Space Agency through grant ST/N00180X/1.

LV acknowledges support of FPA2014-57816-P and MDM-2014-0369 of ICCUB (Unidad de Excelencia Maria de Maeztu).

CC acknowledges support from the Spanish MICINN?s Consolider-Ingenio 2010 Programme under grant MultiDark CSD2009-00064 and AYA2010-21231-C02-01 grant.
CC was also supported by the Comunidad de Madrid under grant HEPHACOS S2009/ESP-1473. C.C. was supported as a MultiDark fellow.
	
Funding for SDSS-III has been provided by the Alfred P. Sloan
Foundation, the Participating Institutions, the National Science
Foundation, and the U.S. Department of Energy Office of Science. The
SDSS-III web site is http://www.sdss3.org/.

SDSS-III is managed by the Astrophysical Research Consortium for the
Participating Institutions of the SDSS-III Collaboration including the
University of Arizona,
the Brazilian Participation Group,
Brookhaven National Laboratory,
University of Cambridge,
Carnegie Mellon University,
University of Florida,
the French Participation Group,
the German Participation Group,
Harvard University,
the Instituto de Astrofisica de Canarias,
the Michigan State/Notre Dame/JINA Participation Group,
Johns Hopkins University,
Lawrence Berkeley National Laboratory,
Max Planck Institute for Astrophysics,
Max Planck Institute for Extraterrestrial Physics,
New Mexico State University,
New York University,
Ohio State University,
Pennsylvania State University,
University of Portsmouth,
Princeton University,
the Spanish Participation Group,
University of Tokyo,
University of Utah,
Vanderbilt University,
University of Virginia,
University of Washington,
and Yale University.
This research used resources of the National Energy Research Scientific
Computing Center, which is supported by the Office of Science of the
U.S. Department of Energy under Contract No. DE-AC02-05CH11231.

Numerical computations were done on the Sciama High Performance Compute (HPC) cluster which is supported by the ICG, SEPNet and the University of Portsmouth. The simulations for $N$-body haloes were performed at the National Energy Research Scientific Computing Center,  the Shared Research Computing  Services  Pilot  of  the  University  of  California  and  the  Laboratory  Research  Computing project at Lawrence Berkeley National Laboratory.





%
%
%


\def\jnl@style{\it}
\def\aaref@jnl#1{{\jnl@style#1}}

\def\aaref@jnl#1{{\jnl@style#1}}

\def\aj{\aaref@jnl{AJ}}                   
\def\araa{\aaref@jnl{ARA\&A}}             
\def\apj{\aaref@jnl{ApJ}}                 
\def\apjl{\aaref@jnl{ApJ}}                
\def\apjs{\aaref@jnl{ApJS}}               
\def\ao{\aaref@jnl{Appl.~Opt.}}           
\def\apss{\aaref@jnl{Ap\&SS}}             
\def\aap{\aaref@jnl{A\&A}}                
\def\aapr{\aaref@jnl{A\&A~Rev.}}          
\def\aaps{\aaref@jnl{A\&AS}}              
\def\azh{\aaref@jnl{AZh}}                 
\def\baas{\aaref@jnl{BAAS}}               
\def\jrasc{\aaref@jnl{JRASC}}             
\def\memras{\aaref@jnl{MmRAS}}            
\def\mnras{\aaref@jnl{MNRAS}}             
\def\pra{\aaref@jnl{Phys.~Rev.~A}}        
\def\prb{\aaref@jnl{Phys.~Rev.~B}}        
\def\prc{\aaref@jnl{Phys.~Rev.~C}}        
\def\prd{\aaref@jnl{Phys.~Rev.~D}}        
\def\pre{\aaref@jnl{Phys.~Rev.~E}}        
\def\prl{\aaref@jnl{Phys.~Rev.~Lett.}}    
\def\pasp{\aaref@jnl{PASP}}               
\def\pasj{\aaref@jnl{PASJ}}               
\def\qjras{\aaref@jnl{QJRAS}}             
\def\skytel{\aaref@jnl{S\&T}}             
\def\solphys{\aaref@jnl{Sol.~Phys.}}      
\def\sovast{\aaref@jnl{Soviet~Ast.}}      
\def\ssr{\aaref@jnl{Space~Sci.~Rev.}}     
\def\zap{\aaref@jnl{ZAp}}                 
\def\nat{\aaref@jnl{Nature}}              
\def\iaucirc{\aaref@jnl{IAU~Circ.}}       
\def\aplett{\aaref@jnl{Astrophys.~Lett.}} 
\def\apspr{\aaref@jnl{Astrophys.~Space~Phys.~Res.}}
\def\bain{\aaref@jnl{Bull.~Astron.~Inst.~Netherlands}} 
\def\fcp{\aaref@jnl{Fund.~Cosmic~Phys.}}  
\def\gca{\aaref@jnl{Geochim.~Cosmochim.~Acta}}   
\def\grl{\aaref@jnl{Geophys.~Res.~Lett.}} 
\def\jcp{\aaref@jnl{J.~Chem.~Phys.}}      
\def\jgr{\aaref@jnl{J.~Geophys.~Res.}}    
\def\jqsrt{\aaref@jnl{J.~Quant.~Spec.~Radiat.~Transf.}}
\def\memsai{\aaref@jnl{Mem.~Soc.~Astron.~Italiana}}
\def\nphysa{\aaref@jnl{Nucl.~Phys.~A}}   
\def\physrep{\aaref@jnl{Phys.~Rep.}}   
\def\physscr{\aaref@jnl{Phys.~Scr}}   
\def\planss{\aaref@jnl{Planet.~Space~Sci.}}   
\def\procspie{\aaref@jnl{Proc.~SPIE}}   
\def\jcap{\aaref@jnl{J. Cosmology Astropart. Phys.}}

\let\astap=\aap
\let\apjlett=\apjl
\let\apjsupp=\apjs
\let\applopt=\ao

\newcommand{\etal}{et al.\ }

\newcommand{\mpc}{\, {\rm Mpc}}
\newcommand{\kpc}{\, {\rm kpc}}
\newcommand{\hmpc}{\, h^{-1} \mpc}
\newcommand{\ihmpc}{\, h\, {\rm Mpc}^{-1}}
\newcommand{\ikms}{\, {\rm s\, km}^{-1}}
\newcommand{\kms}{\, {\rm km\, s}^{-1}}
\newcommand{\hkpc}{\, h^{-1} \kpc}
\newcommand{\lya}{Ly$\alpha$\ }
\newcommand{\lyb}{Lyman-$\beta$\ }
\newcommand{\lyaf}{Ly$\alpha$ forest}
\newcommand{\lr}{\lambda_{{\rm rest}}}
\newcommand{\bF}{\bar{F}}
\newcommand{\bS}{\bar{S}}
\newcommand{\bC}{\bar{C}}
\newcommand{\bB}{\bar{B}}
\newcommand{\vdF}{{\mathbf \delta_F}}
\newcommand{\vdS}{{\mathbf \delta_S}}
\newcommand{\vdf}{{\mathbf \delta_f}}
\newcommand{\vdn}{{\mathbf \delta_n}}
\newcommand{\vdC}{{\mathbf \delta_C}}
\newcommand{\vdX}{{\mathbf \delta_X}}
\newcommand{\xrei}{x_{rei}}
\newcommand{\lrmin}{\lambda_{{\rm rest, min}}}
\newcommand{\lrmax}{\lambda_{{\rm rest, max}}}
\newcommand{\lmin}{\lambda_{{\rm min}}}
\newcommand{\lmax}{\lambda_{{\rm max}}}
\newcommand{\hi}{\mbox{H\,{\scriptsize I}\ }}
\newcommand{\heii}{\mbox{He\,{\scriptsize II}\ }}
\newcommand{\vp}{\mathbf{p}}
\newcommand{\vq}{\mathbf{q}}
\newcommand{\vxperp}{\mathbf{x_\perp}}
\newcommand{\vkperp}{\mathbf{k_\perp}}
\newcommand{\vrperp}{\mathbf{r_\perp}}
\newcommand{\vx}{\mathbf{x}}
\newcommand{\vy}{\mathbf{y}}
\newcommand{\vk}{\mathbf{k}}
\newcommand{\vR}{\mathbf{r}}
\newcommand{\tdtwo}{\tilde{b}_{\delta^2}}
\newcommand{\tstwo}{\tilde{b}_{s^2}}
\newcommand{\tbthree}{\tilde{b}_3}
\newcommand{\tadtwo}{\tilde{a}_{\delta^2}}
\newcommand{\tastwo}{\tilde{a}_{s^2}}
\newcommand{\tabthree}{\tilde{a}_3}
\newcommand{\vnabla}{\mathbf{\nabla}}
\newcommand{\tpsi}{\tilde{\psi}}
\newcommand{\vv}{\mathbf{v}}
\newcommand{\fnl}{{f_{\rm NL}}}
\newcommand{\tfnl}{{\tilde{f}_{\rm NL}}}
\newcommand{\gnl}{g_{\rm NL}}
\newcommand{\orderfour}{\mathcal{O}\left(\delta_1^4\right)}
\newcommand{\SDSSPF}{\cite{2006ApJS..163...80M}}
\newcommand{\PF}{$P_F^{\rm 1D}(k_\parallel,z)$}
\newcommand\ionalt[2]{#1$\;${\scriptsize \uppercase\expandafter{\romannumeral #2}}}%
\newcommand{\vxone}{\mathbf{x_1}}
\newcommand{\vxtwo}{\mathbf{x_2}}
\newcommand{\vRot}{\mathbf{r_{12}}}
\newcommand{\cm}{\, {\rm cm}}

\bibliographystyle{mnras}
\bibliography{dr12_bis.bib}


\appendix

\section{Impact of the systematic weights in the bispectrum}\label{appendixa}
In this appendix we aim to show the effect of applying the systematic weights correction to the bispectrum measurement. The effect on the power spectrum monopole and quadrupole are shown in appendix A of \citep{gil-marin15_rsd}. 
\begin{figure}
\centering
\includegraphics[scale=0.3]{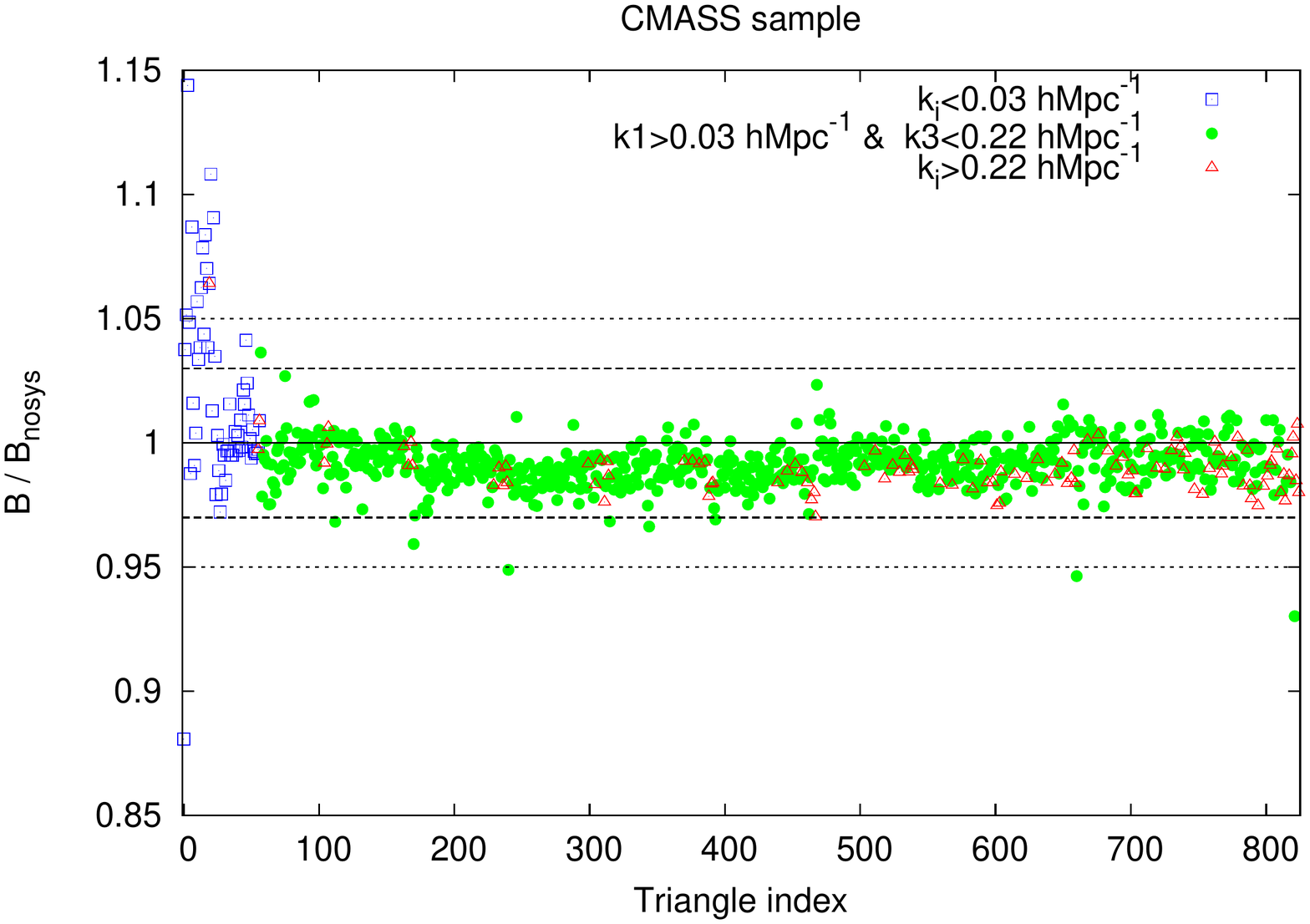}
\caption{Relative impact of the systematic weights in the isotropic bispectrum signal of the CMASS sample. The $y$-axis shows the ratio between the measured CMASS bispectrum with and without the systematic weights correction as a function of the triangle index. The different colours represent triangles corresponding to different scales: with one of its side $<0.03\,h{\rm Mpc}^{-1}$ in empty blue squares; with all their sides between $0.03\,h{\rm Mpc}^{-1}$ and $0.22\,h{\rm Mpc}^{-1}$  in green circles, and with all of its sides $>0.22\,h{\rm Mpc}^{-1}$ in red triangles. The black dashed and dotted lines mark the $3\%$ and $5\%$ deviation, respectively.}
\label{fig:sys}
\end{figure}
From the power spectrum results, we expect that the systematic correction has a strong effect at very large scales and decreases as we move to smaller scales. This behaviour is partly seen in Fig.~\ref{fig:sys}, where those triangles with at least one side corresponding to a large scale mode (blue squares) are more affected by the correction (up to $\sim10\%$), whereas those triangles with all of their sides corresponding to small scales modes (those points with a large triangle index number) are less affected. Among the triangles used for the RSD analysis (green symbols)  the correction due to the systematic weights is less than $3\%$, but in general we see that there is a remaining offset of around $\sim1\%$ that does not vanish at small scales. We also see that the systematic correction varies from shapes with similar $k$-vectors: the correction for triangles with similar triangle index can vary from 0 to 3\%, and in some few cases up to $5\%$. The correction that the systematic weights produce in the bispectrum is therefore larger than the observed in the power spectrum. However, on has to bear in mind that the statistical errors in the bispectrum are also larger than those observed in the power spectrum. 
\begin{figure}
\centering
\includegraphics[scale=0.3]{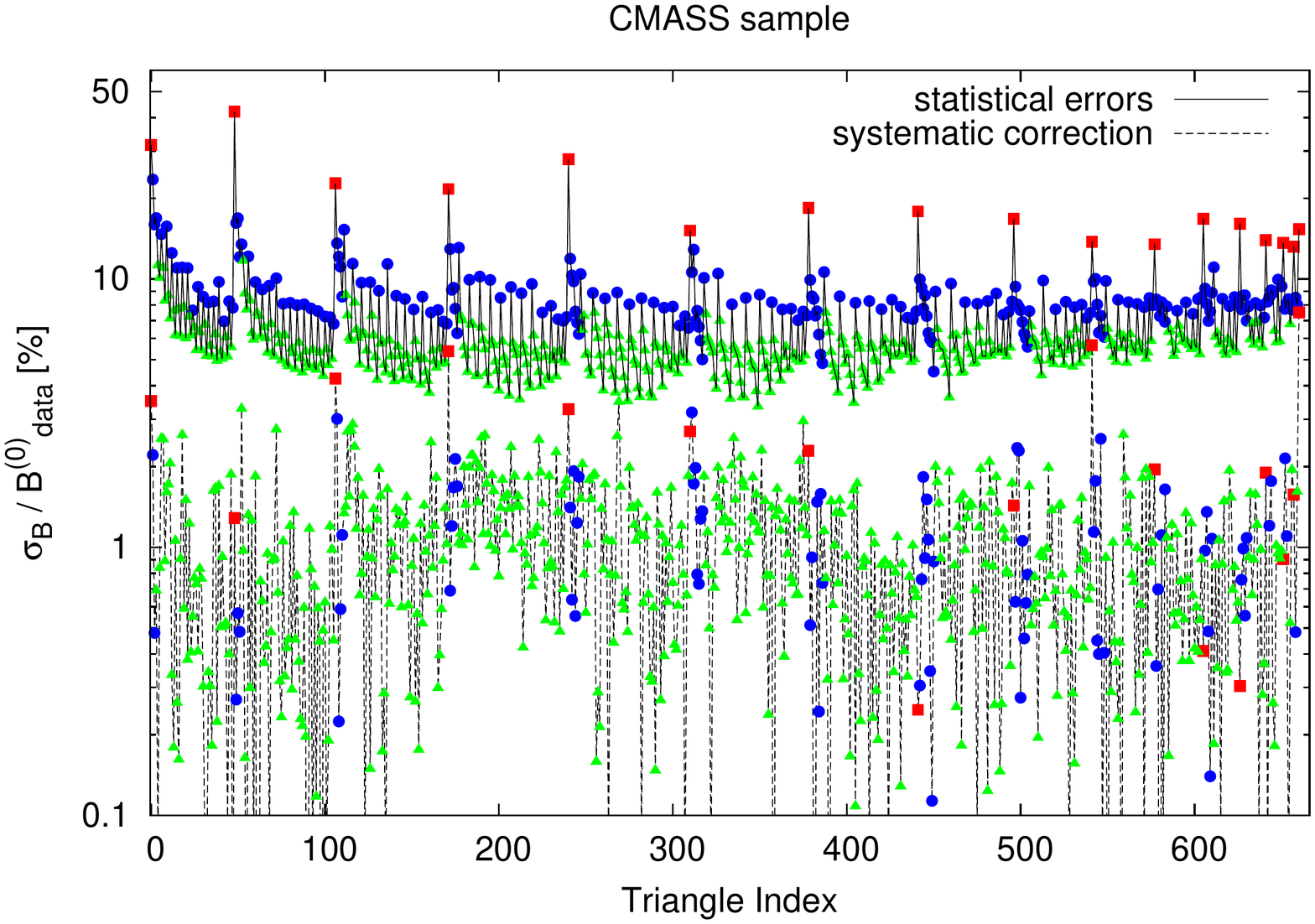}
\caption{Relative impact of systematic weights relative to the statistical errors for the CMASS sample. The black solid line shows the diagonal errors ({\it rms} from the MD-Patchy mocks) of the isotropic bispectrum with respect the systematic correction:$1-B_{\rm nosys}/B_{\rm sys}$ in dashed lines, both in percent deviation. On the top of the lines a colour symbol according to the shape of the triangle is displayed: red squares for equilateral, blue circles for isosceles and green triangles for scalene.}
\label{fig:sys2}
\end{figure}
This is shown in Fig.~\ref{fig:sys2}, where the black solid line stands for the relative statistical errors and the dashed black line for the relative systematic correction, corresponding to the bispectrum whose triangles have $0.03\,h{\rm Mpc}^{-1}\leq k_i\leq 0.22\,h{\rm Mpc}^{-1}$ (which corresponds to the green symbols in Fig.~\ref{fig:sys}). On the top of both lines we plot a colour symbol according to the shape of the triangle: red squares for equilateral, blue circles for isosceles and green triangles for scalene. The systematic correction is always below the statistical error and usually is $\sim5$ times smaller. We note that the most sensitive shape to the systematic correction is the equilateral, which is also the shape with higher statistical error as predicted by perturbation theory (see section A2 of \citealt{Scoccimarroetal:1998} for more details). 

Therefore, we conclude that for the triangles used for the analysis presented in this paper, $0.03\,h{\rm Mpc}^{-1}\leq k_i\leq 0.22\,h{\rm Mpc}^{-1}$, the systematic weights do not have a large impact on the bispectrum signal given the statistical uncertainty, and therefore,  the impact of inaccuracies in the systematic correction has a minor effect on the estimation of parameters.

\section{Impact of close pairs weights in the bispectrum and best-fitting parameter estimation}\label{appendixb}
In this appendix we study the effect of fibre collisions weights on the isotropic bispectrum, and more precisely on the $f\sigma_8$ measurements when the power spectrum monopole, quadrupole and bispectrum monopole are considered, as it is the case in \S \ref{sec:results}. The fibre collision weights are included in order to account for those galaxies that are too close to each other ($<62''$) to put two or more fibres. Details about the fraction of collided galaxies in the dataset and in the mocks can be found in table B1 of \cite{gil-marin15_rsd}. In short we say that the collided fraction is about 3.5 times higher in  the CMASS sample than in the LOWZ sample. Therefore all the upper limits found for CMASS are automatically  equal or lower for the LOWZ sample. Also, the fraction of collided galaxies in the CMASS sample of the  \textsc{MD-Patchy} mocks has been found to be lower than the one from the data, whereas the \textsc{qpm} mocks have a same fraction than in the data. This is due to a resolution problem for the \textsc{MD-Patchy} haloes that is expected to be solved in future version of the mocks. Because all this, we perform a test using only the CMASS sample of the \textsc{qpm} mocks. 

\begin{figure*}
\centering
\includegraphics[scale=0.3]{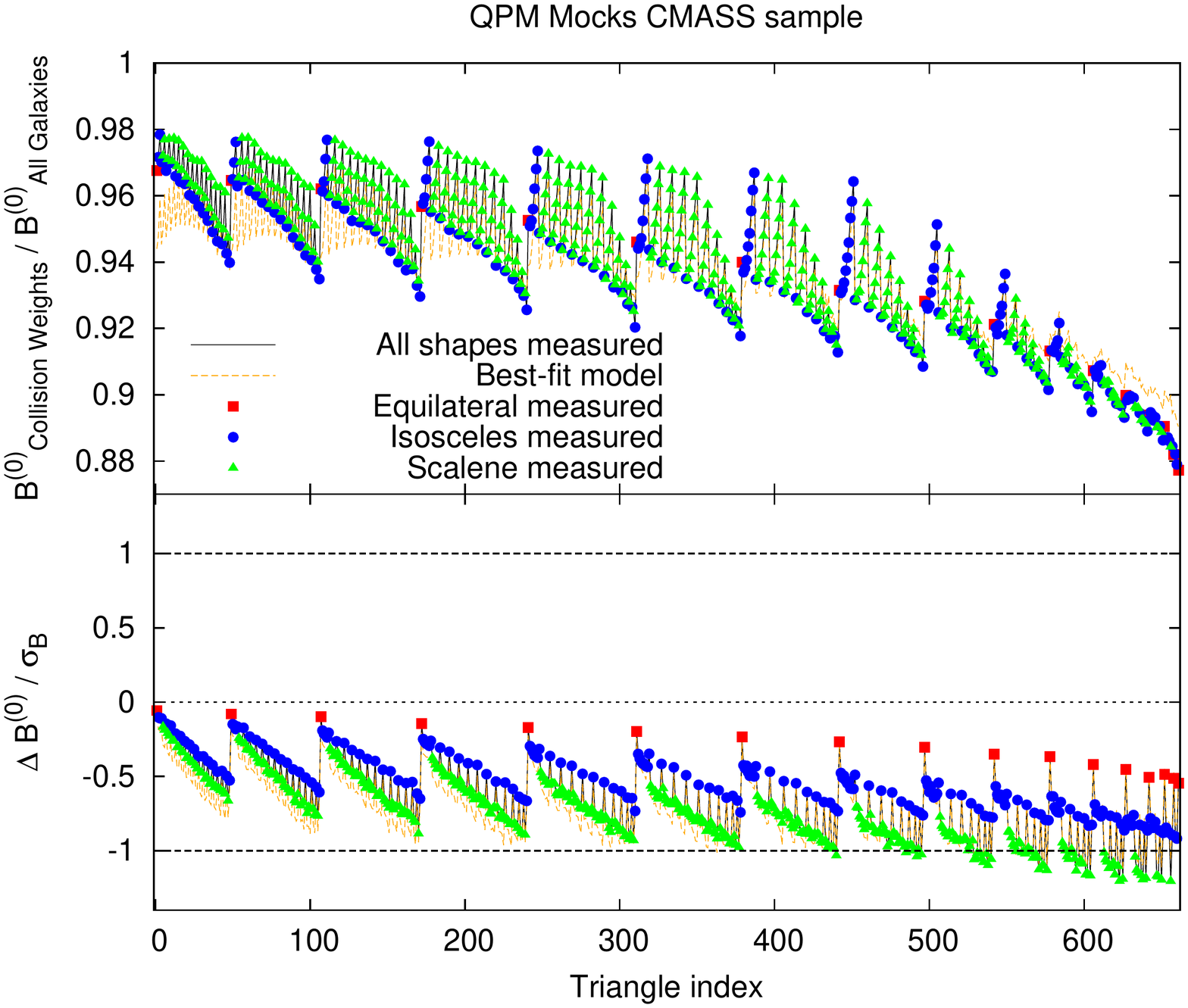}
\includegraphics[scale=0.3]{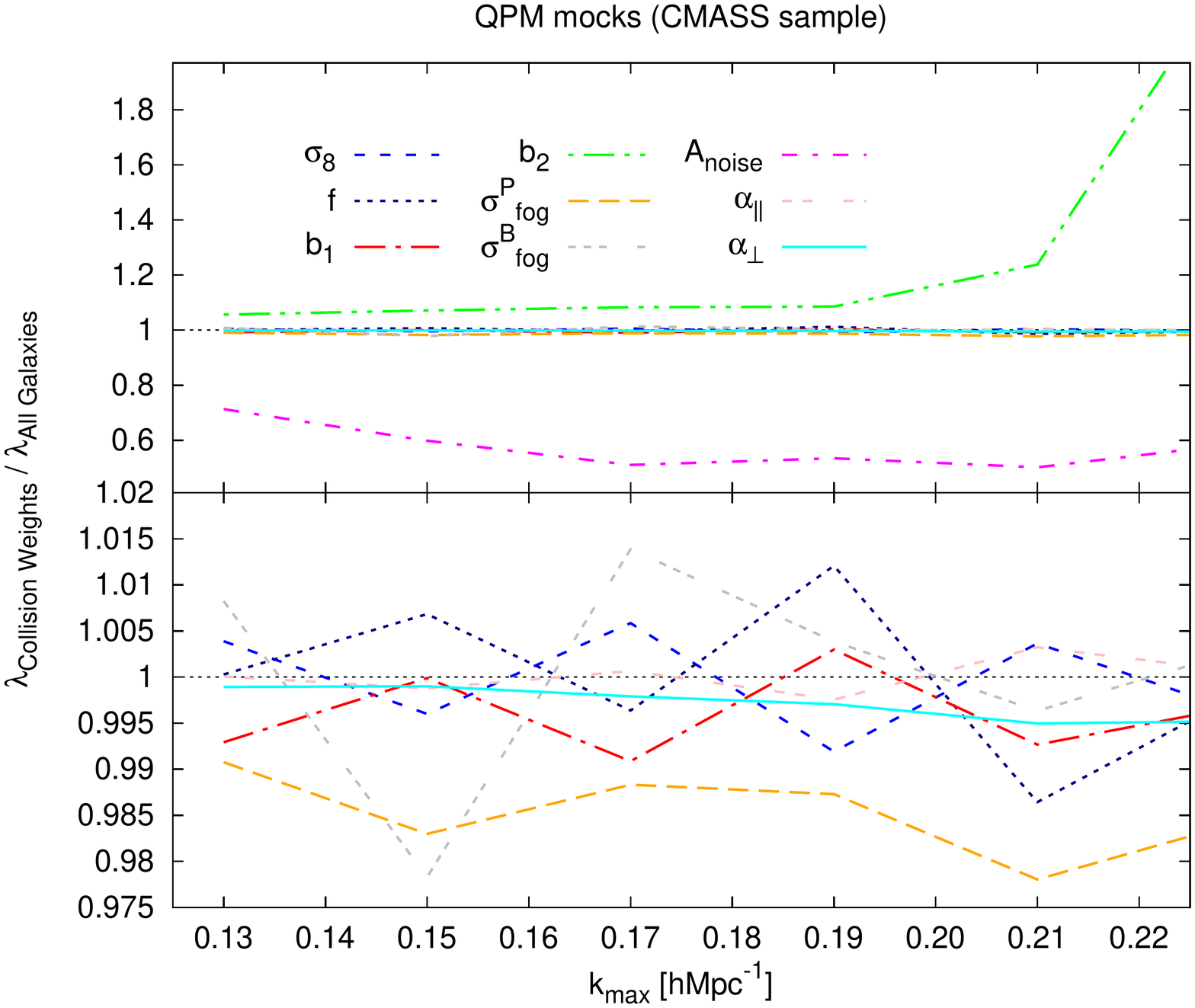}
\caption{Effect of fibre collision weights in the \textsc{qpm} CMASS mocks. The left top sub-panel display the ratio between the bispectrum monopole of cases i) and ii) (see the text for description) as a function of the triangle index for those triangles with $0.03\,h{\rm Mpc}^{-1}\leq k_i\leq 0.22\,h{\rm Mpc}^{-1}$. The black solid lines are the ratio of the bispectrum of the measured galaxies, whereas the orange dashed line is the ratio between the best-fitting theoretical models to the cases i) and ii). The coloured symbols represent the triangle shape as indicated. The left bottom sub-panel displays the difference between the same quantities of the top panel divided by the statistical error expected for the data ({\it rms} of the mocks). The right panel displays the ratio of the best-fitting variables for the cases i) and ii) as a function of the truncation scale, $k_{\rm max}$, where the bottom sub-panel has been simply zoomed in with respect to the top sub-panel. Table~\ref{table:cp} summarises the observed shifts on the variables of cosmological interest for $k_{\rm max}=0.22\,h{\rm Mpc}^{-1}$. }
\label{fig:wcol}
\end{figure*}

\begin{table*}
\begin{center}
\begin{tabular}{|c|c|c|cc}
Parameter & statistical error [\%]  &  systematics of the model [\%] & systematics due to fibre collisions [\%]   \\
\hline
\hline
$f\sigma_8$ & $ 6\%\, /\,  4\%$  & $5.5\%$ & $\lesssim1\%$  \\
$\alpha_\parallel$ & $ 1.7\%\, /\, -$ & $1\%$ & $\lesssim0.5\%$  \\
$\alpha_\perp$ & $ 1.2\%\, /\, -$ & $1\%$ & $\lesssim0.1\%$ \\
$\sigma_8$ & $ 8\%\, /\,  6\%$ & $8\%$ & $\lesssim0.5\%$  \\
$f$ & $ 19\%\, / \,  12\%$ &$4\%$ & $\lesssim1\%$  \\
\end{tabular}
\end{center}
\caption{Summary of the statistical errors and systematic shifts caused by both the fibre collision weights and the systematics of the model in the parameters of cosmological interest for the model of \S\ref{sec:model}. All the results for the CMASS sample at $k_{\rm max}=0.22\,h{\rm Mpc}^{-1}$. For all the cosmological parameters the systematic shifts are sub-dominant and the total error budget is dominated by the statistical errors. In the ``statistical error''  column, the values separated by the dash symbol correspond to the relative errors when the AP test is performed and not performed, respectively. }
\label{table:cp}
\end{table*}%

In order to test the effect of the fibre collision in the bispectrum and in the parameter estimation we measure the power spectrum monopole, quadrupole and bispectrum monopole of 1000 \textsc{qpm} mock realisations and take their mean in order to reduce the statistical uncertainty of modes. We proceed in two ways,
\begin{enumerate}

\item We treat the galaxies as in the real data: when two or more galaxies present an angular separation of $<62''$, on the them is weighted by the number of galaxies within the $<62''$ angular radius, and the rest is weighted by 0. By doing this we mimic what it is done in the real dataset. 

\item We consider all the galaxies resolved in the mocks and weight them equally. This correspond to the case we would have if all the targeted galaxies in the survey where analysed spectroscopically. 

\end{enumerate}

The case ii) has the correct clustering and anisotropic signal, which we use as a reference to test the effects of fibre collision in case i). Therefore, we need to analyse the power spectrum and bispectrum of \textsc{qpm} mocks under these two conditions. We fit the 9 parameter model of \S\ref{sec:model} to the power spectrum monopole, quadrupole and bispectrum monopole using the same technique described in  \S\ref{sec:tests}\footnote{For simplicity we apply the covariance extracted from the MD-\textsc {Patchy} mocks to the \textsc{qpm} mocks}. By comparing the best-fitting parameters obtained in the cases i) and ii) we can quantify by how much every specific parameter is affected by the fibre collision weights.

The left top sub-panel of Fig.~\ref{fig:wcol} displays in black solid lines the ratio between the measured isotropic bispectrum of cases i) and ii) as a function of the triangle index. The coloured symbols represent the different triangle shapes: equilateral, isosceles and scalene as indicated. The orange pink dashed line represents the ratio between the models that better describe the power spectrum monopole, quadrupole and bispectrum monopole of cases i) and ii). The ratio between the corresponding best-fitting parameters is shown in the different colour lines of the right panel of Fig.~\ref{fig:wcol}. In the left bottom sub-panel of Fig.~\ref{fig:wcol} we show the corresponding difference between measured isotropic bispectrum of cases i) and of case ii), $\Delta B^{(0)}$, relative to the statistical error predicted by the {\it rms} of the 2048 realisations of the MD-\textsc{Patchy} mocks. 

We see that the collision weights has an effect of $\sim2\%$ to $\sim6\%$ on the bispectrum amplitude at large scales, which increases up to $12\%$ in those triangles with their $k$-vectors close to $0.22\,h{\rm Mpc}^{-1}$. We also see that at small scales the correction is barely independent of the triangle shape, whereas at large scales the dependence is of $1-2\%$, being the equilateral and isosceles shapes the most affected, and some scalene the less affected. In terms of the statistical error expected for the bispectrum data, the shift between the cases i) and ii) oscillates between 0 and $1\sigma$, depending strongly on the shape and scale of each triangle. However, we see that the best-fitting model to the bispectrum of cases i) and ii) is able to account for these effect by modifying the value of the best-fitting parameters of the model.

The right panel of Fig.~\ref{fig:wcol} display the effect of the fibre collision weights on the variables of the model presented in \S\ref{sec:model}. The most affected variable by the collision weights is the deviation from the Poissonian shot noise, $A_{\rm noise}$, which is reduced by $\sim40\%$, followed by the $b_2$, which increases by a factor that strongly depends on the truncation scale, $k_{\rm max}$, and the $\sigma^P_{FoG}$, which is reduced by $\sim2\%$. The rest of the variables are considerably less affected. In particular $f$ and $\sigma_8$ are modified by $\sim1\%$ and can be either increased or reduced depending on $k_{\rm max}$. $b_1$ is typically reduced by less than $0.5\%$, as well as the AP parameters, $\alpha_\parallel$ and $\alpha_\perp$. Table~\ref{table:cp} summarises the typical systematic shift due to fibre collision on the cosmology variables and compare it with the statistical error of the data as well as the systematic errors of the model.

\section{Covariance matrix for the combined data-set vector}\label{appendix_matrix}
In this appendix we report the $8\times8$ full covariance matrix corresponding to the data-vectors displayed by Eq. \ref{datatotlowz} and \ref{datatotcmass}. Eq. \ref{covtotlowz} and \ref{covtotcmass} display these matrices for the LOWZ and CMASS samples, respectively.

These matrices are generated using the correlation coefficients extracted from the 2048 MD-\textsc{Patchy} mocks,  displayed in Fig.~\ref{fig:covariance_parameters}, in combination with the variance elements for each individual parameter extracted from the likelihood of the data: for the {\it RSD P} analysis parameters this correspond to the diagonal elements of eq 22 and 23 of \cite{gil-marin15_rsd}; for the {\it RSD P+B} analysis to the diagonal elements of the matrices of Eq. \ref{cov1} and \ref{cov2}; for the BAO parameters to the error-bars displayed in table 4 of \cite{gil-marin15_bao} (only those extracted from the MD-\textsc{Patchy} mocks).

  \begin{widetext}
   \begin{equation}
 \label{covtotlowz}
C^{{\rm LOWZ}} = 10^{-3}
 \begin{pmatrix}
4.5028 & 1.3845 & 32.764 & 5.1012 & 4.7978 & 11.179 & 3.8316 & 3.1742\\
 & 5.0837 & 7.0735 & 23.462 & 4.2329 & 3.4808 & 10.699 & 1.9994\\
 &  & 330.03 & 81.557 & 69.444 & 78.104 & 26.945 & 17.117\\
 &  &  & 300.30 & 52.598 & 16.632 & 81.195 & 8.3435\\
 &  &  &  & 302.87 & 12.142 & 16.195 & 41.092\\
 &  &  &  &  & 36.239 & 15.720 & 12.124\\
 &  &  &  &  &  & 47.493 & 9.9699\\
 &  &  &  &  &  &  & 22.203

 \end{pmatrix},
 \end{equation}
\begin{equation} 
\label{covtotcmass}
C^{{\rm CMASS}} = 10^{-3}
 \begin{pmatrix}
 1.7424 & 0.27016 & 15.841 & 1.1124 & 2.9037 & 5.6326 & 1.1388 & 1.7120\\
 & 1.3046 & 1.8199 & 7.9190 & 0.65478 & 1.0489 & 4.6649 & 0.39842\\
 &  & 206.68 & 22.625 & 50.092 & 50.625 & 10.580 & 11.358\\
 &  &  & 77.713 & 10.091 & 6.2962 & 33.648 & 1.3916\\
 &  &  &  & 126.37 & 10.626 & 3.9494 & 17.127\\
 &  &  &  &  & 26.094 & 7.0482 & 10.207\\
 &  &  &  &  &  & 21.700 & 3.2283\\
 &  &  &  &  &  &  & 16.071\\

 \end{pmatrix}.
\end{equation}
\end{widetext}

We conclude that the effect of fibre collisions are absorbed mainly by $A_{\rm noise}$ and $b_2$. Because of this, the systematic shifts on the parameters of cosmological interest, such as $f$, $\sigma_8$, $\alpha_\parallel$ and $\alpha_\perp$ is reduced to percent or sub-percent. Since the effects of the fibre collisions are sub-dominant we  do not consider to add these shifts to the total error budget in \S\ref{sec:results}, as they would not produce any significant change.

 \section{Breaking the degeneracy among \lowercase{$b_1$}, \lowercase{$f$} and $\sigma_8$}\label{appendixc}
As a proof of principle of how the large scale degeneration between $f$ and $\sigma_8$ (as well as $b_1$ and $\sigma_8$) is broken we take a toy model where $b_1\sigma_8$ and $f\sigma_8$ are kept fixed under changes of $f$, $\sigma_8$ and $b_1$. For simplicity we set $b_2=0$ as well the fingers of God parameters and the amplitude of noise is set to poisson prediction. We take $b_1=f\equiv x$, whereas $\sigma_8\equiv 1/x$, and therefore $f\sigma_8$ and $b_1\sigma_8$ are set to 1. We aim to see the effect on the amplitude and shape on the power spectrum and bispectrum predicted by the model of \S\ref{sec:model} when $x$ is varied.
\begin{figure}
\centering
\includegraphics[scale=0.3]{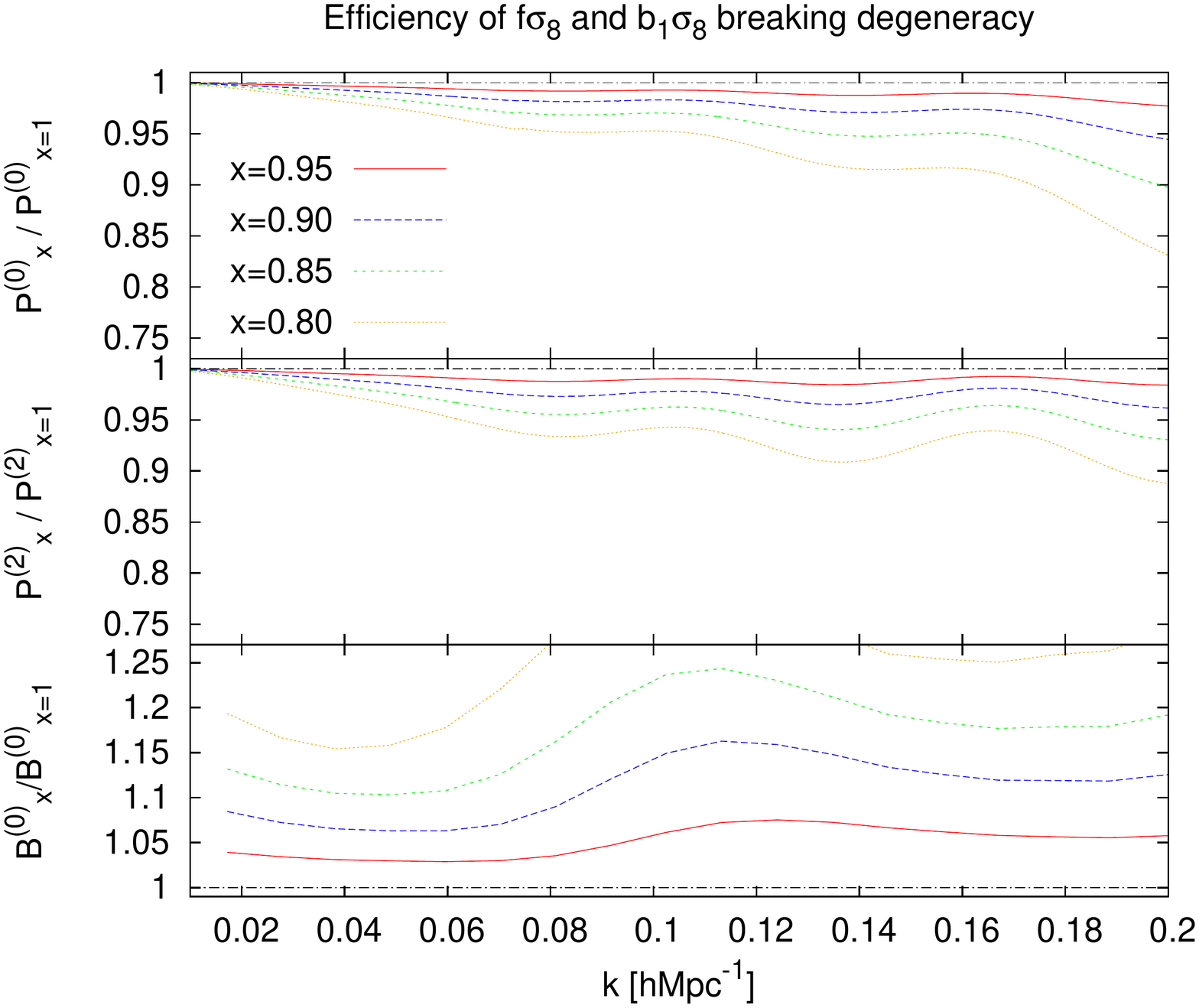}
\caption{Relative signal of the power spectrum monopole (top panel), quadrupole (middle panel) and equilateral bispectrum (bottom panel) according to the model presented in \S\ref{sec:model}, for different values of $x$ (where $x\equiv f=b_1=\sigma_8^{-1}$), relative to the signal when $x=1$. Red, blue, green and orange lines display the results for $x=0.95,\, 0.90,\, 0.85,\, 0.80$, respectively. Whereas the power spectrum multipoles are very degenerated under changes of $x$ (especially at large scales), the bispectrum signal is much more sensitive to such changes.}
\label{fig:fs8degbreak}
\end{figure}
This is shown in Fig.~\ref{fig:fs8degbreak}, where the top, middle and bottom panel show the amplitude of the power spectrum monopole, quadrupole and equilateral bispectrum monopole, respectively,  relative to the amplitude of the corresponding statistic when $x=1$. The different colours show the results for different values of $x$, as indicated. For clarity, the scales in the $y$-axis of the 3 sub-panels have been kept the same, so the relative change in the power spectra multipoles and on the bispectrum can be directly compared. At large scales we see how the changes on $x$ do not produce any significant change on the amplitude of the power spectrum multipoles.  This is because the model described in \S\ref{sec:model} tends to the Kaiser prediction at these scales, where $f$ and $\sigma_8$ are perfectly degenerated. On the other hand, this is not the case for the amplitude of the bispectrum where even at large scales the amplitude of the equilateral bispectrum is sensitive to different values of $x$. As we explore smaller scales the amplitude of the power spectrum multipoles start to be sensitive to changes of $\sigma_8$, whereas for the equilateral bispectrum is similar to its large scale value. For example for $x=0.90$, the amplitude in the power spectrum monopole and quadruple at $k=0.20\,h{\rm Mpc}^{-1}$ changes just by 5\% and 4\%, respectively, relative to that with $x=1$. For the equilateral bispectrum this change reaches $15\%$ at $k=0.12\,h{\rm Mpc}^{-1}$ and $\geq10\%$ for $k\geq 0.08\,h{\rm Mpc}^{-1}$.  In practice the degeneracy between $f$ and $\sigma_8$ cannot be broken only using the power spectrum at small scales because at such scales the amplitude and shape of the power spectrum is sensitive to other variables that here we have set constant for simplicity, such as, $b_2$ and the Fingers-of-God dispersion parameter. However, using the bispectrum signal allows us split $f$ and $\sigma_8$ into two independent (and correlated) variables, as it is shown in \S\ref{sec:degeneracy}.

\bsp	
\label{lastpage}
\end{document}